\documentclass[12pt]{article}
\usepackage{cite}
\usepackage{amsmath, amsthm, amssymb,slashed,upgreek,url}
\usepackage{ifpdf}
\ifpdf
  \usepackage[pdftex]{graphicx}
  \usepackage{epstopdf}
\else  
  \usepackage[dvips]{graphicx}
\fi
\usepackage[utf8]{inputenc}
\usepackage[T1]{fontenc}
\parskip=\baselineskip
\def\Bbb{\mathbb}
\def\Tr{{\rm Tr}}
\def\16{{\bf 16}}
\def\S{S}
\def\1{{\bf 1}}
\def\2{{\bf 2}}
\def\3{{\bf 3}}
\def\4{{\bf 4}}

\def\sv{{\sf v}}
\def\sx{{\sf x}}
\def\y{{\sf y}}
\def\I{{\mathcal I}}

\def\\SPT{\sf{\SPT}}
\def\a{{\sf a}}
\def\sv{{\sf v}}

\def\D{{\sf Y}}
\def\D{{\sf D}}
\def\DD{{\mathcal D}}

\def\ds{{\mathrm{ds}}}

\def\veps{\varepsilon}

\def\tr{{\mathrm{tr}}}
\def\\SRT{{\sf{\SRT}}}
\def\CC{\mathcal C}
\def\h{\widehat}
\def\bar{\overline}

\def\ra{\rangle}

\def\bp{\begin{pmatrix}}
\def\ep{\end{pmatrix}}
\def\la{\langle}
\def\R{{\Bbb{R}}}\def\Z{{\Bbb{Z}}}

\def\hat{\widehat}
\font\teneurm=eurm10 \font\seveneurm=eurm7 \font\fiveeurm=eurm5
\newfam\eurmfam
\textfont\eurmfam=\teneurm \scriptfont\eurmfam=\seveneurm
\scriptscriptfont\eurmfam=\fiveeurm

\font\teneusm=eusm10 \font\seveneusm=eusm7 \font\fiveeusm=eusm5
\newfam\eusmfam
\textfont\eusmfam=\teneusm \scriptfont\eusmfam=\seveneusm
\scriptscriptfont\eusmfam=\fiveeusm

\font\tencmmib=cmmib10 \skewchar\tencmmib='177
\font\sevencmmib=cmmib7 \skewchar\sevencmmib='177
\font\fivecmmib=cmmib5 \skewchar\fivecmmib='177
\newfam\cmmibfam
\textfont\cmmibfam=\tencmmib \scriptfont\cmmibfam=\sevencmmib
\scriptscriptfont\cmmibfam=\fivecmmib

\numberwithin{equation}{section}

\def\d{\mathrm d}

\def\L{{\mathcal L}}
\def\\S{{\Bbb \S}}
\def\Z{{\Bbb Z}}
\def\bar{\overline}
\def\A{{\mathcal A}}
\def\min{{\mathrm{min}}}
\def\bar{\overline}

\begin{document}
\begin{titlepage}
\begin{flushright}
\end{flushright}
\vskip 1.5in
\begin{center}
{\bf\Large{Light Rays, Singularities, and All That}}
\vskip
0.5cm {Edward Witten} \vskip 0.05in {\small{ \textit{School of
Natural Sciences, Institute for Advanced Study}\vskip -.4cm
{\textit{Einstein Drive, Princeton, NJ 08540 USA}}}
}
\end{center}
\vskip 0.5in
\baselineskip 16pt
\abstract{This article is an introduction to causal properties of General Relativity.   Topics include the Raychaudhuri equation,
singularity theorems of Penrose and Hawking,  the black hole area theorem, 
topological censorship, and the Gao-Wald theorem.  The article is based on lectures
at the 2018  summer program Prospects in Theoretical Physics at the Institute for Advanced Study, and also at the 2020 New Zealand Mathematical Research Institute summer
school in Nelson, New Zealand.}
\date{May, 2018}
\end{titlepage}
\def\Hom{\mathrm{Hom}}
\def\H{{\mathcal H}}
\def\d{{\mathrm d}}
\def\t{\widetilde}
\def\U{{\mathcal U}}
\def\UU{{\mathrm U}}
\def\V{{\mathcal V}}
\def\st{{\sf t}}
\def\O{{\mathcal O}}
\def\i{{\mathrm i}}
\def\A{{\mathcal A}}
\def\be{\begin{equation}}
\def\ee{\end{equation}}

\tableofcontents

\def\aa{{\sf a}}
\def\bb{{\sf b}}
\def\a{{\bf a}}
\def\x{{\bf x}}
\def\y{{\bf y}}
\def\z{{\bf z}}
 \def\sym{{\mathrm{sym}}}
 \def\zotimes{{\otimes N}}
 \def\cl{{\mathrm{cl}}}
\def\h{\widehat}
\def\t{\widetilde}
\def\tt{\sf t}
\def\tr{{\mathrm{tr}}}
\def\d{{\mathrm{d}}}
\def\H{{\mathcal H}}
\def\O{{\mathcal O}}
\def\Tr{{\mathrm{Tr}}}
\def\diag{{\mathrm{diag}}}
\def\la{\langle}
\def\ra{\rangle}

\section{Introduction}\label{intro}

Here are some questions about the global properties of classical General Relativity:
\begin{enumerate}
\item Under what conditions can one predict the formation of a black hole?
\item Why can the area of a classical black hole horizon only increase?
\item Why, classically, is it not possible to travel through a ``wormhole'' in spacetime?
\end{enumerate}

These are questions of Riemannian geometry in Lorentz signature.
They involve the causal structure of spacetime: where can one get, from a given starting point, along a worldline
that is everywhere within the local light cone?

 Everyday life gives us some intuition about Riemannian
geometry in the ordinary case of Euclidean signature.   We live in  three-dimensional Euclidean space, to an excellent
approximation, and we are quite familiar 
 with two-dimensional curved surfaces.   A two-dimensional curved surface is a reasonable
prototype for Riemannian geometry in Euclidean signature, though Riemannian geometry certainly has important features that only
appear in higher dimension.

By contrast, everyday life  does not fully prepare us for   Lorentz signature geometry.   What is fundamentally different about Lorentz signature is the constraint
of causality:  a signal cannot travel outside the light cone.  Because the speed of light is so large on a human scale, the constraints of relativistic causality
are not apparent in everyday life.

These constraints are most interesting in the context of gravity.   Black holes -- regions of spacetime from which no signal can escape
to the outside world -- provide a dramatic manifestation of how the constraints of relativistic causality play out in the context of General Relativity.

The field equations of classical General Relativity are notoriously difficult, nonlinear equations, from which it can be hard to extract insight.
But it turns out that by rather simple arguments involving
 a fascinating interplay of causality, positivity of energy, and the Einstein equations,  it is possible to gain a great deal of qualitative understanding
of cosmology, gravitational collapse, and spacetime singularities.

The aim of the present article is to introduce this subject as readily as possible, with a minimum of formalism.  
 This will come at the cost
of cutting a few mathematical corners and omitting  important details -- and further results -- 
 that can be found in more complete treatments.    Several classical accounts of this material were written by the original pioneers,
  including Penrose \cite{Penrose} and Hawking \cite{HawkingEllis}.   Some  details and further results omitted in the present article can be found in
   the classic textbook by Wald \cite{Wald}, especially chapters 8, 9, and 12.   Several helpful sets of lecture notes are  \cite{Galloway, Chrusciel, Aretakis}.    A detailed mathematical
   reference on Lorentz signature geometry is \cite{BEE}.

 To a reader wishing to become familiar with these matters, one can offer some
good news:   there are many interesting results, but  a major role is played by a few key ideas that date back to the 1960's.
  So one can become  conversant with a significant body of material in a relatively short span of time.

The article is organized as follows.
Some basics about causality are described in sections \ref{classic} and \ref{glohyper}.  In section \ref{geofocal}, we explore the properties of timelike geodesics and
navigate towards  what is arguably
the most easily explained singularity theorem,
namely a result of Hawking about the Big Bang \cite{Hawkingsing}.   In  section \ref{ng}, we analyze the somewhat subtler problem of null geodesics  and present the original and most important modern
singularity theorem, which is  Penrose's theorem about gravitational collapse \cite{Penrosesing}.  
Section \ref{blackholes} describes some basic properties of black holes which can be understood once one is familiar
with the ideas that go into Penrose's proof.  A highlight is the Hawking area theorem \cite{Hawkingarea}. Section \ref{addtop} is devoted
to some additional matters, notably topological censorship \cite{topocensor,topocensor2}, the Gao-Wald theorem \cite{Gao-Wald}, and their extension to the case that one
assumes an averaged null energy condition (ANEC) rather than the classical null energy condition.    Finally, in section \ref{newlook}
we re-examine null geodesics in a more precise way, with fuller explanations of some important points from section \ref{ng} plus some further results.

For the most part, this article assumes only a basic knowledge of General Relativity.   
At some points, it will be helpful to be familiar with the Penrose diagrams of some standard spacetimes.
The most important example -- because of its role in motivating Penrose's work --  is simply the Schwarzschild solution.
The other examples that appear in places 
(Anti de Sitter space, de Sitter space, and the Reissner-Nordstr\"{o}m solution) provide illustrations that actually can be omitted on first reading; the article will remain comprehensible.  
Much of the relevant background to the Anti de Sitter and de Sitter
examples is explained in Appendices \ref{ads} and \ref{dSitter}.

\section{Causal Paths}\label{classic}

To understand the causal structure of spacetime,\footnote{ \label{recall} In this article, a ``spacetime'' $M$ is a $\D$-dimensional manifold,
 connected, with a smooth Lorentz signature metric.  We assume that $M$ is time-oriented, which means
that at each point in $M$, there is a preferred notion of what represents a ``future'' or ``past'' timelike direction.   We allow arbitrary $\D$ -- rather than specializing to $\D=4$ -- since
this does not introduce any complications.   It is interesting to consider the generalization to arbitrary $\D$, since any significant dependence on $\D$
might shed light on why we live in $\D=4$, at least macroscopically.   Moreover, 
 the generalization to arbitrary $\D$ is important in contemporary research on quantum gravity.}
 we will have to study causal paths.   We usually describe a path in parametric form as 
  $x^\mu(s)$, where $x^\mu$ are local coordinates in spacetime and $s$ is a parameter. (We require that the tangent vector $\d x^\mu/\d s$ is nonzero for all $s$, and  we consider two paths to be equivalent if they differ only by a reparametrization $s\to \t s(s)$.)   A path $x^\mu(s)$ is causal if its tangent vector $\frac{\d x^\mu}{\d s}$ is everywhere timelike or null. 
 
 \begin{figure}
 \begin{center}
   \includegraphics[width=2in]{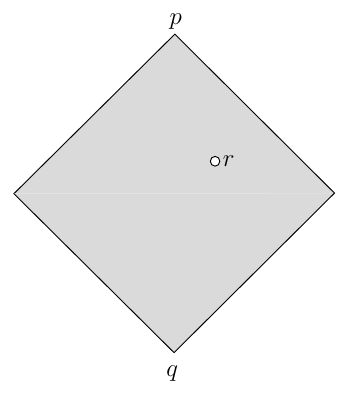} 
 \end{center}
\caption{\small  For two spacetime points $q,p$, with $p$ to the future of $q$, the causal diamond $D_q^p$ consists of all points
that are in the causal future of $q$ and the causal past of $p$.   
Sometimes it is of interest to consider a spacetime in which a point $r$ is omitted from the causal diamond, as sketched here.  \label{Fig2}}
\end{figure}
    
We will often ask questions along the lines of ``What is the optimal causal path?'' for achieving some
given objective.   For example, what is the closest one can come to escaping from the black hole or traversing the wormhole?
The answer to such a question usually involves a geodesic with special properties.   So geodesics will play an important role.

We start simply by considering causal paths in Minkowski space  from a point $q$ to a point $p$ in its causal future (the points inside  or on the future light cone of $q$).
Such a path will lie within a subset of spacetime that we will call the ``causal diamond'' $D_q^p$ (fig. \ref{Fig2}).  This diamond
  is the intersection of the causal
future of\footnote{\label{detail} An important detail -- here and 
 later -- is that we consider $q$ itself to be in its own causal future (or past).
 Thus $q$ (or $p$) is contained in $D_q^p$.   Related to this, in the definition of a causal path, we allow the case of a trivial
 path that consists of only one point.  The purpose of this is to simplify various statements about closedness or compactness.} $q$  with the causal past of $p$.

The first essential point is that the space of causal paths from $q$ to $p$ is in a suitable sense compact.  
Causality is essential here.  Without it, a sequence of paths, even if confined to a compact region of spacetime (like $D_q^p$),
could oscillate more and more wildly, with no convergent subsequence.   For example, in two-dimensional Minkowski space $M$ with
metric $\mathrm{ds}^2=-\d t^2+\d x^2$ (where we will sometimes write $\x$ for the pair $(t,x)$), here is a sequence of non-causal paths\footnote{To put these paths in parametric
form, one would simply write $t(s)=s$, $x(s)=\sin (\pi n s)$.} from $q=(0,0)$ to $p=(1,0)$:
\be\label{first} x=\sin(\pi n t),~~~~0\leq t\leq 1.\ee      Though restricted to a compact portion of $M$,
these paths oscillate more and more wildly with no limit for $n\to\infty$.  Taking a subsequence does not help, so the space of all paths from
$q$ to $p$ is not compact in any reasonable sense.

Causality changes things because  it gives a constraint $|\d  x/\d t|\leq 1$.  
To understand why this leads to compactness,  it is convenient to flip the sign of the $\d t^2 $	term in the metric (in our chosen coordinate system) and define
the Euclidean signature metric 
\be\label{second} \ds_E^2=\d t^2+\d x^2. \ee    
A straight line from $q=(0,0)$ to $p=(1,0)$ has Euclidean length 1, and an arbitrary causal path between those points has Euclidean length no more than $\sqrt 2$.
(The reader should be able to describe an example of a causal path of maximal Euclidean length.)

Once we have an upper bound on the Euclidean length, compactness follows.   Parametrize a causal path of Euclidean
length $\lambda \leq \sqrt 2$ by a parameter $s$ that measures  the arclength divided by $\lambda$, so $s$ runs from 0 to 1.  Suppose
that we are given a sequence of such causal paths $\x_n(s), ~~n=1,2,3,\cdots$.   Since each of these paths begins
at $q$, ends at $p$, and has total Euclidean length $\leq \sqrt 2$, there is a compact subset $D$ of Minkowski space that contains all of them.

The existence of a convergent subsequence of the sequence of paths $\x_n(s)$ follows by an argument that might be familiar.  First consider what happens at $s=1/2$.  
Since $\x_n(1/2)$ is contained, for each $n$, in the compact region $D$, there is a subsequence of the paths $\x_n(s)$ such that $\x_n(1/2)$ converges
to some point in $D$.  Extracting a further subsequence, one ensures that $\x_n(1/4)$ and $\x_n(3/4)$ converge.   Continuing in this way, one ultimately
extracts a subsequence of the original sequence with the property  that $\x_n(s)$ converges for any rational number $s$ whose denominator is a power of 2.
 The bound on the length ensures that wild fluctuations are not possible, so actually for this subsequence, $\x_n(s)$ converges for all $s$.
So  any sequence of causal paths from $q$ to $p$ has a convergent subsequence, and thus the space of such causal paths is compact.\footnote{\label{humph}In this footnote
and the next one, we give the reader a taste of the sort of mathematical details that will generally be elided in this article.    The topology in which the space of causal
paths is compact is the one for which the argument in the text is correct:   a sequence $\x_n(s)$ converges to $\x(s)$ if it converges for each $s$.   In addition, to state properly
the argument in the text, we have to take into account that such a pointwise limit of smooth curves is not necessarily smooth.   One defines a continuous causal curve as the pointwise
limit of a sequence of smooth causal curves.   (A simple example of a continuous causal curve that is not smooth is a piecewise smooth causal curve.)
The argument in the text has to be restated to show that a sequence of continuous causal curves has a
subsequence that converges to a continuous causal curve. Basically, if $\gamma_1,\gamma_2,\cdots$ is a sequence of continuous causal curves, then after passing to a subsequence, one can assume as in the text
that the $\gamma_i$ converge to a continuous curve $\gamma$, and we have to show that $\gamma$ is a continuous causal curve.   Each $\gamma_j$, being a continuous causal curve, is the pointwise limit of a sequence $\gamma_{j,1},\gamma_{j,2},\cdots$
of smooth causal curves.   After possibly passing to subsequences, we can assume that $\gamma_j$ and $\gamma_{j'}$ are close to each other for large $j,j'$ (say within Euclidean distance $1/n$ if $j,j'\geq n$)
and similarly that $\gamma_{j,k}$ is close to $\gamma_j$ for $k\geq j$ (say within a Euclidean distance $1/j$).  Then the diagonal sequence $\{\gamma_{j,j}\}$ is a sequence of smooth
causal curves that converges to $\gamma$, so $\gamma$ is a continuous causal curve.}

This argument carries over without any essential change to $\D$-dimensional Minkowski space with metric $\ds^2=-\d t^2+\d \vec x^2$.
Thus if $p$ is to the causal future of $q$, the space of causal paths from $q$ to $p$ is compact.

 \begin{figure}
 \begin{center}
   \includegraphics[width=3in]{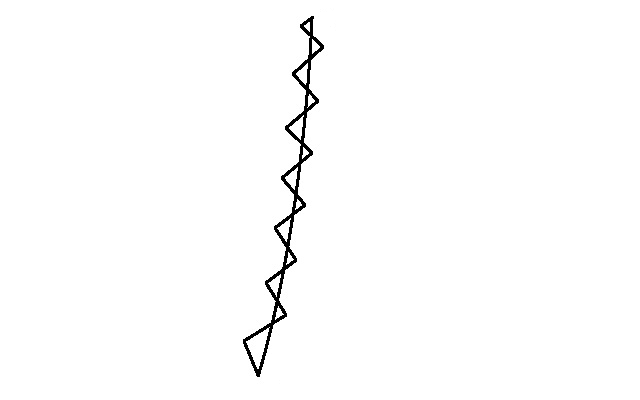} 
 \end{center}
\caption{\small  The purpose of this figure is to illustrate the fact that a smooth timelike curve $\gamma$ can be approximated by a sequence of causal curves
$\gamma_n$ that fluctuate rapidly near $\gamma$ in lightlike or almost lightlike directions, such that the $\gamma_n$ converge to $\gamma$ but the proper time elapsed along
the $\gamma_n$ does not converge to the proper time elapsed along $\gamma$.   Rather the proper time elapsed along the $\gamma_n$ can be very small or even zero.
  \label{Wild}}
\end{figure}

Here is a consequence that turns out to be important.
The proper time elapsed along a causal path is 
\be\label{third}\tau =\int_0^1\d s\sqrt{ \left(\frac{\d t}{\d s} \right) ^2- \left(\frac{\d \vec x}{\d s}  \right)^2   }.\ee
Assuming that $p$ is in the causal future of $q$ -- so that causal paths from $q$ to $p$ do exist -- the compactness of the space of causal
paths from $q$ to $p$ ensures that there must be such a path 
 that maximizes $\tau$.
In more detail, there must be an upper bound on $\tau$ among all causal paths from $q$ to $p$, because a sequence of causal paths $\x_n(s)$
whose elapsed proper time $\tau_n$ grows without limit for $n\to \infty$ could not have a convergent subsequence.   If $\tau_0$ is the least upper
bound of the proper time for any causal path from $q$ to $p$, then a sequence of paths $\x_n(s)$ of proper time $\tau_n$ such that $\tau_n\to\tau_0$ for $n\to\infty$
 has a convergent subsequence,
and the limit of this subsequence is a causal path with proper time\footnote{The proper time is not actually a continuous function on the space of paths, in the
topology defined in footnote \ref{humph}.   That is because a sequence of causal paths $\gamma_n$ might converge pointwise to a causal path $\gamma$
but with wild short-scale oscillations in lightlike (or almost lightlike) directions.  See fig.  \ref{Wild}.  The correct statement is that if a sequence $\gamma_n$ converges
to $\gamma$, then the proper time elapsed along $\gamma$ is equal to or greater than the limit (or if this limit does not exist, the lim sup) of the proper
time elapsed along $\gamma_n$.   In other words, upon taking a limit, the proper time can jump upwards but cannot jump downwards.   Technically this
is described by saying that the proper time function is an upper semicontinuous function on the space of causal paths.   Upward jumps  do not spoil the argument
given in the text (though downward jumps would spoil it), because by the way $\tau_0$ was defined,  the  limit of a subsequence of the $\x_n(s)$ cannot
have a proper time greater than $\tau_0$.    One could modify the notion of convergence to make the elapsed proper time a continuous function on the space of causal
paths, but this is inconvenient because then compactness would fail.} $\tau_0$.  

A causal path that maximizes -- or just extremizes -- the elapsed proper time is a geodesic.   So if $p$ is in the future of $q$,
there must be a proper time maximizing geodesic from $q$ to 
$p$.   In the particular case of Minkowski space, we can prove this more trivially.    
There is a unique geodesic from $q$ to $p$, namely a straight line, and it maximizes the proper time.   (The fact that a timelike geodesic in Minkowski space
maximizes the proper time between its endpoints is sometimes called the twin paradox.  A twin who travels from $q$ to $p$ along a non-geodesic path, accelerating
at some point along the way, comes back younger than a twin whose trajectory from  $q$ to $p$ is a geodesic.   In a suitable Lorentz frame, the twin whose path
is a geodesic was always at rest, but there is no such frame for the twin whose trajectory involved acceleration.)

The only fact that we really needed about Minkowski space to establish the compactness of the space of causal paths from $q$ to $p$ was
that the causal diamond $D_q^p$ consisting of points that such a path can visit is compact.   If $q$ and $p$ are points in any Lorentz signature spacetime,
we can define a generalized causal diamond $D_q^p$ that consists of points 
 that are in the causal
future of $q$ and the causal past of $p$.    Whenever $D_q^p$ is compact,   the same reasoning as before will show that
the space of causal paths from $q$ to $p$ is also compact, and therefore that there is a geodesic from $q$ to $p$ that maximizes the elapsed proper time.

A small neighborhood  $U$ of  a point $q$ in any spacetime $M$ can always
be well-approximated by a similar small open set in Minkowski space.     A precise statement (here and whenever we want to compare a small neighborhood of a point in some
spacetime
to a small open set in Minkowski space) is that $q$ is contained in  a convex normal neighborhood $U$, in which
there is a unique geodesic between any two points.\footnote{For more detail on this concept,
 see \cite{HawkingEllis}, especially p. 103, or \cite{Wald}, p. 191.}    Roughly, in such a neighborhood, causal relations are as they are in a similar neighborhood in Minkowski space.  We will
 call such a neighborhood a local Minkowski neighborhood.
 
  If $p$ is just slightly to the future of $q$,  one might hope that $D^p_q$ will be compact, just like a causal diamond in Minkowski space.  
 This is actually true if $M$ satisfies a physically sensible condition of causality.  A causality condition is needed for the following reason.   In order to compare $D^p_q$ to a causal diamond
 in Minkowski space, we want to know that if $p$ is close enough to $q$, then $D^p_q$ is contained in a local Minkowski neighborhood of $q$.   That is not true
 in a spacetime with closed causal curves (if $\gamma_0$ is a closed causal curve from $q$ to itself, then a causal path from $q$ to $p$ can traverse $\gamma_0$ followed by any causal
 path from $q$ to $p$; such a path will not be contained in a local Minkowski neighborhood of $q$, since there are no closed causal curves in such a neighborhood).   The condition that we need is
  a little stronger than absence of closed causal curves and is called ``strong causality'' (see section \ref{causality} for more
 detail).     In a strongly causal spacetime, causal paths from $q$ to $p$ are contained in a local Minkowski neighborhood of $q$ if $p$ is close
 enough to $q$.   $D^p_q$ is then compact, as in Minkowski space.   In this article, we always assume strong causality.

As $p$ moves farther into the future, compactness of $D^p_q$ can break down.   We will give two examples.
  The first example is simple but slightly artificial.  The second example is perhaps more natural.

For the first example, start with Minkowski space $M$, and make a new spacetime $M'$  by omitting from $M$ a point
$r$ that is in the interior of $D_q^p$ (fig. \ref{Fig2}).    $M'$ is a manifold, with a smooth Lorentz signature metric tensor, so we can regard
it as a classical spacetime in its own right.   But in $M'$,  the causal diamond $D_q^p$ is not compact, since the point $r$ is missing.
Accordingly the space of causal paths from $q$ to $p$ in $M'$ is not compact.  A sequence of causal paths in $M'$
whose limit in $M$ would pass through $r$ does not have a limit among paths in $M'$.   If in $M$, $r$ happens to lie on the geodesic from $q$ to $p$,
then in $M'$ there is no geodesic from $q$ to $p$.  Of course, in this example $D_q^p$ is compact if $p$ is to the past of $r$.

We can make this example a little less artificial by using the fact that the space of causal paths is invariant under 
a Weyl transformation of the spacetime metric, that is, under multiplying the metric by a positive  function $e^\phi$, for any real-valued
function $\phi$ on spacetime.   The reason
for this is that two metrics $\ds^2$ and $e^\phi\ds^2$ that differ by a Weyl transformation  have the same local light cones and so the same spaces of causal paths.
With this in mind,  replace the usual Minkowski space metric $-\d t^2+\d \vec x^2$ with a Weyl-transformed metric
$e^\phi(-\d t^2+\d\vec x^2) $.  If the function $\phi$ is chosen to blow up at the point $r$, producing a singularity,
this provides a rationale for omitting that point from the spacetime. This gives a relatively natural example of a spacetime in which causal diamonds are not compact.

\begin{figure}
 \begin{center}
   \includegraphics[width=3in]{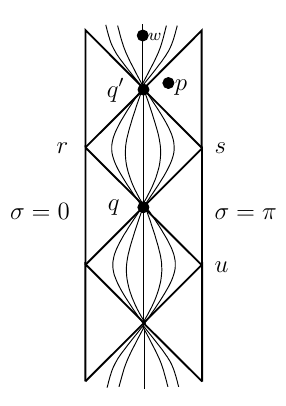} 
 \end{center}
\caption{\small The Penrose diagram of AdS$_2$ spacetime.    The causal structure is that of the strip $0<\sigma<\pi$ in Minkowski spacetime with metric $\ds^2=-\d t^2+\d\sigma^2$. Causal
curves make an angle no greater than $\pi/4$ from the vertical.
   A causal curve from $q$ can travel
to the right edge of the figure, linger for a while very near the boundary, and then proceed on to $p$.   A causal curve of this kind can have an arbitrarily large
elapsed proper time.  By contrast, if $p$ is replaced by a point $p'$ inside the quadrilateral $qsq'r$, then a causal curve from $q$ to $p'$ cannot reach
the boundary and has a maximum possible elapsed proper time.   The curved lines in the figure are the timelike geodesics through the point $q$; every
point $p'$ inside the quadrilateral $qsq'r$ is on such a geodesic (reflecting compactness of $D_q^{p'}$ for such $p'$).
These timelike geodesics all focus at the point $q'$, as shown.  Their continuation to the future of $q'$ or the past of $q$ is indicated. No geodesic from $q$
reaches $p$. There is a timelike geodesic from $q$ to $w$, but it does not maximize the proper time.  Some of these details will be important in section \ref{loran}.
 (Compare with Fig. 7 of \cite{Penrose}.)  \label{Fig1}}
\end{figure}
 
For a second example,\footnote{\label{noted}This example can be omitted on first reading.   The background required to understand it is largely explained in Appendix \ref{ads}.} we consider Anti de Sitter (AdS) spacetime, which is the maximally homogeneous spacetime with negative cosmological constant.   For simplicity, we consider two-dimensional Anti de Sitter spacetime AdS$_2$.    As explained in detail in Appendix \ref{ads}, this spacetime
 can be described by the metric
 \be\label{wombox} \ds^2=\frac{R^2}{\sin^2\sigma}\left(-\d t^2+\d\sigma^2\right),~~-\infty<t<\infty, ~~~0<\sigma<\pi. \ee
 The causal structure is not affected by the factor $R^2/\sin^2\sigma$, which does not affect which curves are timelike or null.  
 So from a causal point of view, we can drop this factor and just consider the strip $0<\sigma<\pi$ in a Minkowski space with metric $\ds^2=-\d t^2+\d\sigma^2$.
 This strip is depicted in the Penrose diagram
 of fig. \ref{Fig1}.    A causal curve in the Penrose diagram is one whose tangent vector everywhere makes an angle no greater than $\pi/4$ with the vertical.  Null geodesics
 are straight lines for which the angle is precisely $\pi/4$; some of these are drawn in the figure.
 
 The boundaries of the Penrose diagram at $\sigma=0,\pi$ are not part of the AdS$_2$ spacetime. They are infinitely far away along, for example, a spacelike hypersurface
 $t={\mathrm{constant}}$.    Nevertheless, a causal curve that is null or asymptotically null can reach $\sigma=0$ or $\pi$ at finite $t$.    Accordingly, it is 
sometimes useful to partially compactify
AdS space by adding in  boundary points at $\sigma=0,\pi$.  Those points make up what is called the conformal boundary of AdS$_2$.   But importantly,
the conformal boundary is not actually part of the AdS$_2$ spacetime.   We cannot include the boundary points because (in view of the $1/\sin^2\sigma$ factor) the metric blows up along the conformal
boundary.

 As in any spacetime (satisfying a reasonable causality condition), if  $q$ is a point in AdS spacetime, and $p'$ is a point slightly to its future, then $D_q^{p'}$ is compact.   But in AdS spacetime,
$D_q^{p'}$ fails to be compact if $p'$ is sufficiently far to the future of $q$ that a causal path from $q$ to $p'$  can reach from $q$ all the way to the conformal
boundary on its way to $p'$.  In this case, the fact that the conformal boundary points are not actually part of the AdS spacetime means that $D_q^{p'}$
is not compact.  The ``missing'' conformal boundary points play a role similar to the missing point $r$ in the previous example.

In practice, if a point $p'$ is  contained in the quadrilateral  $qsq'r$ in the figure, then a causal path from $q$ to $p'$ cannot reach the conformal
boundary en route, and 
 $D_q^{p'}$ is compact.   But if $p'$ is to the future of $q$ and not in this quadrilateral, then a causal path from $q$ to $p'$ can reach the conformal
 boundary on its way, and compactness of $D_q^{p'}$ fails.   Concretely, a sequence of causal curves from $q$ within the AdS space 
 that go closer and closer to the conformal boundary
 on their way to $p'$ will have no convergent subsequence.
  
 When $D_q^{p'}$ is not compact, there may not exist a causal path from $q$ to $p'$ of maximal elapsed proper time.
An example is the point labeled $p$ in the figure, which is to the future of $q$ but is not contained in the quadrilateral.   
There is no upper bound on the elapsed proper time of a causal path from $q$ to $p$.
A causal path that starts at $q$, propagates  very close to the right edge of the figure, lingers there for a while, and then continues on to $p$,
can have an arbitrarily long elapsed  proper time.   This statement reflects the factor of $1/\sin^2\sigma$ in the AdS$_2$ metric (\ref{wombox}).   Since a causal path
from $q$ to $p$ can linger for a positive interval of $t$ in a region of arbitrarily small $\sin\sigma$, there is no upper bound on its elapsed proper time.

  In fact, there is no geodesic from $q$ to $p$.   All  future-going timelike geodesics
from $q$ actually converge at a focal point $q'$ to the future of $q$, whose importance will become clear in section \ref{geofocal}.    From $q'$, these geodesics continue
``upward'' in the diagram, as shown, never reaching $p$.  Lightlike or spacelike geodesics from $q$ terminate on the conformal boundary in the past of $p$.   

In view of examples such as these, one would like a useful criterion that ensures the compactness of the generalized causal diamonds  $D_q^p$.
Luckily, there is  such a criterion.

\section{Globally Hyperbolic Spacetimes}\label{glohyper}

 \subsection{Definition}\label{def}
  
In a traditional understanding of physics, one imagines specifying initial data on an initial value hypersurface\footnote{By definition, a hypersurface
is a submanifold of codimension 1.} $\S$  and then
one defines equations of motion that are supposed to determine what happens to the future (and past) of $\S$.  

To implement this idea in General Relativity, we require $S$ to be a spacelike hypersurface.\footnote{More generally,
one can define initial data on a hypersurface that has some null portions, as long as it is achronal (see below).   We will consider only the more intuitive case
of a spacelike hypersurface.}
 Saying that $\S$ is spacelike
means that the nearby points in $\S$ are spacelike separated; more technically, the Lorentz signature metric of the full spacetime manifold $M$ induces on $\S$
a Euclidean signature metric.   A typical example is the hypersurface $t=0$ in a Minkowski space with metric   $\ds^2=-\d t^2+\d \vec x^2$.
The induced metric on the surface $t=0$ is simply the Euclidean metric $\d\vec x^2$.

\begin{figure}
 \begin{center}
   \includegraphics[width=1.4in]{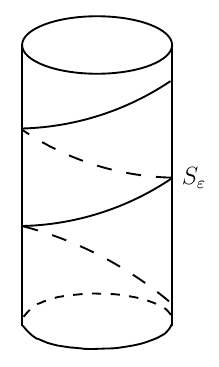} 
 \end{center}
\caption{\small A spacelike hypersurface $\S_\varepsilon$ that is closed but not achronal.  \label{Fig3}}
\end{figure}

We actually need a further condition, which is that $\S$ should be {\it achronal}.   In general, a subset $\S$ of a spacetime  $M$ is called achronal if there is
no timelike path in $M$ connecting distinct points $q,p\in \S$.  If there is such a path, then data at $q$ will influence what happens at $p$ and it is not sensible
to specify ``initial conditions'' on $\S$ ignoring this.

To see that a spacelike hypersurface is not necessarily achronal, consider the two-dimensional cylindrical spacetime with flat metric
\be\label{zorfo} \ds^2=-\d t^2+\d\phi^2,\ee
where $t$ is a real variable, but $\phi$ is an angular variable, $\phi\cong\phi+2\pi$.
The hypersurface $\S_\varepsilon$ defined by
\be\label{morfo} t=\varepsilon\phi, \ee
for nonzero $\varepsilon$, wraps infinitely many times around the cylinder (fig. \ref{Fig3}).   $\S_\varepsilon$ is  spacelike if $\varepsilon$ is small, but it is not achronal;
for example, the points $(t,\phi)=(0,0)$ and $(t,\phi)=(2\pi\varepsilon,0)$ in $\S_\varepsilon$ can be connected by an obvious timelike geodesic.   Thus a spacelike
hypersurface is not necessarily achronal: in a spacelike hypersurface $S$,  sufficiently near points in $\S$ are not connected by a nearby timelike
path, while in an achronal hypersurface, the same statement holds without the condition that the points and paths should be sufficiently near.   (Conversely, an achronal
hypersurface may not be spacelike as it may have null regions.)

\begin{figure}
 \begin{center}
   \includegraphics[width=1.6in]{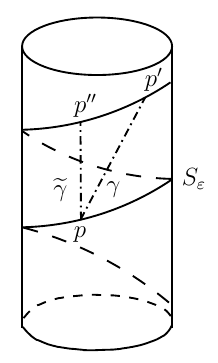} 
 \end{center}
\caption{\small If $\S$ is a spacelike hypersurface and $\gamma$ is a null curve 
 connecting points $p,p'\in \S$, then by moving
$p'$ along $\S$ ``towards'' $\gamma$, one gets a point $p''\in \S$ such that there is a strictly timelike curve $\t\gamma$ from $p$ to $p''$.
This is illustrated using the hypersurface $\S_\varepsilon$ of fig. \ref{Fig3}.   \label{Fig3a}}
\end{figure}

If an achronal set $\S$ is also a spacelike hypersurface, the statement of achronality can be sharpened.
The definition of achronal says that there is no timelike path $\gamma\subset M$ connecting distinct points $p,p'\in \S$, but if $\S$ is of codimension 1 in $M$, 
it follows that there actually is no causal path from $p$ to $p'$.  In other words, $\gamma$ does not exist even if it is allowed to be null  rather than timelike.
To see this, let us think of the directions in which a point $p\in \S$ can be displaced, while remaining in $\S$, as the ``spatial'' directions.
For $\S$ a spacelike  hypersurface, these directions do constitute, at every point $p\in \S$, a complete set of spatial directions, in some local Lorentz frame at $p$.
If $p,p'\in \S$ are connected by a  causal path $\gamma$ that is null (in whole or in part), then after displacing $p$ or $p'$ along $\S$
in the appropriate  direction, $\gamma$ can be deformed to
 a timelike path; see fig. \ref{Fig3a}.    
 
 Here it is important that $\S$ is of  codimension 1.   Otherwise the necessary displacement  may not be possible. 

 \begin{figure}
 \begin{center}
   \includegraphics[width=2.5in]{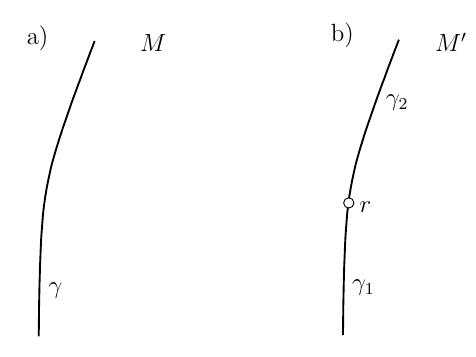} 
 \end{center}
\caption{\small a) The causal curve $\gamma$ in a spacetime $M$ can be extended indefinitely into the past and future. b) A point $r$ has been removed from $M$ to
make a new spacetime $M'$.  $\gamma$ splits into two causal curves $\gamma_1$ and $\gamma_2$.  $\gamma_1$ is inextendible in the future and $\gamma_2$ is inextendible
in the past.  \label{Fig3M}}
\end{figure}

To complete the definition of an initial value hypersurface in General Relativity, we need the notion of an {\it inextendible} causal curve.   A causal
curve is extendible if it can be extended further.  Otherwise it is inextendible. For example, in Minkowski space, the timelike geodesic $(t(s),\vec x(s))=(s,0)$
is inextendible if $s$ is regarded as a real variable.   But if we arbitrarily restrict $s$ to a subset of the real line, for example $0<s<1$,
we get an extendible causal curve.  If we remove from some spacetime $M$ a point $r$ to get the spacetime $M'$ of fig. \ref{Fig2}, then an inextendible
causal path in $M$ that passes through $r$ breaks up in $M'$ into two separate causal paths, one to the future of $r$ and one to the past of $r$ (fig. \ref{Fig3M}).
Each of them is  inextendible.   A sufficient, but not necessary, condition for a timelike path $\gamma$ to be inextendible is that the  proper time elapsed
along $\gamma$  diverges
both in the past and in the future.    

Let $\gamma$ be a causal curve $x^\mu(s)$.   With a suitable choice of the parameter $s$, we can always assume that $s$ ranges over the unit interval,
with or without its endpoints.   Suppose for example that $s$ ranges over a closed interval $[0,1]$ or a semi-open interval $(0,1]$.   Then we define $p=x^\mu(1)$ as the future
endpoint\footnote{\label{technical} We use the term ``endpoint'' in this familiar sense, but we should warn the reader that the same term is used in mathematical
relativity with a somewhat different and more technical meaning.   See \cite{Wald}, p. 193.}
 of $\gamma$.   Such a $\gamma$ is always extendible to the future beyond $p$; the extension can be made by adjoining to $\gamma$ any future-going causal path from $p$.
Even if $\gamma$ is initially defined on an open or semi-open interval $(0,1)$ or $[0,1)$  without a future endpoint, if the limit $p=\lim_{s\to 1}x^\mu(s)$ exists,
we can add $p$ to $\gamma$ as a future endpoint, and then continue $\gamma$ past $p$ as before.   So if $\gamma$ is inextendible to the future, this means that
$\gamma$ has no future endpoint and it is not possible to add one.    (Informally, this might mean, for example, that $\gamma$ has already been extended infinitely to the future,
or at least as far as the spacetime goes, or that $\gamma$ ends at a singularity.)
Similarly, if $\gamma$ cannot be extended to the past, this means that $\gamma$ has no past endpoint and it is not possible to add one.

Finally we can define the appropriate concept of an initial value surface in General Relativity, technically called a Cauchy hypersurface.
A Cauchy hypersurface or initial value hypersurface in $M$ is an achronal spacelike hypersurface $\S$ with the property that if $p$ is a point in $M$ not in $\S$, then every
inextendible causal path $\gamma$ through $p$ intersects $\S$.
  A spacetime $M$ with a Cauchy hypersurface $\S$ is said to be {\it globally hyperbolic}.

 This definition  requires some explanation.  
The intuitive idea is that any signal that one observes at $p$ can be considered to have arrived at $p$ along some causal path.
If $p$ is to the future of $\S$ and every sufficiently extended past-going causal path
through $p$ meets $\S$, then what one will observe at $p$ can be predicted from a knowledge of what there was on $\S$, together with suitable dynamical equations.
If there is a past-going causal path $\gamma$ through $p$ that cannot be extended until it meets  $\S$, then a knowledge of what there was on $\S$ does not 
suffice to predict the physics at $p$;
one would also need an initial condition along $\gamma$.   (Obviously, if $p$ is to the past of $\S$, then similar statements hold, after exchanging ``past'' with ``future''
and ``initial condition'' with ``final condition.'')   Thus, globally hyperbolic spacetimes are the ones in which the traditional idea of predicting the future from the past is
applicable.

 \begin{figure}
 \begin{center}
   \includegraphics[width=2.5in]{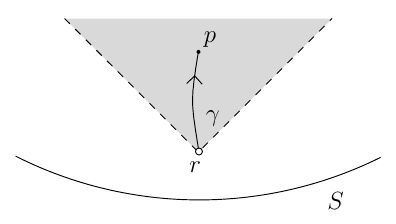} 
 \end{center}
\caption{\small $S$ is a spacelike hypersurface in a spacetime that is not globally hyperbolic because a point $r$ in the future of $S$ has been removed.
Initial data on $S$ do not suffice to predict what will be observed at a point $p$ that is to the future of $r$.  To make such a prediction, one would need
to know something about possible signals that might emerge from the point $r$.   Technically, $S$ is not a Cauchy hypersurface for this spacetime  because there is a timelike curve
from $p$, as shown, that cannot be extended in the past until it meets $S$.  Rather, it gets ``stuck'' at the missing point.   \label{Fig7M}}
\end{figure}

 For example, the hypersurface $t=0$ in Minkowski space is an initial value hypersurface, and Minkowski space is globally hyperbolic.  
 On the other hand, omitting a point $r$  from a globally hyperbolic spacetime gives a spacetime that is not globally hyperbolic.   See fig. \ref{Fig7M}.  More
 sophisticated examples of  spacetimes that are not globally hyperbolic arise when certain black hole spacetimes (such as the Reissner-Nordstr\"{o}m or Kerr
 solutions) are extended beyond their horizons (see fig. \ref{Fig41M} in section \ref{ch}).

 An inextendible causal
path $\gamma\subset M$ will always intersect an initial value hypersurface $\S$, by the definition of such a hypersurface, but actually it will
always intersect $\S$ in precisely one point.  If $\gamma$ intersects $\S$ in two points $p$ and $p'$,
then the existence of the causal path $\gamma$ connecting $p$ and $p'$ contradicts the achronal nature of $\S$.    A Cauchy hypersurface $\S$ always divides $M$ into
a ``future'' and a ``past,'' for the following reason.   Suppose that $p\in M$ is not contained in $\S$, and let $\gamma$ be any inextendible causal path through $p$.
As we have just seen, such a path will intersect $\S$ at a unique point $q$.   We say that $p$ is to the future (or past) of $\S$ if it is to the future (or past) of $q$
along $\gamma$.   (As an exercise, the reader can check that this definition is consistent: if $\gamma, \gamma'$ are causal curves through $p$, meeting
$\S$ at points $q,q'$, then $q'$ is to the past of $p$ along $\gamma'$ if and only if $q$ is to the past of $p$ along $\gamma$.)  

It is not completely obvious that in General
Relativity, only globally hyperbolic spacetimes are relevant. Perhaps physics will eventually transcend the  idea of predictions
based on initial conditions.  But this will certainly involve more exotic considerations.  The study of globally hyperbolic spacetimes
is surely well-motivated.  

\subsection{Some Properties of Globally Hyperbolic Spacetimes}\label{somep}

 The following are useful facts that will also help one become
familiar with globally hyperbolic spacetimes.  

A globally hyperbolic spacetime $M$ can have no closed causal curves.  
 If there is a closed causal curve $\gamma\subset M$, then in parametric
form $\gamma$ is represented by a curve $x^\mu(s)$ with, say,  $x^\mu(s+1)=x^\mu(s)$.   We could think of $s$ as a periodic variable $s+1\cong s$, but
for the present argument, it is better to think of $s$ as a real variable, in which case $x^\mu(s)$ is an inextendible causal curve that repeats the same
spacetime trajectory infinitely many times. This curve will have to intersect  a Cauchy hypersurface 
$\S$, and it will do so infinitely many times.  But the existence of a causal curve that intersects $S$ more than once contradicts the achronality of $S$.  

Actually, if  $M$ is a globally hyperbolic spacetime with Cauchy hypersurface $\S$, then any other Cauchy hypersurface $\S'\subset M$
is topologically equivalent to $\S$.   The intuitive idea
is that $\S'$ can be reached from $\S$ by flowing backwards and forwards in ``time.''  Of course, for this idea to make sense in General Relativity, we have
to make a rather arbitrary choice of what we mean by time.  Let $\ds^2=g_{\mu\nu}\d x^\mu\d x^\nu$ be the Lorentz signature metric of $M$.  Pick an arbitrary Euclidean
signature metric $\ds_E^2=h_{\mu\nu}\d x^\mu\d x^\nu$.   At any point $p\in M$, one can pick coordinates with $h_{\mu\nu}=\delta_{\mu\nu}$.   Then
by an orthogonal rotation of that local coordinate system, one can further diagonalize $g_{\mu\nu}$ at the point $p$, putting it in the form $\mathrm{diag}(\lambda_1,\cdots,\lambda_\D)$.  Since
$g_{\mu\nu}$ has Lorentz signature, precisely one of the eigenvalues is negative.   The corresponding eigenvector $V^\mu$ is a timelike vector at $p$.  We  can normalize
this eigenvector up to sign by\footnote{Here and later, a summation over repeated indices is understood, unless otherwise stated.}
\be\label{morrox}h_{\mu\nu}V^\mu V^\nu=1,\ee  and, since $M$ is assumed to be time-oriented, we can fix its sign by requiring that it is future-pointing in the Lorentz signature metric $g$.

Having fixed this timelike vector field $V^\mu$ on $M$, we now construct timelike curves on $M$ by flowing in the $V^\mu$ direction.
For this, we look at the ordinary differential equation
\be\label{morro}\frac{\d x^\mu}{\d s}=V^\mu(x(s)), \ee
whose solutions, if we take the range of $s$ to be as large as possible, are inextendible causal curves.\footnote{If it is possible to add a past or future endpoint  $p$ to such a curve $\gamma$, then the
solution of 
 equation (\ref{morro}) can be continued for at least a short range of $s$ beyond $p$ (this is obvious in a local Minkowski neighborhood of $p$), so that $\gamma$ was not defined by solving the equation in the largest possible range of $s$. If it is not
 possible to add a past or future endpoint, then $\gamma$ is inextendible.}
   These curves are called the integral curves of the vector
field $V^\mu$.   
Every point $p\in M$ lies on a unique\footnote{We consider two parametrized causal paths $x^\mu(s)$ to be equivalent if they differ
 only by the choice of the parameter. In the present context, this means that two solutions of eqn. (\ref{morro}) that differ by an additive shift of the parameter
 $s$ correspond to the same integral curve.}
 such curve, namely the one that starts at $p$ at $s=0$.  Since the integral curves are inextendible causal curves, each such curve $\gamma$ intersects $\S$
 in a unique point.   If the integral curve through $p\in M$ meets $\S$ at a point $q$, then we define the ``time'' at $p$ by $t(p)=s(p)-s(q)$.   The 
 function defined this way vanishes for $p\in \S$, and increases towards the future along the integral curves. 
 
  The normalization condition (\ref{morrox}) means that the
 parameter $t$ just measures  arclength  in the Euclidean signature metric $h$.    The metric $h$ can always be chosen so that the arclength along
 an inextendible  curve in $M$ is divergent in both directions, roughly by making  a Weyl rescaling by a factor that blows up at infinity  \cite{NO}.  (See Appendix \ref{moredetail}
 for more detail on this statement.)
  Once this is done,   $t$ runs over the full range $-\infty<t<\infty$ for every  integral curve.

 \begin{figure}
 \begin{center}
   \includegraphics[width=4.5in]{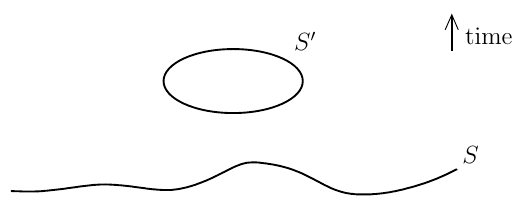} 
 \end{center}
\caption{\small If $\S\subset M$ is an initial value hypersurface, then every achronal set $\S'\subset M$ is equivalent topologically to a subset of $\S$.
The equivalence is found, roughly, by ``flowing'' in the time direction from $\S'$ to $\S$. This picture is meant to represent a typical impossible
situation, in which $\S$ is not
compact and $\S'$ is a compact achronal hypersurface.  The ``flow'' from $\S'$ to $\S$ would map more than one
point in $\S'$ to the same point in $\S$.  \label{Fig4}}
\end{figure}
 
 Now let $\S'\subset M$ be another Cauchy hypersurface.  Every $p\in \S'$ is on a unique integral curve, which intersects $\S$ at a unique point $q$.
We define a map $\varphi:\S'\to \S$ by  $\varphi(p)=q$.   Then $\varphi$ maps $\S'$ onto all of $\S$, because conversely, for any $q\in \S$, the integral curve through $q$  intersects
$\S'$ at some point $p$, so that $\varphi(p)=q.$    So $\varphi$ 
is  a 1-1 smooth mapping between $\S'$ and $\S$, showing that they are equivalent topologically, as claimed.  

 If $\S'$ is achronal,
 but not Cauchy, we can still map $\S'$ to $\S$ by $\varphi(p)=q$.   This map is an embedding but is not necessarily an isomorphism (as $\varphi(\S')$ may
 not be all of $\S$).   So we learn that any achronal set  $\S'\subset M$ is equivalent topologically to a portion of $\S$.  A special case of this is important
 in the proof of Penrose's singularity theorem for black holes (section \ref{penroseproof}).
 If $\S$ is not compact (but connected), then an achronal codimension 1 submanifold $\S'\subset M$ cannot be compact (fig. \ref{Fig4}).  For a compact $(\D-1)$-manifold $\S'$ cannot
 be embedded as part of a noncompact (connected) $(\D-1)$-manifold $\S$.

We can now also deduce  that topologically, $M=\S\times \R$.   Indeed, we can continuously parametrize $p\in M$ 
 by $t(p)$ 
and $\varphi(p)$.   Since $\varphi(p)$ is any point in $\S$, and $t(p)$ is any real number (assuming the metric $h$ is chosen to be complete, as discussed above),
this shows that $M=\S\times \R$.

\begin{figure}
 \begin{center}
   \includegraphics[width=3in]{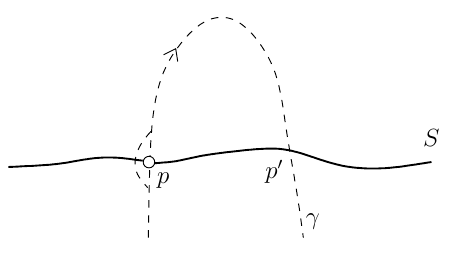} 
 \end{center}
\caption{\small  This picture aims to illustrate an argument in the text showing that a Cauchy hypersurface $\S\subset M$  is always closed in $M$. \label{Singul4}}
\end{figure}

A Cauchy hypersurface $\S\subset M$ is always a closed subspace of $M$. This immediately follows from the fact that $M=S\times \R$, since $S\times \{0\}$ (for any point $0\in\R$) is
closed in $S\times \R$.    For a more direct proof,  suppose that $p\in M$ is in the closure of $\S$ but not in $\S$ (fig. \ref{Singul4}).
Let $\gamma$ be any inextendible timelike path through $p$.  $\gamma$ meets $\S$ at some other point $p'$, since $\S$ is Cauchy.
But then, since $p$ is in the closure of $\S$, it is possible to slightly modify $\gamma$ only near $p$ to get a timelike path from $\S$ to itself,
showing that $\S$ is not achronal.

Although the function $t(p)$ that we have defined increases towards the future along the integral curves, it does not necessarily increase towards the future
along an arbitrary causal curve.   For some purposes, one would like a function, known as a time function, with this property.   A simple construction was given in \cite{Geroch}.

\subsection{More On Compactness}\label{cpct}

Globally hyperbolic spacetimes have the property that spaces of causal paths with suitable conditions on the endpoints
are compact.\footnote{This was actually the original definition of a globally hyperbolic spacetime \cite{Leray}.   The definition in terms of intextendible causal curves
came later \cite{Geroch}.}   For example, for $\S$ an initial value hypersurface and $q$ a point to the past of $\S$, let $\CC_q^\S$
be the space of causal paths from $q$ to $\S$.
The space of such paths is compact, as one can see by considering a sequence of causal
paths $\gamma_n\in \CC_q^\S$.

\begin{figure}
 \begin{center}
   \includegraphics[width=3in]{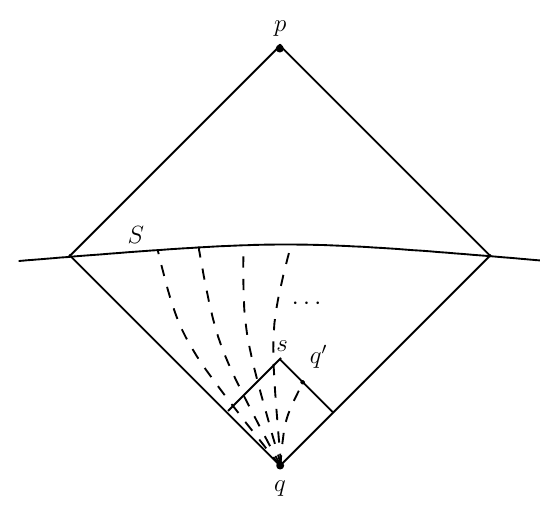} 
 \end{center}
\caption{\small This picture illustrates the argument showing that $\CC_q^\S$ is compact.   At the bottom of the figure is a small causal diamond $D_q^s$ that
can be approximated by a causal diamond in Minkowski space.
 \label{Singul2}}
\end{figure}

If the point $s$ is slightly to the future of $q$, then the causal diamond $D_q^s$  looks like a causal diamond in Minkowski
spacetime, with $q$ at its past vertex, and in particular is compact.\footnote{As explained in section \ref{classic}, this statement requires  strong causality.    For now we take this as a physically
well-motivated assumption, but in Appendix \ref{detailedproof}, 
we show that globally hyperbolic spacetimes are strongly causal.  See also section \ref{causality} for more on strong causality.}  If we restrict the paths $\gamma_n$ to that
diamond, we can make the same argument as in Minkowski space, showing that
a subsequence $\gamma_{n,[1]}$ of the $\gamma_n$ (when restricted to the diamond)
converges to some causal path $\gamma_{[1]}$ from $q$ to a point $q'$ on
one of the future boundaries of the diamond, as in fig. \ref{Singul2}.    This much does not require the assumption of global hyperbolicity.   Now we start at $q'$,
and continue in the same way, showing that a  subsequence $\gamma_{n,[2]}$ of the $\gamma_{n,[1]}$
 converges in a larger region to a causal curve $\gamma_{[2]}$ that continues
past $q'$.    We keep
going in this way and eventually learn that a subsequence of the original sequence converges to a causal curve $\overline\gamma$ from
$q$ to   $\S$.   For more details on this argument, see  Appendix \ref{detailedproof}.   

The role of global hyperbolicity in the argument is to ensure that we never get ``stuck.''
Without global hyperbolicity, after iterating the above process,
we might arrive at a subsequence of the original sequence that has a limit path $\gamma_*$ that does not reach $\S$
and cannot be further extended.  To give an example of how this would actually happen, suppose that the original sequence $\gamma_n$ converges
to a causal curve $\gamma_*\in \CC_q^\S$, and let $p$ be a point in $\gamma_*$  that is not in any of the $\gamma_n$.   Then in the spacetime $M'$
that is obtained from $M$ by omitting the point $p$, the sequence $\gamma_n\in\CC_q^\S$ has no subsequence that converges to any causal curve from
$q$ to $\S$.  

Compactness of the space of causal curves from $q$ to $\S$ implies in particular that the set of points $D_q^\S$ that can be visited by such a curve
is also compact.   Clearly, if $p_1,p_2,\cdots$ is a sequence of points in $D_q^\S$ with no convergent subsequence, and $\gamma_1,\gamma_2,\cdots
\in \CC_q^\S$ is a sequence of causal curves from $q$ to $\S$ with $p_i\in\gamma_i$, then the sequence of curves $\gamma_1,\gamma_2,\cdots $ can have
no convergent subsequence.  Conversely, if one knows that $D_q^\S$ is compact, compactness of $\CC_q^\S$ follows by essentially the same argument that we
used in section \ref{classic} for curves in Minkowski space.    Pick an auxiliary Euclidean signature metric on $D_q^\S$.   Given a sequence $\gamma_1,\gamma_2,\cdots\in D_q^\S$,
parametrize the $\gamma_i$ by a parameter $s$ that is a multiple of the Euclidean arclength, normalized to run over the interval $[0,1]$, with $s=0$ corresponding to the initial
endpoint at $q$, and $s=1$ corresponding to a final endpoint in $\S$.   The $\gamma_i$ already coincide at $s=0$.   Using the compactness of $D_q^\S$, one can extract
a subsequence of the $\gamma_i$ that converges at $s=1$, a further subsequence that converges at $s=1/2$, and eventually, as in section \ref{classic},
 a subsequence that converges at a dense set of values of $s$.
Because the $\gamma_i$  are causal curves, wild fluctuations are impossible and this subsequence converges for all values of $s$.

If $q$ is to the future of $\S$, we write $\CC_\S^q$ for the space of causal paths from $\S$ to $q$ and $D_\S^q$ for the space of points that can be visited by such
a path.   The above reasoning has an obvious mirror image to show that $\CC_\S^q$ and $D_\S^q$ are compact.  

 \begin{figure}
 \begin{center}
   \includegraphics[width=2.5in]{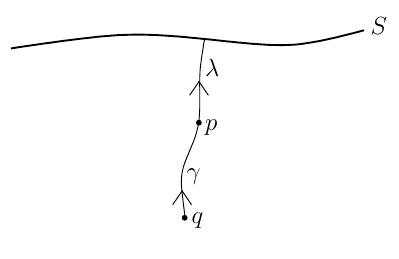} 
 \end{center}
\caption{\small  Suppose that $p$ is to the past of a hypersurface $\S$ and $q$ is to the past of $p$.   A causal path $\gamma$ from $q$ to $p$ and a causal path $\lambda$ from
$p$ to $\S$ can be conjoined or ``composed'' to make a causal path $\gamma *\lambda$ from $q$ to $\S$. \label{Fig5M}}
\end{figure}
Now let $p$ and $q$ be points in $M$ with $p$ to the future of $q$.   Let $\CC_q^p$ be the space of causal curves from $q$ to $p$, and $D_q^p$
the space of points that can be visited by such a curve (thus $D_q^p$ is the intersection of the causal past of $p$ with the causal future of $q$).   
We want to show that $D_q^p$ and $\CC_q^p$ are compact.   Suppose, for example, that $q$ and $p$ are both to the past of $\S$.
Let $\lambda$ be some fixed causal path from $p$ to $\S$.   If $\gamma $ is any causal path from from $q$ to $p$, then we define
$\gamma *\lambda$ to be the ``composition'' of the two paths (fig. \ref{Fig5M}).   Then $\gamma\to \gamma * \lambda$ gives an embedding of $\CC_q^p$ in $\CC_q^\S$.  Given a sequence $\gamma_1,\gamma_2,\cdots$
in $\CC_q^p$, compactness of $\CC_q^\S$ implies that the sequence $\gamma_1* \lambda,\gamma_2*\lambda,\cdots$ has a convergent
subsequence, and this determines a convergent subsequence of the original sequence $\gamma_1,\gamma_2,\cdots$.   So $\CC_q^p$ is compact.
As in our discussion of $\CC_q^\S$, this implies as well compactness of $D_q^p$.   

Clearly, nothing essential changes in this reasoning if $q$ and $p$ are both to the future of $\S$.   What if $q$ is to the past of $\S$ and $p$ to the future?
One way to proceed is to observe that a causal path $\gamma$ from $q$ to $p$ can be viewed as the composition $\mu *\lambda$
of a causal path $\mu$ from $q$ to $\S$ and a causal path $\lambda$ from $\S$ to $p$.   So a sequence $\gamma_n\in \CC_q^p$
can be viewed as a sequence $\mu_n*\lambda_n$, with $\mu_n\in \CC_q^\S$, $\lambda_n\in \CC_\S^p$.   Compactness of $\CC_q^\S$ and $\CC_\S^p$
means that after restricting to a suitable subsequence, we can assume that $\mu_n$ and $\lambda_n$ converge.   This gives a convergent
subsequence of the original $\gamma_n$, showing compactness of $\CC_q^p$.  Accordingly $D_q^p$ is also compact.  

Compactness of the spaces of causal paths implies that in a globally hyperbolic
spacetime $M$, just as in Minkowski space, there is a causal path of maximal elapsed proper time from $q$ to any point $p$ in its future.
This path will be a geodesic, of course.    Assuming that $p$ can be reached from $q$ by a causal path that is not everywhere null, this geodesic will be timelike.
(If every causal path from $q$ to $p$ is null, then every such path is actually a null geodesic.   This important case will be analyzed in section \ref{ng}.)

  Similarly, if $\S$ is a Cauchy  hypersurface, and $q$ is a point not on $\S$,  then there is a causal path
of maximal elapsed proper time from $q$ to $\S$.  Such a path will be a timelike geodesic $\gamma$ that satisfies some further conditions that we will discuss in section \ref{loran}.
 
 \subsection{Cauchy Horizons}\label{ch}

 \begin{figure}
 \begin{center}
   \includegraphics[width=2.5in]{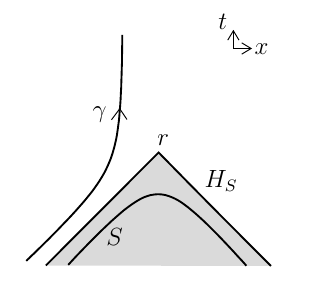}\caption{\small    Let $\S$ be the past hyperboloid $t^2=\vec x^2+R^2$, $t<0$ in Minkowski spacetime.  It is spacelike and achronal, but it is not
   a Cauchy hypersurface because, as shown, it is possible for a causal curve $\gamma$ that is infinitely extended to both the past and future to never intersect $\S$.    The domain of
   dependence of $\S$ is the interior of the past light cone of the origin (the point labeled $r$).   The Cauchy horizon $H_S$ is the past light cone of the origin.   \label{Fig6M}}\end{center}
\end{figure} 
 
Sometimes in a spacetime $M$, one encounters a spacelike hypersurface $\S$ that is achronal, but is not an initial value hypersurface because
there are inextendible causal paths in $M$ that never meet $\S$.  For an example, see fig. \ref{Fig6M}.  

 Nevertheless, it will always be true that if $p$ is a point just slightly to the future
of  $q\in \S$, then any inextendible causal path through $p$ will meet $\S$.   This is true because a very small neighborhood of $q$ can be
approximated by a small open set in Minkowski space, with $\S$ approximated by the spacelike hyperplane $t=0$.  The statement is true in that model example,
so it is true in general.

\begin{figure}
 \begin{center}
   \includegraphics[width=4in]{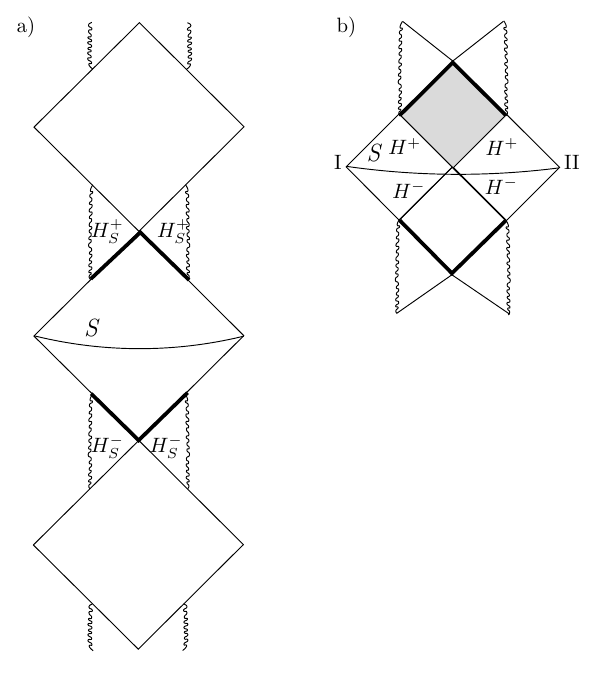}\caption{\small  (a) This is the Penrose diagram of the maximal analytic extension of the Reissner-Nordstr\"{o}m spacetime.
  It  describes a black hole carrying electric and/or magnetic charge.   The full spacetime $\h M$ is not globally hyperbolic, but
each  ``diamond,'' if considered by itself, is globally hyperbolic.   For example, the diamond
   shown in the center  represents a globally hyperbolic spacetime $M$ with Cauchy hypersurface $\S$.   The null surfaces labeled $H_S^+$ and $H_S^-$ (thick black lines)
    are the future and past Cauchy horizons of $\S$; the spacetime continues
   to the future of $H_S^+$ or to the past of $H_S^-$, but the continuation is not uniquely determined by initial data on $\S$.   Sketched here is the unique  continuation that is real analytic; 
   beyond $H_S^\pm$,
it has timelike singularities (represented by the wiggly vertical lines)   and repeats indefinitely into the past and future,
   with multiple globally hyperbolic ``diamonds.''   (b) We take a closer look at one causal diamond $M$.   It contains two asymptotically flat regions, labeled  I and II, connected by a ``wormhole''
   (see section \ref{cc}),
   and separated by  future and past horizons, labeled $H^\pm$.  To the future of $H^+$, from the standpoint of an observer in region I or II,
    is a black hole region.   The part of this region that is before the future Cauchy horizon is represented by the small shaded diamond.
    To the past of $H^-$ is a white hole region whose portion in the globally hyperbolic spacetime is represented by the small unshaded diamond.  
    (Much of this structure is also present in the simpler case of Schwarzschild spacetime, where it is exhibited in fig. \ref{Fig40M} of section \ref{cc}.)  In particular, the Cauchy horizons (again
    shown as thick black lines) are respectively to the future of $H^+$ and the past of $H^-$.  This structure
     was omitted in (a) to avoid clutter.   \label{Fig41M}}\end{center}
\end{figure}

This suggests the following
definition.    The {\it domain of dependence} $\DD_\S$ of $\S$ consists of all points $p\in M$ with the property that every inextendible causal curve
through $p$ meets $\S$.    $\DD_\S$ is the largest region in $M$ in which the physics can be predicted from a knowledge of initial conditions
on $\S$.   
 
$\DD_\S$ can be regarded as a spacetime in its own right.   (The only nontrivial point here is that $\DD_\S$ is open in $M$, so that it is a manifold.)
As such, $\DD_\S$ is globally hyperbolic, with $\S$ as an initial value surface.   This is clear from the way that $\DD_\S$ was defined; we threw
away from $M$ any point that lies on an inextendible causal curve that does not meet $\S$.     Therefore, all results for globally hyperbolic spacetimes
apply to $\DD_\S$.

In particular, $\S$ divides $\DD_\S$ into a ``future'' and a ``past,'' which are known as the future and past domains of dependence of $\S$, denoted $\DD_\S^+$ and
$\DD_\S^-$.

The boundary of the closure of $\DD_\S$ (or of $\DD_\S^+$ or $\DD_\S^-$) is called the Cauchy horizon, $H_\S$ (or the future or past Cauchy horizon $H_\S^+$ or $H_\S^-$).
For an additional but more complicated example of a spacetime with Cauchy horizons, see fig. \ref{Fig41M}.  (Details of this example are not important in the rest of the article.  There is a much simpler Penrose diagram in section \ref{trapped}, and in that context we will give a very short explanation of the meaning of such diagrams.
To appreciate the more complicated example of fig. \ref{Fig41M} requires familiarity with the analytic continuation of the Reissner-Nordstr\"{o}m solution
of General Relativity,  which is  described for example in \cite{Wald} or \cite{MTW}.)

\subsection{Causality Conditions}\label{causality}

The most obvious causality condition in General Relativity is the absence of closed causal curves: there is no (nonconstant) closed causal curve from a point $q$ to itself.

It turns out that to get a well-behaved theory, one needs a somewhat stronger causality condition.   The causality condition that we will generally use in the present article
is global hyperbolicity.   It is the strongest causality condition  in wide use. We have seen that it implies the absence of closed causal curves.

A somewhat weaker condition, but still stronger than the absence of closed causal curves,
 is ``strong causality.'' A spacetime $M$ is strongly causal if every point $q\in M$ has an arbitrarily small neighborhood $V$ with the property that any causal
curve between points $p,p'\in V$ is entirely contained in $V$. (Arbitrarily small means that if $U$ is any open set containing $q$, then $q$ is contained in a smaller open
set $V\subset U$ that has the stated property.)  The absence of closed causal curves
can be regarded as the special case of this with $p=p'=q$, since if every closed causal curve from a point $q$ to itself is contained in an arbitrarily small open set around $q$,
this means that there is no nontrivial causal curve from $q$ to itself.   Strong causality roughly says that there are no causal curves that are arbitrarily close to being closed.  

For an example of a spacetime that has no closed timelike curves but is not strongly causal, consider the two-dimensional spacetime $M$ with metric tensor
\be\label{vunxo}\d s^2=-\d v \d u+ v^2 \d u^2,\ee
where $v$ is real-valued, but $u$ is an angular variable, $u\cong u+2\pi$.   A short calculation shows that the only closed causal curve in $M$ is the curve $v =0$.     Suppose
that we remove a point $p$ on that curve, to make a new spacetime $M'$.   Then strong causality is violated at any point $q\in M'$ that has $v=0$.   There is no closed causal curve
from $q$ to itself (since the point $p$ was removed), but there are closed causal curves from $q$ that come arbitrarily close to returning to $q$; these are curves that remain everywhere
at very small $v$.   It is reasonable to consider such behavior
to be unphysical.

We already met a typical application of strong causality in section \ref{classic}: it  ensures that a causal diamond $D_q^p$, with $p$ just slightly to the future of $q$,
is compact. To justify this statement in somewhat more detail, let $U$ be a local Minkowski neighborhood of $q$.   If strong causality holds
at $q$, then $q$ is contained in an open neighborhood $V\subset U$ with the property that any causal curve between two points in $V$ is entirely contained in $V$.
So if $p\in V$, the causal diamond $D_q^p$ is contained in $V$ and therefore in $U$. Because $U$ is a local Minkowski neighborhood, $D_q^p$ is compact.  
    Another typical application  will appear in section \ref{prompt}.

Minkowski space is strongly causal, because if $U$ is any neighborhood of a point $q$ in Minkowski space, then $q$ is contained in some causal diamond $D_r^{s}$ (with $r$ just
to the past of $q$, and $s$ just to its future) that
is in turn contained in $U$.   Moreover, any causal curve in Minkowski space between two points in $D_r^s$ is entirely contained in $D_r^s$.  Thus we can use the interior
of $D_r^s$ as
the open set $V\subset U$ in the definition of strong causality.

More generally, globally hyperbolic spacetimes are strongly causal.  The proof  is deferred to Appendix \ref{detailedproof}.   
In reading this article, one may assume that spacetime is globally hyperbolic and strongly causal.  It is not important on first reading to understand that the first assumption
implies the second.  The assumption of strong causality is not always stated
explicitly, as it does follow from global hyperbolicity. 

It turns out there are also advantages to a causality condition somewhat stronger than  strong causality, known as ``stable causality.''   For the definition, and a proof that
globally hyperbolic spacetimes are stably causal, see \cite{Wald}, p. 198.

\subsection{Maximal Extensions}\label{maximal}

To avoid artificial examples, one usually places one more technical condition on a globally hyperbolic spacetime.

To explain what we wish to avoid, let $M$ be Minkowski space, and let $M'$ be the subset of $M$ defined by a condition $-1<t<1$ on the ``time'' coordinate $t$ (in some Lorentz
frame).    Then $M'$ is a globally hyperbolic spacetime, with Cauchy hypersurface  $t=0$.  A timelike path in $M'$ will end after a finite proper time,
but only because of the way we truncated the spacetime.

 In discussing a globally hyperbolic spacetime $M$ with Cauchy hypersurface $\S$, to avoid such artificial examples,
 one typically asks that $M$ be maximal under the condition that physics in $M$ can be predicted
based on data in $\S$.   In other words, one asks that $M$ be maximal\footnote{A theorem of Choquet-Bruhat and Geroch \cite{CG} states
that, given initial data on a Cauchy hypersurface $S$, a maximally extended globally hyperbolic
spacetime $M$ obeying the Einstein equations always exists and is unique up to diffeomorphism.}  under the condition that it is globally hyperbolic with Cauchy hypersurface $\S$.

It can happen that $M$ can be further extended to a larger spacetime $\h M$, but that this larger spacetime is not globally hyperbolic.   If so, there is a Cauchy horizon in $\h M$ and
$M$ is the domain of dependence of $\S$ in $\h M$.   

 A well-known example is the maximal analytic extension of the Reissner-Nordstr\"{o}m solution, describing a spherically symmetric
 charged black hole.  Let $\h M$ be this spacetime; the Penrose diagram is in fig. \ref{Fig41M}.    The spacelike surface $\S$ is a Cauchy hypersurface for a spacetime $M\subset \h M$
that includes the region exterior to the black hole (in each of two asymptotically flat worlds, in fact) as well as a portion of the black hole interior.   Behind the black hole
horizon are past and future Cauchy horizons, so $\S$ is not a Cauchy hypersurface for all of $\h M$.

\section{Geodesics and Focal Points }\label{geofocal}

From here, we will navigate to the easiest-to-explain non-trivial result about singularities in General Relativity.  This means not following
the historical order.  The easiest result to explain is a theorem by Hawking \cite{Hawkingsing}
about the Big Bang singularity in traditional cosmology without inflation.
It is easier to explain because it involves only timelike geodesics, while
more or less all our other topics  involve the slightly subtler
case of null geodesics.

\subsection{The Riemannian Case}\label{riemann}

 \begin{figure}
 \begin{center}
   \includegraphics[width=3in]{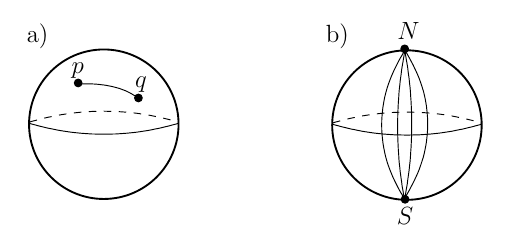} 
 \end{center}
\caption{\small  (a) A geodesic between two points $p$ and $q$ in a two-sphere that goes less than half way around the sphere minimizes
the length between those two points.   (b)   Any geodesic through the north pole $N$ reaches the south pole $S$ after going half-way around the sphere;
the point $S$ is called a focal point for these geodesics.   The geodesics from $N$ to $S$ are lines of constant longitude, as drawn.  When continued more than half-way around
the sphere, these geodesics are no longer length minimizing.   \label{Fig13M}}
\end{figure}

 \begin{figure}
 \begin{center}
   \includegraphics[width=2.4in]{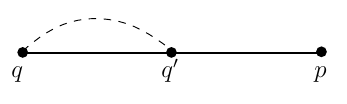} 
 \end{center}
\caption{\small  If the geodesic segment $qq'$ can be deformed -- at least to first order --  to a nearby geodesic connecting the same two points,
then $q'$ is called a focal point for geodesics that emanate from $q$.    \label{Fig14M}}
\end{figure}

We will start in ordinary Riemannian geometry, where we have more intuition,\footnote{For more detail on this material, see for example \cite{Jost}, chapter 4.}
and then we will go over to the Lorentz signature case.    Here is a question:   in Riemannian geometry, is a geodesic the shortest
distance between two points?    The answer is always ``yes'' for a sufficiently short geodesic, but in general if one follows a geodesic
for too far, it is no longer length minimizing.   An instructive and familiar example is the two-sphere
with its round metric.  A geodesic between two points $q$ and $p$
that goes less than half way around the sphere is the unique shortest path between those two points (fig. \ref{Fig13M}(a)).
But any geodesic that leaves $q$ and goes more than half way around the sphere is no
longer length minimizing.    What happens
is illustrated  in fig. \ref{Fig13M}(b) for the case that $q$ is the north pole $N$.  The geodesics that 
emanate from $N$ initially separate, but after going half way around the sphere, they reconverge at the south pole $S$.  The point of reconvergence is called a focal point or  a conjugate point.
  A geodesic that is continued past a focal point and thus has gone more than half way around the sphere is no longer length minimizing.   By ``slipping it around the
  sphere,'' one can replace it with a shorter geodesic between the same two points that goes around the same great circle on the sphere in the opposite direction.

This phenomenon does not depend on any details of the sphere.  Consider geodesics that originate at a point $q$ in some Riemannian manifold $M$.  
 Let $qp$
be such a geodesic and suppose   (fig. \ref{Fig14M}) that the $qq'$ part of this geodesic can be deformed slightly to another nearby
geodesic that also connects the two points $q$ and $q'$.  This displaced geodesic automatically has the same length as the first one since geodesics are stationary points of the length function.
 Then the displaced path $qq'p$ has a ``kink'' and
its length can be reduced by rounding out the kink.   So the original geodesic $qp$ was not length minimizing.

 \begin{figure}
 \begin{center}
   \includegraphics[width=3in]{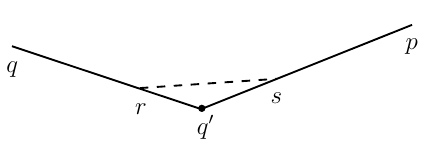} 
 \end{center}
\caption{\small    This is a new picture of the displaced path $qq'p$ that was introduced in
 fig. \ref{Fig14M}.   Here we emphasize that this path is made of two geodesic segments $qq'$ and $q'p$ that meet
at a ``kink.''    A
small neighborhood of $q'$, containing the nearby points $r$ and $s$, can be approximated by
a portion of flat Euclidean space.   The triangle inequality of Euclidean space says that the portion $rq's$ of $qq'p$ can be shortened by replacing it by a straight line
$rs$, shown as a dashed line.  So ``rounding out the kink''  of $qq'p$ reduces its length. \label{Drawing}}
\end{figure}

The fact that rounding out the kink reduces the length is basically the triangle inequality of Euclidean space, as explained in fig. \ref{Drawing}.  Indeed, a small
neighborhood of the point $q'$ can be approximated by a corresponding neighborhood in Euclidean space, and in Euclidean space, the triangle inequality says that rounding out
a kink reduces the length.    Quantitatively, if the displaced path $qq'p$ ``bends'' by a small
angle $\alpha$ at the point $q$, then rounding out the kink can reduce the length by an amount of order $\alpha^2$.  One can verify this with a little plane geometry, using
the fact that the geometry near the kink  can be embedded in Euclidean space.  Since
$\alpha$ is proportional to the amount by which the $qq'$ segment of the original geodesic $qp$ was displaced, this means that rounding out the kink reduces the length by an amount that is of second
order in the displacement.

It is not important here  that {\it all} geodesics from $q$ are focused to $q'$ (as happens in the case of a sphere).   To ensure that the geodesic
$qq'p$ is not length minimizing, it is sufficient that there is {\it some} direction in which the $qq'$ part can be displaced, not changing its endpoints.
  We do not even need to know that the  geodesic segment $qq'$ can be displaced {\it exactly} as a geodesic.    We only need to know that it can be displaced
while still solving the geodesic equation in first order.     That ensures that the displacement does not change the length function in second order. 
Rounding off the kink in $qq'p$ does reduce the length in second order, so displacing the $qq'$ segment and rounding off the kink will reduce
the length if the displacement caused no increase in second order.   

We will explain this important point in a little more detail.  A geodesic is a curve that extremizes the length between its endpoints, so any displacement of the geodesic segment $qq'$ to a nearby
path from $q$ to $q'$ will not change the length of this segment
in first order.   A displacement that obeys the geodesic equation in first order will leave fixed the length in second order.\footnote{To orient oneself to this
statement, let
 $f(y)$ be a smooth function of a real variable $y$, and suppose an equation $\d f/d y=0$ is satisfied at $y=y_0$.   We expand near $y=y_0$ by $y=y_0+\delta y$.
Since $f'(y_0)=0$, the general form of the expansion is $f(y)=f(y_0)+\frac{1}{2}\delta y^2 f''(y_0)+\O(\delta y^3)$.  But suppose that the equation $f'(y)=0$
is still satisfied at linear order in $\delta y$.   Since $f'(y)=\delta y f''(y_0)+\O(\delta y^2)$, this statement is equivalent to $f''(y_0)=0$, so it implies that
$f(y)$ is independent of $\delta y$ up to  order  $\delta y^3$.  In our application, what plays the role of $y$ is a path $x(s)$ between given points $q$ and $q'$,
and what plays the role of $f$ is the length function $L$ on the space of such paths.  The geodesic equation $\delta L/\delta x(s)=0$ is the analog of $f'(y)=0$.   A first order deformation  $\delta x(s)$ of a geodesic that preserves the geodesic equation $\delta L/\delta x(s)=0$ is the analog of a deformation $y=y_0+\delta y$
that satisfies the equation $f'(y)=0$ to first order in $\delta y$. So such a deformation leaves $L$ fixed in second order in $\delta x(s)$.  First order deformations
of geodesics are discussed in detail (in the null case) in  section \ref{newlook}.  }  
But rounding out the kink reduces the length in second
order, as we have noted. 
So    if a displacement obeys the geodesic equation in first order, then making a small displacement and rounding off the kink
will reduce the length.  All this will have an analog for timelike geodesics in Lorentz signature.

To summarize, if a geodesic segment $qq'$ can be displaced, at least in first order, to a nearby geodesic from $q$ to $q'$, we call the point $q'$ a focal point (or conjugate point).
A geodesic that emanates from $q$ is no longer length minimizing once it is continued past its first focal point.   The absence of a focal point is, however, only a necessary condition for a geodesic
to be length minimizing, not a sufficient one.   For example, on a torus with a flat metric,  geodesics have no focal points no matter how far they are extended.
On the other hand, any two points on the torus can be connected by infinitely many
different (and homotopically inequivalent) geodesics.   Most of those geodesics are not length minimizing.

 \begin{figure}
 \begin{center}
   \includegraphics[width=4in]{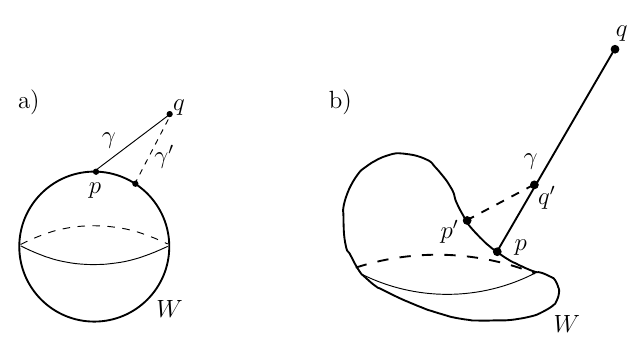} 
 \end{center}
\caption{\small This figure illustrates necessary (not sufficient)  conditions for a geodesic $\gamma$ to be the shortest path from a point $q$ to a submanifold $W$. (a) $\gamma$ must be orthogonal
to $W$ at the point $p$ at which they meet, or its length can be reduced by moving $p$ as shown, to make an orthogonal geodesic $\gamma'$.
(b) There must be no focal point along $\gamma$.   Here a point $q'\in\gamma$ is called a focal point if the segment $pq'$ of $\gamma$ can be displaced, at least to first order,
to another geodesic from $W$ to $q'$ that is also orthogonal to $W$.  This is not possible for ``outgoing'' geodesics orthogonal to a convex body in Euclidean
space, but it can occur
in a more general situation, as shown. \label{Fig15M}}
\end{figure}
Often, we are interested in a length minimizing path, not from a point $q$ to a point $p$, but from $q$ to some given set $W$. 
(This will be the situation when we are proving Hawking's singularity theorem.)   The simple case is that $W$ is a submanifold, without boundary.  
 A path
that {\it extremizes} the distance from $q$ to $W$ is now a geodesic that is orthogonal to $W$.   The condition here of orthogonality is familiar from elementary
geometry.   If $\gamma$ is a geodesic from $q$ to a point $p\in W$ but $\gamma$ is not orthogonal to $W$ at $p$, then the length of $\gamma$
can be reduced by moving slightly the endpoint $p$ (fig. \ref{Fig15M}(a)).

Assuming that $\gamma$ is orthogonal to $W$ at $p$, it will be length minimizing (and not just length extremizing) if $q$ is close enough to $W$.
Once again, however, if $q$ is sufficiently far from $W$, it may develop a focal point, and in that case it will no longer be length minimizing.
Now, however, the appropriate definition of a focal point is slightly different (fig. \ref{Fig15M}(b)).   
A geodesic $\gamma$ from $q$ to $W$, meeting $W$ orthogonally at some
point $p\in W$, has a focal point $q'$ if the $q'p$ portion of this geodesic can be displaced -- at least to first order -- to a nearby  geodesic, also connecting $q'$
to $W$ and meeting $W$ orthogonally.   In this situation, just as before, by displacing the $q'W$ portion of $\gamma$ and then rounding
out the resulting ``kink,'' one can find a shorter path from $q$ to $W$.

\subsection{Lorentz Signature Analog}\label{loran}

Now we go over to Lorentz signature.    What we have said has no good analog for {\it spacelike} geodesics. 
A spacelike geodesic in a spacetime of Lorentz signature is never a minimum or a maximum of the length function, since oscillations in spatial
directions tend to increase the length and oscillations in the time direction tend to reduce it.    Two points at spacelike separation can be
separated by an everywhere spacelike path that is arbitrarily short or arbitrarily long.

 \begin{figure}
 \begin{center}
   \includegraphics[width=.5in]{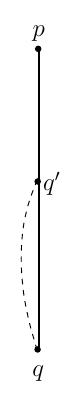}\caption{\small  This figure is a Lorentz signature analog of fig. \ref{Fig14M}.  (Time runs vertically here and in fig. \ref{Drawing2}.)   A timelike geodesic that originates at a point $q$
    does not maximize the elapsed proper time along its path
   if it contains a focal point $q'$.   For in that case, displacing the $qq'$ segment of the given geodesic and smoothing out the resulting kink will increase the elapsed proper time.    \label{dufo}}\end{center}
\end{figure} 
However, what we have said does have a close analog for {\it timelike} geodesics.      Here we should discuss the elapsed proper time
of a geodesic (not the length) and spatial fluctuations tend to reduce it.    So a sufficiently short segment of any timelike geodesic $\gamma$ {\it maximizes} the elapsed
proper time.\footnote{In any spacetime, this is true at least locally, meaning that if  $q,p\in\gamma$ with $p$ slightly to the future of $q$, then $\gamma$ has a greater proper time than
any nearby causal path from $q$ to $p$.  If $M$ is strongly causal, then the local statement is enough, since causal curves between sufficiently nearby points do not make large excursions.}   But if we continue a timelike geodesic past a focal point, it no longer maximizes the proper time.    

\begin{figure}
 \begin{center}
   \includegraphics[width=1in]{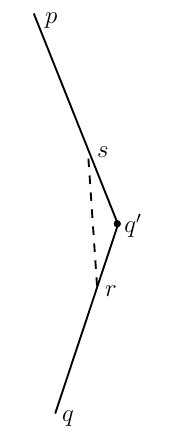}\caption{\small The displaced path $qq'p$ of fig. \ref{dufo} has been redrawn in a way that emphasizes
   that the segments $qq'$ and $q'p$ are timelike geodesics.  A small neighborhood of $q'$ containing the points $r$ and $s$ can be approximated by a portion of Minkowski space.
   The ``twin paradox'' of special relativity says that the proper time elapsed 
   along the portion $rq's$ of $qq'p$ is less than the proper time elapsed along the geodesic $rs$, which is shown as a dashed line.  
   (In other words, the twin who takes a trip on the worldline
   $rq's$ comes back younger than the twin who stays home on the worldline $rs$.)
 Thus, the proper time of $qq'p$ can be increased by rounding out the kink, replacing $rq's$ with $rs$. \label{Drawing2}}\end{center}
\end{figure}

 The appropriate definition of a focal point is the same as before.  
Consider a future-going timelike geodesic that originates at a point $q$ in spacetime (fig. \ref{dufo}).
Such a geodesic is said to have a  focal point at $q'$ if the $qq'$ part of the geodesic can be slightly displaced to another timelike geodesic connecting $q$ to $q'$.
This displacement produces a kink at $q'$, and rounding out the kink will increase the proper time.   
That rounding out the kink increases the proper time is basically the ``twin paradox'' of Special Relativity (fig. \ref{Drawing2}), in the same sense that the analogous statement in Euclidean signature
is the triangle inequality  of Euclidean geometry  (fig. \ref{Drawing}).

As in the Euclidean signature case and for the same reasons, to ensure that the original geodesic $qp$ does not maximize the proper time, 
it is not important here that the $qq'$ segment of $qp$ can be displaced exactly as a geodesic from $q$ to $q'$.  It is sufficient if this displacement can be made to first order.   

Here 
are two examples of timelike geodesics that do not maximize the proper time between initial and final points.  The first arises in the motion of the Earth
around the Sun.  Continued over many orbits, this motion is a geodesic that does not maximize the proper time.  One can do better
by launching a spaceship with almost the escape velocity from the Solar System, with the orbit so adjusted that the spaceship
falls back to Earth after a very long time, during which the Earth makes many orbits.  The elapsed proper time is greater for the spaceship than for the Earth
because it is less affected both by the gravitational redshift and by the  Lorentz time dilation.

A second example arises in Anti de Sitter spacetime.\footnote{This example can be omitted on first reading; alternatively, see Appendix \ref{ads} for background.}  Here we may refer back to fig. \ref{Fig1} of section \ref{classic}.   Future-going timelike geodesics from $q$ meet at the focal point $q'$.  These
geodesics fail to be proper time maximizing when continued past $q'$.   For example, the timelike geodesic $qw$ shown in the figure is not proper time
maximizing.  Indeed, there is no upper bound on the proper time of a  causal path from $q$ to $w$, since a timelike path from $q$ that travels very close to the edge of the figure,
lingers there for a while, and then goes on to $w$ can have an arbitrarily large elapsed proper time.

The absence of a focal point is only a necessary condition for a timelike geodesic $\ell$ from $q$ to $p$ to maximize the proper time, not a sufficient condition.
The presence of a focal point means that $\ell$ can be slightly deformed to a timelike path with greater elapsed proper time.  But even if this is not possible, there might be another
timelike path from $q$ to $p$, not a small deformation of $\ell$, with greater proper time.  Apart from examples we have already given, this  point can be illustrated using the cylindrical spacetime of eqn. (\ref{zorfo}).   This spacetime is flat,
and the timelike geodesics in it do not have focal points, no matter how far they are continued.   If the point $p$ is sufficiently far to the future of $q$, then there are multiple timelike geodesics
from $q$ to $p$, differing by how many times they wind around the cylinder en route from $q$ to $p$.   These timelike geodesics have different values of the elapsed proper time,
so most of them are not proper time maximizing, even though they have no focal point.

 \begin{figure}
 \begin{center}
   \includegraphics[width=4in]{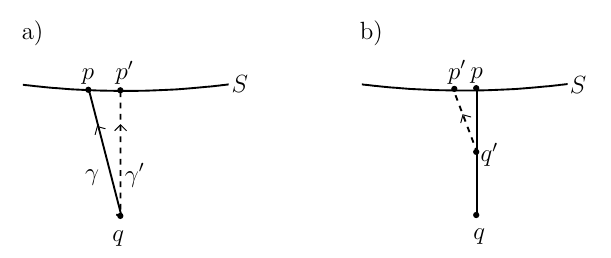}\caption{\small   This is a Lorentz signature analog of fig. \ref{Fig15M}, showing necessary conditions in order for a timelike geodesic
   $\gamma$ from a point $q$ to a spacelike hypersurface $S$ to maximize the elapsed proper time.  (a)  $\gamma$ must be orthogonal to $S$ at the point $p$ at which they meet, or else the
   proper time could be increased by moving $p$ along $S$ in the appropriate direction, to $p'$ as shown. (b)  And $\gamma$ must contain no focal point $q'$.   Here $q'$ is a focal point if the $q'p$ segment of
   $\gamma$ can be displaced (at least in first order) 
   to a nearby timelike geodesic $q'p'$, also orthogonal to $S$.    In that case, by rounding out the kink of the composite path $qq'p'$, one would get a timelike
   path from $q$ to $S$ with a proper time greater than that of $qp$.   Thus, in the picture, $q'$ is a focal point on the geodesic $qp$ if $q'p'$ is a geodesic
    orthogonal to $S$ at $p'$.   \label{lufo}}\end{center}
\end{figure} 

We can also consider a causal path $\gamma$ from a point $q$ to a spacelike submanifold $S$ in its future.   To maximize the elapsed proper time, $\gamma$ must satisfy conditions that parallel
what we found in the Euclidean case (fig. \ref{lufo}).  First, $\gamma$ must be a timelike geodesic from $q$ to $S$.  Second, $\gamma$ must be orthogonal to $S$ at the point $p$ at which it meets $S$.
Third, there must be no focal point $q'$ on $\gamma$.   $q'$ is a focal point if the $q'p$ segment of $\gamma$ can be slightly displaced to a nearby timelike geodesic, also connecting
$q'$ to $S$ and orthogonal to $S$.  

\subsection{Raychaudhuri's Equation}\label{rayeq}

To prove a singularity theorem, we need a good way to predict the occurrence of focal points on timelike geodesics.   Such a method is provided
by Raychaudhuri's equation.  (In fact, what is relevant here is Raychaudhuri's original timelike equation \cite{Raychaudhuri}, not the slightly more
subtle null version that was described later by Sachs  \cite{sachs} and that we will discuss in due course.)   Raychaudhuri's equation shows
that focal points are easy to come by, roughly because gravity tends to focus nearby geodesics.  

In $\D=d+1$ dimensions, we consider a spacetime $M$ with an initial value surface $S$ with local coordinates $\vec x= (x^1,\cdots, x^{d})$.   By looking at timelike geodesics
orthogonal to $S$, we can construct a coordinate system in a neighborhood of $S$.   If a point $p$ is on a timelike 
geodesic that meets $S$ orthogonally at $\vec x$, and the proper time from $S$ to $p$ (measured along the geodesic) is $t$,
then we assign to $p$ the coordinates $t,\vec x$ if $p$ is to the future of $S$, or $-t,\vec x$ if it is to the past.

In this coordinate system, the line element of $M$ is
\be\label{zolbo}\ds^2=-\d t^2+g_{ij}(t,\vec x) \d x^i\d x^j. \ee   
We can verify this as follows.  First of all, in this coordinate system, $g_{tt}=-1$, since $t$ was defined to 
measure the proper time along any path with constant $x^i$.  Further,  the geodesic equation can be written
\be\label{polbo}0=\frac{D^2 x^\mu}{D\tau^2}=\frac{\d^2 x^\mu}{\d\tau^2}+\Gamma^{\mu}_{\alpha\beta}\frac{\d x^\alpha}{\d\tau}\frac{\d x^\beta}{\d\tau}. \ee
In our coordinate system, this equation is supposed to have a solution with $t=\tau$, and with the $x^i$ equal to arbitrary constants.   For that to be so, we
need $\Gamma^\mu_{tt}=0$ for all $\mu$.  From this, it follows that $\partial_t g_{t i}=0$.   But $g_{ti}$ vanishes at $t=0$ (since the coordinate system is
constructed using geodesics that are orthogonal to $S$ at $t=0$) so $g_{ti}=0$ for all $t$.

Thus the coordinate system constructed using the orthogonal geodesics  could be obtained  by merely asking for coordinates in which
the metric tensor satisfies $g_{tt}=-1$, $g_{ti}=0$.
The advantage of the more geometric language of orthogonal geodesics is that this will help us understand how the coordinate
system can break down.  The conclusions we draw will be manifestly independent of the local coordinate system on $S$, which was chosen for convenience.

 \begin{figure}
 \begin{center}
   \includegraphics[width=1.8in]{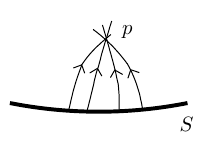}\caption{\small   If orthogonal geodesics from a spacelike hypersurface $S$ are focused at a point $p$ to the future of $S$, then the coordinate
   system based on the orthogonal geodesics will break down at $p$.  \label{Fig29M}}\end{center}
\end{figure}

Even if $M$ remains nonsingular, our coordinate system breaks down if orthogonal geodesics that originate at different points on $S$ meet at the same point $p\in M$.
For in this case, we do not know what $\vec x$ value to assign to $p$.   A related statement is that the coordinate system breaks down at focal points.  For if orthogonal geodesics
from nearby starting points converge at $p$ (fig. \ref{Fig29M}), then the starting points of the orthogonal geodesics will not be part of a good coordinate system near $p$.

Since $g_{ij}(t,\vec x)$ measures the distance between nearby orthogonal geodesics, a sufficient criterion for a focal point is
\be\label{holbo}\det\,g_{ij}(t,\vec x)=0.\ee
This condition is actually necessary as well as sufficient.
That is not immediately obvious, since 
it might appear that  if one eigenvalue of $g_{ij}(t,\vec x)$ goes to 0 and one to $\infty$, then $\det g_{ij}(t,\vec x)$ could remain fixed while
a focal point develops.    However, as long as $M$ remains smooth, the point in $M$ that a geodesic that is  
orthogonal to $S$ at a point $q\in S$ reaches after a proper time $t$ is a smooth function of $t$
and $q$.  Hence matrix elements and  eigenvalues of $g_{ij}(t,\vec x)$ never diverge except at a singularity of 
$M$, and so (with $M$ being smooth)   the determinant of $g_{ij}(t,\vec x)$ vanishes if and only if one
of its eigenvalues vanishes.

 Raychaudhuri's equation gives a useful criterion for predicting that $\det g_{ij}$ will go to 0 within a known time.   In general, 
 this will represent only a breakdown of the coordinate
system, not a true spacetime singularity, but we will see that the criterion provided by Raychaudhuri's equation is a 
useful starting point for predicting spacetime singularities.

Raychaudhuri's equation is just the Einstein equation
\be\label{raych} R_{tt}=8\pi G\left(T_{tt}-\frac{1}{\D-2}g_{tt}T^\alpha_\alpha\right)\ee
in the coordinate system defined by the orthogonal geodesics. 
A straightforward computation in the metric (\ref{zolbo}) shows that 
\begin{align}\label{wobbo} R_{tt} & = -\partial_t \Gamma^i_{ti}-\Gamma^i_{tj}\Gamma^j_{ti} =
-\frac{1}{2}\partial_t(g^{ik}\partial_t g_{ik})-\frac{1}{4}\left(g^{ik}\partial_t g_{kj}\right)\left(g^{jm}\partial_t g_{mi}\right)\cr
&=-\frac{1}{2}\partial_t \Tr \,g^{-1} \dot g -\frac{1}{4} \Tr (g^{-1}\dot g)^2,
\end{align}
where the dot represents a derivative with respect to $t$.

It is convenient to define
\be\label{naych} V=\sqrt {\det g} \ee
which measures the volume occupied by a little bundle of geodesics.    The quantity 
\be\label{aych}\theta= \frac{\dot V}{V}=\frac{1}{2}\Tr\, g^{-1} \dot g\ee
 is called the expansion. 

It is convenient to also define the traceless part of $g^{-1}\dot g$ (the ``shear'')
\be\label{shear}\sigma^i_j=\frac{1}{2}\left(g^{ik}\dot g_{kj}-\frac{1}{d}\delta^i_j\Tr \,g^{-1}\dot g\right),\ee
where the factor of $1/2$ is conventional. So
\be\label{rzero}R_{tt}=-\partial_t \left(\frac{\dot V}{V}\right)-\frac{1}{d}\left(\frac{\dot V}{V}\right)^2 -\Tr\, \sigma^2= -\dot\theta-\frac{\theta^2}{d}-\Tr\,\sigma^2 .\ee
If we define
\be\label{strest}\h T_{\mu\nu}=T_{\mu\nu}-\frac{1}{\D-2}g_{\mu\nu}T^\alpha_\alpha,\ee
then the Einstein-Raychaudhuri equation $R_{tt}=8\pi G\h T_{tt}$ becomes
\be\label{ER}\partial_t\left(\frac{\dot V}{V}\right)+\frac{1}{d} \left(\frac{\dot V}{V}\right)^2 =-\Tr\,\sigma^2 -8\pi G\h T_{tt}.\ee

The {\it strong energy condition} is the statement that
\be\label{minxo}\h T_{tt}\geq 0\ee
at every point and in every local Lorentz frame.    It is satisfied by the usual equations of state of ordinary radiation and matter.
It is also satisfied by a negative cosmological constant.  The outstanding example that does {\it not} satisfy the strong energy condition is a positive cosmological constant.\footnote{More generally, in a  theory -- such as the Standard Model of particle physics  -- with an elementary scalar field $\phi$ and a potential energy
function $U(\phi)$, the strong energy condition
is violated unless  $U(\phi)$ is negative-definite (this is not true in the Standard Model even if we assume the cosmological constant to vanish).}    If we assume the
strong energy condition, then all the terms on the right hand side of the Einstein-Raychaudhuri equation are negative and so we get an inequality
\be\label{yeg}\partial_t\left(\frac{\dot V}{V}\right)+\frac{1}{d} \left(\frac{\dot V}{V}\right)^2\leq 0.\ee
Equivalently,
\be\label{zeg}\partial_t\left(\frac{1}{\theta}\right)= \partial_t\left( \frac{1}{\dot V/V}\right)\geq \frac{1}{d}.\ee

Now we can get a useful condition for the occurrence of focal points.  Let us go back to our initial value surface $S$ 
and assume that
$\theta<0$ at some point on this surface, say $\theta=-w$, $w>0$.      So the initial value of $1/\theta$ is $-1/w$ and
the lower bound on $\partial_t(1/\theta)$ implies that $1/\theta\geq -1/w+t/d$ or
\be\label{pleg} \frac{\dot V}{V}\leq-\left(\frac{1}{w}-\frac{t}{d}\right)^{-1}. \ee
Since $\dot V/V={\d}\log V/\d t$, we can integrate this to get
\be\label{mleg} \log V(t)-\log V(0)\leq   {d}\bigl(\log (1/w -t/d)-\log (1/w)\bigr), \ee
showing that $\log V(t)\to -\infty$ and thus $V(t)\to 0$ at a time no later than $t=d/w$.

For $V(t)$ to vanish signifies a focal point, or possibly a spacetime singularity.   So (assuming the strong energy condition) 
an  orthogonal geodesic that departs from $S$ at $t=0$ at a point at which $\theta=-w<0$
will reach a focal point, or possibly a singularity, after a proper time $t\leq d/w$.

In many situations the vanishing of $V$ predicted by the Raychaudhuri equation represents only a focal point, a breakdown of the coordinate
system, and not a spacetime singularity.
The following example may help make this obvious.   For $M$, take Minkowski space, which certainly has no singularity.   For an initial value surface $S\subset M$, consider first
the flat hypersurface $t=0$.  For this hypersurface, $\theta$ vanishes identically, and the orthogonal geodesics are simply lines of constant $\vec x$; they do not meet at focal points.
    Now perturb $S$ slightly to $t=\varepsilon f(\vec x)$ for some
function  $f(\vec x)$ and small real  $\varepsilon$.  
The reader should be able to see that in this case,
 $\theta$ is not identically zero, and the orthogonal geodesics will reach focal points, as predicted by the Raychaudhuri equation.   In general, these focal points occur
 to the past or future of $S$, depending on which way $S$ ``bends'' in a given region.
 Clearly these focal points have nothing to do with a spacetime singularity.

Thus, to predict a spacetime singularity requires a more precise argument, with some input beyond Raychaudhuri's equation.

\subsection{Hawking's Big Bang Singularity Theorem}\label{hbb}

In proving a singularity theorem,
Hawking assumed that the universe is globally hyperbolic with Cauchy hypersurface $\S$.
  He also assumed the strong energy condition, in effect assuming that the stress tensor is  made of ordinary matter and radiation.   (The inflationary universe, which gives
a way to avoid Hawking's conclusion because a positive cosmological constant does not satisfy the strong energy condition, was still in the future.)
 If the universe is perfectly homogeneous and isotropic, it is described by the Friedmann-Lema\^itre-Robertson-Walker (FLRW)  solution and emerged
from the Big Bang at a calculable time in the past. 

Suppose, however, more realistically, that the universe is not perfectly homogeneous
but that the local Hubble parameter is everywhere positive.  Did such a universe emerge from a Big Bang?   One could imagine that following the Einstein
equations back in time, the inhomogeneities become more severe, the FLRW solution is not a good approximation, and part or most (or maybe even all)
of the universe did not really come from an initial singularity.

Hawking, however, proved that assuming the strong energy condition and assuming that the universe is globally hyperbolic (and of course assuming the classical
Einstein equations), this is not the case. 
 To be more exact, he showed that if the local Hubble parameter
has a positive minimum value $h_\min$ on an initial value surface $\S$,
then there is no point in spacetime that is a proper time more than $1/h_\min$ to the past of $\S$, along any causal path.

Here we should explain precisely what is meant by the local Hubble parameter.   For a homogeneous isotropic expansion such as
the familiar FLRW cosmological model 
\be\label{incob}\ds^2=-\d t^2+a^2(t) \d \vec x^2 ,\ee where for simplicity we ignore the curvature of the spatial sections,
 one usually defines
the Hubble parameter by $h=\dot a/a$.     We can view the line element (\ref{incob}) as a special case of eqn. (\ref{zolbo}),
and from that point of view, we have  $h=\dot V/dV$ (since $V=\sqrt {\det g} $ is the same as $a^{d}$ in the homogeneous isotropic case).  We will
use that definition in general, so the assumption on the Hubble
parameter is that in the coordinate system of eqn. (\ref{zolbo}), $\dot V/V\geq dh_\min$, assuming time is measured towards the future.   But we will measure time towards the past and so instead we write
the assumption as
\be\label{mumwot}\frac{\dot V}{V}\leq-d h_\min.\ee

Hawking's proof consists of comparing two statements.   (1)  Since the universe is globally hyperbolic, every point $p$ is connected to $\S$ by a causal path of maximal
proper time, as explained in section \ref{cpct}.
 As we know from the discussion of fig. \ref{lufo}, such a path is a timelike geodesic without focal points that is orthogonal to $\S$.
(2) But the assumption that the initial value of $\dot V/V$ on the surface $\S$ is everywhere $\leq -dh_\min$ implies that any past-going timelike geodesic
orthogonal to $\S$ develops a focal point within a proper time at most $1/h_\min$.

  Combining the two statements, we see that there is no point in spacetime that is to the past of $\S$ by a proper time more than $1/h_\min$, along any causal path.
 Thus (given Hawking's assumptions) the minimum value of the 
local Hubble parameter gives an upper bound on how long anything in the universe could
have existed in the past.  This is Hawking's theorem about the Big Bang.

An alternative statement of Hawking's theorem is that no timelike geodesic $\gamma$ from $\S$ can be continued into the past for a proper time greater than $1/h_\min$.    Otherwise, there would be a point
$p\in \gamma$ that is to the past of $\S$ by a proper time measured along $\gamma$
that is greater than $1/h_\min$, contradicting what was just proved.     

In Euclidean signature, a Riemannian manifold is said to be geodesically complete if all geodesics can be continued up to an arbitrarily large
distance in both directions.   In Lorentz signature, there are separate notions of completeness for timelike, spacelike, or null geodesics, which say that any timelike, spacelike, or null
geodesic can be continued in both directions,  up to arbitrarily large values of the proper time, the distance, or the affine parameter, respectively.    Hawking's theorem
shows  that a Big Bang spacetime that satisfies certain hypotheses is timelike geodesically incomplete, in a very strong sense: {\it no} timelike geodesic from $\S$ can be continued
into the past for more than a bounded proper time.   (Penrose's theorem, which we explore in section \ref{penroseproof}, gives a much weaker statement of null geodesic incompleteness:
under certain conditions,  not all null geodesics can be continued indefinitely.)

Although Hawking's theorem is generally regarded as a statement about the Big Bang singularity, singularities are not directly involved in the statement or proof of the theorem.
 In fact, to the present day, one has only a limited understanding of the implications of Einstein's equations
concerning singularities.  In the classic singularity theorems, going back to Penrose \cite{Penrosesing}, only the
 smooth part of  spacetime is studied, or to put it differently, ``spacetime'' is taken to be, by definition,
a manifold with a smooth metric of Lorentz signature (this is the definition that we started with in footnote \ref{recall} of section \ref{classic}).  Then
``singularity theorems'' are really statements about  geodesic incompleteness of  spacetime.   In the case of Hawking's theorem, one may surmise that the reason that the past-going
timelike geodesics from $\S$ cannot be continued indefinitely is that they terminate on singularities, as in the simple FLRW model, but this goes beyond what is proved.   When we come to 
Penrose's theorem, we will explore this point in more detail.

\section{Null Geodesics and Penrose's Theorem}\label{ng}

\subsection{Promptness}\label{prompt}

Hopefully, our study of timelike geodesics in section \ref{geofocal} was enough fun that the reader is eager for an analogous study of null geodesics.  In this section,
we explain the properties of null geodesics that are needed for applications such as Penrose's theorem and an understanding of black holes.   Some important points are  explained
only informally in this section and are  revisited more precisely in section \ref{newlook}.   A converse of some statements is explained in Appendix \ref{failure}.
 
Causal paths and in particular null geodesics will be assumed to be future-going unless otherwise specified.   Of course, similar statements
apply to past-going causal paths, with the roles of the future and the past exchanged.

 \begin{figure}
 \begin{center}
   \includegraphics[width=2in]{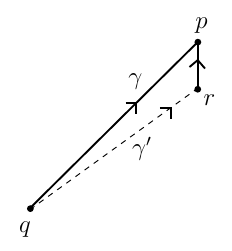}\caption{\small     In the diagram, time runs vertically, so $r$ is to the past of $p$.
    A causal path $\gamma$ from $q$ to $p$ is ``prompt'' if there does not exist
   any causal path $\gamma'$ from $q$ to a point such as $r$ that is to the past of $p$.
      \label{Fig20M}}\end{center}
\end{figure} 
 
 Any null geodesic has zero elapsed proper time.  Nevertheless, there is a good notion that has  properties somewhat similar to ``maximal elapsed proper time'' for timelike paths.
  We will say that a  causal
path from $q$ to $p$ is ``prompt'' if  no causal path from $q$ to $p$   arrives sooner.    To be precise, the path $\ell$ from $q$ to $p$
is prompt if  there is  no  causal path $\ell'$ from $q$ to a point $r$ near $p$ and just to its past (fig. \ref{Fig20M}).     

A prompt causal path from $q$ to $p$ only exists if it is just barely possible to reach $p$ from $q$ by a causal path.  
  To formalize this, we write $J^+(q)$ for the causal future
of $q$, consisting of all points that can be reached from $q$ by a future-going causal path,  and $\partial J^+(q)$ for its boundary.
  If $p$ can be reached from $q$ by a prompt causal path $\ell$, then $p\in J^+(q)$
(because $\ell$ is causal), and $p\in \partial J^+(q)$ (because points slightly to the past of $p$ are not in $J^+(q)$). Thus, almost
by definition, a prompt causal path from $q$ to $p$ can only exist if $p\in\partial J^+(q)$.  

 \begin{figure}
 \begin{center}
   \includegraphics[width=2.5in]{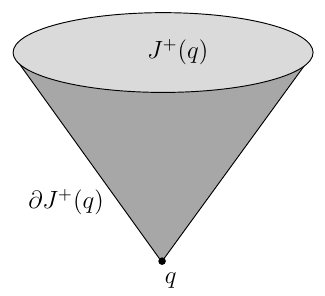}\caption{\small   If $q$ is a point in Minkowski space, then the causal future $J^+(q)$ of $q$ consists of all
   points in or on the future light cone of $q$.   Its boundary $\partial J^+(q)$ consists of the points on the future light cone.   We consider
   $q$ itself to be contained in $J^+(q)$ and in $\partial J^+(q)$.  \label{Fig21M}}\end{center}
\end{figure} 
 For example, if $q$ is a point in Minkowski space, then $J^+(q)$ consists of points inside or on the future light cone (fig. \ref{Fig21M}); $\partial J^+(q)$ consists of the future light cone itself.
 Every point $p\in \partial J^+(q)$ is connected to $q$ by a null geodesic, and this null geodesic is a prompt causal path.

In a spacetime that obeys a suitable causality condition, a short enough segment of a null geodesic  is always prompt as a path between its endpoints, just
as in Minkowski space.  This statement does require a causality condition, because it is untrue in a spacetime with closed timelike curves, as the reader can verify.
At the end of this section, we will show that strong causality is enough.  If
continued far enough, a null geodesic may become non-prompt because of gravitational lensing.    For example, when we see multiple images of the
same supernova explosion, the images do not arrive at the same time and clearly the ones that do not arrive first are not prompt.

\begin{figure}
 \begin{center}
   \includegraphics[width=3.8in]{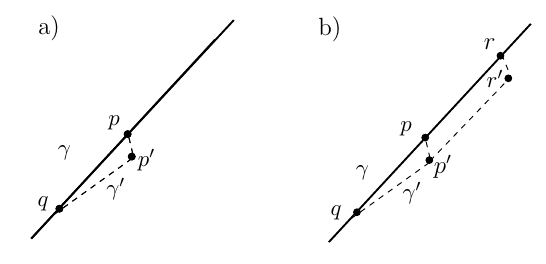}\caption{\small  (a) The null geodesic $\gamma$ is not a prompt path from $q$ to $p$, since $\gamma'$ is a causal path from $q$
   that arrives at the point $p'$ that is slightly to the past of $p$.   (b)     When continued on to a later point $r$, $\gamma$ is not prompt as a path from $q$ to $r$, because $\gamma'$
   can likewise be continued past $p'$ to a point $r'$ that is just to the past of $r$.    \label{46M}}\end{center}
\end{figure}

The prompt part of a null geodesic is always an initial segment, since if a null geodesic is continued until it becomes nonprompt, continuing it farther does not make it prompt again. 
 To see this, suppose that a future-going null geodesic $\gamma$ from $q$ to $p$  is continued
farther in the future until it reaches a point $r$.  Suppose that $\gamma$ is not prompt as a path from $q$ to $p$, meaning that there is a causal curve $\gamma'$ from $q$
that arrives at a point $p'$ just to the past of $p$ (fig. \ref{46M}(a)).    Then continuing $\gamma'$ into the future, always close to $\gamma$ and slightly in its past, we get a causal curve from $q$
to a point $r'$ that is just to the past of $r$, showing that $\gamma$ is not prompt as a path from $q$ to $r$ (fig. \ref{46M}(b)).    Note that a very small neighborhood of a single geodesic, such as $\gamma$, can be embedded
   to good approximation in Minkowski space, so one can visualize the relation between $\gamma$ and $\gamma'$  in the region between $p$ and $r$
   as the relation between a lightlike straight line in Minkowski space and a causal
   path (possibly close to a parallel lightlike line) that is slightly to its past.

Prompt causal paths have very special properties that ultimately make them good analogs of proper time maximizing timelike paths.
A causal path $\ell$ from $q$ to $r$  whose tangent vector is somewhere timelike
(rather than null) cannot be prompt, because by modifying $\ell$ slightly to be everywhere null, we could find a causal path  from $q$ that is slightly to the
past of $\ell$.   
Actually, to be prompt, $\ell$
has to be a null geodesic, since if it ``bends'' anywhere, one could take a shortcut by straightening out the bend (in some small region that can be approximated by Minkowski space)
and again replace $\ell$ with a causal path from $q$ that is slightly to its past.    In Minkowski space, these statements amount to the assertion that to get somewhere as quickly as possible,
one should travel in a straight line at the speed of light.   They are true in general because a sufficiently small portion of any spacetime can be approximated by Minkowski space.  If $\ell$
fails to be a null geodesic at a point $p\in \ell$, then, in a local Minkowski neighborhood of $p$, one can replace $\ell$ with a causal path
$\ell'$ that agrees with $\ell$ almost up to $p$ and thereafter is slightly to its past.  No matter what $\ell$ does to the future of $p$, one then continues $\ell'$ in a similar way, keeping always slightly to the
past of $\ell$.   The relation between $\ell'$ and $\ell$ in the region beyond $p$ is similar to the relation between $\gamma'$ and $\gamma$ in fig. \ref{46M}(b).

We conclude this section by showing that in a strongly causal spacetime $M$, every null geodesic has an initial segment that is prompt.  
Let $\gamma$ be a future-going null geodesic that originates at a point $q$.   If a point $p$ that is to the future of $q$  along $\gamma$ has the property that
$\gamma$ is the only causal path from $q$ to $p$, then the $qp$ segment of $\gamma$ is prompt.
The alternative -- that points on $\gamma$ arbitrarily close to $q$ can be reached from $q$
by some causal path other than $\gamma$ -- is not possible in a strongly causal spacetime.  To see this, 
observe that $q$ has a local Minkowski neighborhood  $U$ that can be well approximated by the interior of a causal diamond in Minkowski space.
This implies that   any null geodesic $qp\subset U$ is prompt among paths in $U$, and in fact it is the only causal path in $U$ from $q$ to $p$.
This much does not require strong causality.   If $M$ is strongly causal, then $q$ has a possibly smaller neighborhood $V\subset U$
with the property that  any causal path in $M$ between two points in $V$ is entirely contained in $V$.
Taking these statements together, we see that if $p\in V$,  a null geodesic segment from $q$ to $p$ is entirely contained in $V\subset U$ and so is prompt.   So if $\ell$ is any
future-going null geodesic from $q$, the initial segment of $\ell$ that lies in $V$ is prompt.

An analogous question for timelike geodesics is the following.
If $\gamma$ is a timelike geodesic, does a small initial segment of $\gamma$ maximize the proper time between its endpoints?   We leave it to the reader to verify that this
statement  is false in a spacetime with closed timelike curves, and  true in a strongly causal spacetime.

\subsection{Promptness And Focal Points}\label{pfp}

 \begin{figure}
 \begin{center}
   \includegraphics[width=2in]{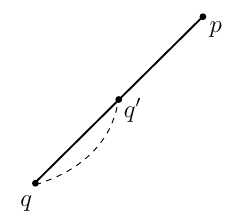}\caption{\small   Depicted is a null geodesic segment $qp$.   The point $q'$ is called a focal point of the null geodesics that
   originate at $q$ if the $qq'$ segment of $qp$ can be displaced, at least to first order, to a nearby null geodesic from $q$ to $q'$.   If it contains such a focal point, then the original
   geodesic $qp$ is not prompt.  \label{Fig22M}}\end{center}
\end{figure}

\begin{figure}
 \begin{center}
   \includegraphics[width=1.5in]{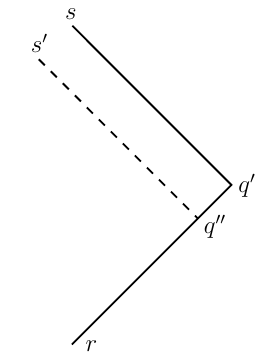}\caption{\small  Here $rq's$ is a neighborhood of the ``kink''  $q'$ in the displaced path $qq'p$ of fig. \ref{Fig22M}.     
   The picture is an analog of fig \ref{Drawing2}, which depicted a similar situation for timelike geodesics,  but in this null case, we only show a local Minkowski  
   neighborhood of the kink.    Thus we think of $rq'$ and $q's$ as noncollinear null geodesics in Minkowski space (represented in the picture by
    straight lines at  $\pi/4$ angles to the vertical;
   in fact, by a Lorentz transformation, two noncollinear null straight lines in Minkowski space can be made ``back to back,'' as drawn here).
  By replacing the null geodesic segment
   $q's$ with the null geodesic segment  $q''s'$ (the dashed line), which is parallel to $q's$ but to its past, one can replace $rq's$ with a more 
   prompt causal path $rq''s'$.   Then no matter what the path
   $q's$ does when continued into the future, one continues $q''s'$ into the future to be parallel to it and just to its past, as in fig. \ref{46M}(b).   
   The result is to replace the original displaced path $qq'p$
   with a causal path that is more prompt.
    \label{Drawing3}}\end{center}
\end{figure}

To be prompt, a null geodesic $\ell$ from $q$ to $p$ must contain no focal point.\footnote{For a detailed explanation of the following based on
formulas rather than pictures,  see section \ref{newlook}.}   Here $q'$ is a focal point if (at least in first order, as discussed shortly)
the $qq'$ segment of the geodesic $\ell=qp$ can be displaced to a nearby null geodesic
from $q$ to $q'$ (fig. \ref{Fig22M}).    If so, then by slightly displacing the $qq'$ segment of $\ell$, one gets a causal path from $q$ to $p$ that is not a null geodesic,
as it has a ``kink'' at $q'$.  Not being a null geodesic, this causal path is not prompt, as we learned in section \ref{prompt}, and can be slightly modified 
to make a causal path from $q$ that arrives slightly to the past of $p$.
In fig. \ref{Drawing3}, we zoom in on the kink $q'$ in the deformed path $qq'p$ and describe this modification concretely.   The details of fig. \ref{Drawing3} are a little different from the analogous
fig. \ref{Drawing} for geodesics in Euclidean signature or fig.  \ref{Drawing2} for timelike geodesics in Lorentz signature; one has to modify the deformed path $qq'p$ not just
near the kink but in a way that continues  to the future.   This is a consequence of the fact that null geodesics do
not minimize or maximize an integral such as  length or proper time; promptness is a more global notion that depends by definition on when the signal arrives.

Just as in the analogous discussion of timelike or Euclidean signature geodesics, it is not necessary here that the null geodesic segment $qq'$ can be displaced {\it exactly} to a nearby null geodesic connecting the same two points.
It suffices if the $qq'$ segment can be displaced in a way that  satisfies the null geodesic condition in first order. A thorough explanation of this point can be found in section \ref{newlook}.
The rough idea is that the amount by which the displaced path $qq'p$ can be made more
prompt by changing the behavior near and beyond the kink is of second order in the displacement, while if the displacement satisfies the geodesic equation in first order, then the fact that it does not
satisfy that equation exactly has no effect in second order.

We can illustrate focal points of null geodesics and their consequences
 by promoting to Lorentz signature an example that we discussed in the Euclidean case in section \ref{riemann}.   Let $M=\R\times S^2$ with the obvious
line element
\be\label{mocc}\ds^2=-\d t^2+R^2\d\Omega^2,\ee
where $\R$ is parametrized by the time $t$, and $R^2\d\Omega^2$ is the line element of a sphere of radius $R$.  A null geodesic in this spacetime is simply 
a Riemannian geodesic on $S^2$
with its arclength parametrized by $t$.  As discussed in section \ref{riemann}, in the Euclidean case, a geodesic on $S^2$ reaches a focal point 
when it has traveled half way around the sphere; once continued past that point it is not length minimizing,
since a geodesic that goes around the sphere in the opposite direction would be shorter.
Likewise a null geodesic on $\R\times S^2$ reaches a focal point when it has traveled half way around the sphere; 
when continued past that point, it is no longer prompt, as one could arrive
at the same destination sooner by traveling around the sphere in the opposite direction.

A remark about timelike geodesics that we made in section \ref{loran} has an analog here.   For a null geodesic $\ell$ from $q$ to $p$ to be prompt, it is necessary but
not sufficient for there to be no focal point in the segment $qp$.  Existence of a focal point implies that $\ell$ can be slightly deformed to a nearby causal path from $q$ to
 a point $p'$ slightly to the past of $p$.
But even if there is no focal point, there may anyway be some causal path $\gamma$ from $q$ to $p'$
  that is not a slight deformation of $\ell$.  
  
   \begin{figure}
 \begin{center}
   \includegraphics[width=2in]{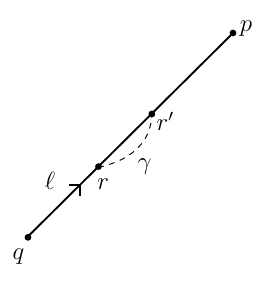}\caption{\small   
    Here $\ell$ is a null geodesic between two points $q$ and $p$. $\ell$ is is not achronal, because its segment $rr'$ can be replaced by a timelike
   path $\gamma$ between the same two points.   Accordingly, the $rr'$ segment of $\ell$ can be replaced with a causal path from $r$ that arrives
   strictly to the past of $r'$, and this can be continued into the future in a way that remains always to the past of $\ell$.   Thus when continued past $r'$, $\ell$ is not prompt.  For later 
   reference, $\ell$ is similarly not prompt on any interval strictly larger than $rr'$ if  $\gamma$  is assumed to be (rather than a timelike path) a null geodesic from $r$ to $r'$ , distinct from the $rr'$ segment of $\ell$.   \label{Fig23M}}\end{center}
\end{figure} 
  
  If such a $\gamma$ exists -- implying that $\ell$ is not prompt -- then $\ell$ is not achronal.
To see this, suppose a causal path $\gamma$ from $q$ to $p'$ exists and let  $\lambda$ be a timelike path from $p'$ to $p$.   Then $\gamma * \lambda$ is a causal path from $q$ to $p$ that is not everywhere
null.  It  can
  be deformed to a strictly timelike path from $q$ to $p$, showing that $\ell$ is not achronal.
   Conversely, if $\ell$ is not achronal, then it is not prompt.   Indeed, if $\ell$ is not achronal, this means by definition that there is  a timelike path $\gamma$ between two points $r,r'$ in the geodesic $\ell=qp$ (fig. \ref{Fig23M}).   Replacing the $rr'$ segment of $\ell$ with $\gamma$, we get a causal path from $q$ to $p$ that is not a null geodesic.    As usual, this can be deformed to a causal path from $q$
   to a point $p'$ slightly to the past of $p$, showing that $\ell$ is not prompt.  
    To summarize this, a null geodesic $\ell$ from $q$ to $p$ is prompt if and only if it is achronal.

\begin{figure}
 \begin{center}
   \includegraphics[width=4.3in]{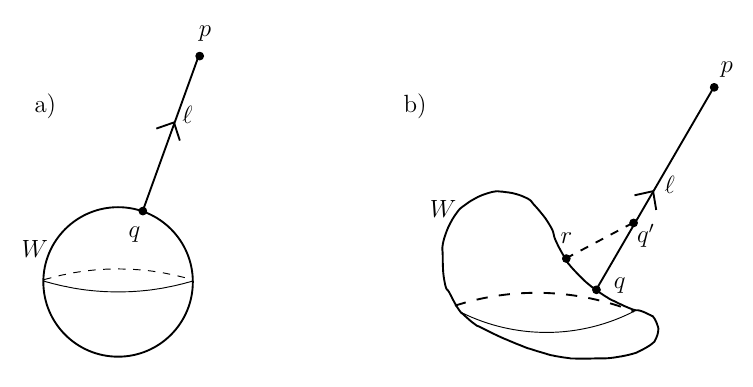}\caption{\small  (a) $\ell$ is a null geodesic from a spacelike submanifold $W$ to a point $p\in M$.  For $\ell$ to be prompt, it must be orthogonal
   to $W$ at the point $q$ at which it departs $W$.   Otherwise, by moving $q$ slightly ``towards'' $p$, one can replace $\ell$ with another causal curve to $p$ that gets
   a ``head start'' and arrives in the past of $p$.
   (The picture can be understood to depict
    a slice of Minkowski four-space at time $t=0$.  $W$ is a two-sphere embedded at $t=0$,
   and the null geodesic $\ell$ has been projected to the $t=0$ slice.   Note that if $\ell$ is othogonal to $W$ in four dimensions, then its projection to three dimensions is also orthogonal to $W$,
   as drawn.)   (b) Even if orthogonal to $W$, $\ell$ is nonetheless not
   prompt if a segment $qq'$ of $\ell$ can be displaced to a nearby null geodesic, meeting $W$ orthogonally at some other point $r$.  This displacement
   gives a causal path from $W$ to $p$ that is not a null geodesic, so it can be modified to a causal path that is more prompt.   In this situation, we say that the point $q'$ is
   a focal point of the orthogonal null geodesics from $W$.  This situation does not arise in Minkowski space  if $W$ is a round sphere supported at $t=0$ and the
   geodesic $\ell$ is ``outgoing,''  but it can arise for a more
   general choice of $W$, as shown, or for a round sphere embedded in a more general spacetime, or for ``incoming'' geodesics from a round sphere in Minkowski
   space.  \label{Fig24M}}\end{center}
\end{figure}

So far we have discussed prompt causal paths between two points.   
More generally, if $W$ is any set in spacetime, we say that a future-going causal path from $W$ to $p$ is prompt if it arrives as soon as possible, in the sense that no causal path from $W$ arrives
 to the past of $p$.   To be prompt, a causal path $\ell$ from $W$ to $p$ has
to be an achronal  null geodesic, just as before.    If $W$ is a spacelike submanifold of spacetime, then in addition $\ell$ has to be orthogonal to $W$ at the point $q$
at which they meet, as depicted in fig. \ref{Fig24M}(a).
   Otherwise, by changing the initial point of $\ell$ a little, one can get a causal path from $W$to $p$ that is not everywhere
null and hence can be deformed to arrive a little sooner.

Orthogonality between a spacelike submanifold $W$ and a null curve that intersects it is only possible if $W$ has real codimension at least 2.
This is a basic although elementary fact about Lorentz signature geometry.   For example, in Minkowski space with line element $\d s^2=-\d t^2+\sum_{i=1}^{\D-1}
(\d x^i)^2$, a codimension 1 spacelike hypersurface  through the origin can be modeled locally, in some Lorentz frame, by the equation $t=0$.   This hypersurface is
orthogonal to the timelike vector $(1,0,0,\cdots,0)$, but not to any null vector.  Since the question really only involves linear algebra, this example
is completely representative.    By contrast, consider the codimension 2 spacelike hypersurface $W$
defined by $t=x^1=0$.    It is orthogonal to the null vectors $u_\pm=(1,\pm 1,0,\cdots,0)$, with either choice of sign.  
These vectors  are both future-directed.   Thus, in the important case that $W$ has codimension 2, there are at each
point in $W$ two different future-directed orthogonal null directions.
 For higher codimension, the condition of orthogonality
to a null vector becomes  less restrictive.

As usual, there is a further condition for promptness that involves focal points.  If $\ell$ is an orthogonal null geodesic from $W$ to $p$, then to be prompt, 
$\ell$ must  have no focal point, where now $q'$ is a focal point if the segment of $\ell$ that connects $W$ to $q'$
can be displaced slightly to a nearby null geodesic, also connecting $W$ to $q'$ and orthogonal to $W$ (fig. \ref{Fig24M}(b)).
The reasoning is analogous to that in previous examples; after making such a displacement and then modifying the path near and beyond the resulting
kink (rather as in fig. \ref{Drawing3}), one gets a causal path from $W$
that arrives strictly to the past of $p$.

Suppose now that a codimension 2 closed spacelike submanifold $W$ is the boundary of a codimension one spacelike submanifold $Z$.  (For example, $W$ might be a sphere,
embedded in Minkowski space in the obvious way by $|\vec x|=R$, $t=0$, and $Z$ could be the closed ball $|\vec x|\leq R$, $t=0$).    What would be a prompt causal path from $Z$
to a point $p\notin Z$?   It will have to be a null geodesic from $Z$ to $p$; if it departs from $Z$ at an interior point of $Z$, it would have to be orthogonal to $Z$, but orthogonality between  a null geodesic
and a codimension 1 spacelike manifold is not possible.  So a prompt causal path from $Z$ to $p$ will have to depart from a boundary point of $Z$, that is a point in $W$. 
It will have to be orthogonal to $W$ just to be prompt as a path from $W$, let alone from $Z$. Even if a causal path  is prompt as a path from $W$, it is  not necessarily prompt as a path from $Z$, but it may be.
  For the case that $Z$ is a  closed ball in Minkowski space, the reader 
might want to figure out which points in Minkowski space
can be reached from $Z$ by prompt causal paths, and what are those paths.   Are all prompt causal paths from $W$ prompt as paths from $Z$?

In general, for any set $W$, we write $J^+(W)$ for the causal future of $W$ -- the set of points that can be reached from $W$ by a causal path.   A point $p\in J^+(W)$ is in the interior of $J^+(W)$ if
a small neighborhood of $p$ is in $J^+(W)$, and otherwise we say that it is in $\partial J^+(W)$, the boundary of $J^+(W)$.   If 
$p\in J^+(W)$  and a point $r$ slightly to the past of $p$ is also in $J^+(W)$,
then a neighborhood of $p$ is in the future of $r$ and therefore of $W$; so $p$ is in the interior of $J^+(W)$.   If, on the other hand, $p$ is in $J^+(W)$ but points slightly to the past of
$p$ are not in $J^+(W)$, then $p$ does not have a neighborhood in $J^+(W)$ so instead $p\in \partial J^+(W)$.  The reader should be able to see what these statements mean if $W$ is
a point $q$ in Minkowski space.

 An important detail is that, 
as in footnote \ref{detail} of section \ref{classic}, we consider $W$ itself to be in $J^+(W)$; in other words, we allow the
case of a future-going causal curve that consists of only one point.  The purpose of this is to ensure that for compact $W$, in a globally hyperbolic spacetime $M$,  $J^+(W)$ and $\partial J^+(W)$ are both closed in $M$.   That will make the discussion of Penrose's theorem simpler.

\begin{figure}
 \begin{center}
   \includegraphics[width=2.3in]{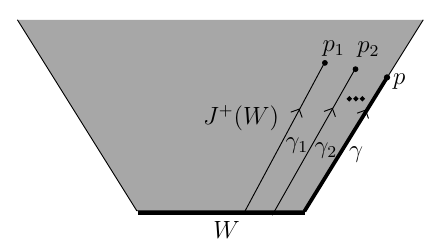}\caption{\small   
   If $p\in \partial J^+(W)$, then there is a sequence of points $p_1,p_2,\cdots$ in the interior of $J^+(W)$ that converge to $p$.  Each of the $p_i$
   can be reached from $W$ by a causal path $\gamma_i$.   In a globally hyperbolic spacetime, a subsequence of the $\gamma_i$ will converge to
   a causal path $\gamma$ from $W$ to $p$.   $\gamma$ will automatically be prompt, since if there is a causal path from $W$ that arrives slightly
   to the past of $p$, this implies that $p$ is in the interior of $J^+(W)$, not in its boundary.  \label{Fig25M}}\end{center}
\end{figure}

To show that $J^+(W)$ is closed in $M$ if $W$ is compact and
 $M$ is globally hyperbolic, we will show that if a sequence $p_1,p_2,\cdots\in J^+(W)$ converges in $M$ to a point $p$, then actually $p\in J^+(W)$.   Indeed,
the points $p_i\in J^+(W)$ can be reached from $W$ by future-going causal curves $\gamma_i$ (fig \ref{Fig25M}).  Because of compactness of spaces of causal curves in a globally hyperbolic spacetime, the $\gamma_i$
have a convergent subsequence, and this will be a future-going  causal curve $\gamma$ 
from $W$ to $p$, showing that $p\in J^+(W)$.  (If we did not consider $W$ to be in $J^+(W)$, then we could have $p\in W$
with $p_i$ strictly to the causal future of $W$, and the argument would fail.) 
To show that also $\partial J^+(W)$ is closed, we will show that
 if the $p_i$ are actually in $\partial J^+(W)$, then $p$ must be likewise in $\partial J^+(W)$, rather than in the interior of $J^+(W)$.   Indeed, if $p$ is in the interior of $J^+(W)$, then a neighborhood
 of $p$ is also in the interior, so if the sequence $p_1,p_2,\cdots$ converges to $p$, then the $p_i$ for sufficiently large $i$ are in the interior of $J^+(W)$, contradicting the hypothesis that they are in $\partial J^+(W)$.
 
 If $p\in \partial J^+(W)$, then $p$ can be reached from $W$ by a  future-going causal curve $\gamma$ (by the definition of $J^+(W)$), and more specifically  any such  $\gamma$ 
  is prompt and hence is a null geodesic without focal points.   For indeed, if $\gamma$ is not prompt, then there is a causal
 curve $\gamma'$ from $W$ to a point $p'$ to the past of $p$, and in that case $p$ is in the interior of $J^+(W)$, not the boundary.  
  
\begin{figure}
 \begin{center}
   \includegraphics[width=2.3in]{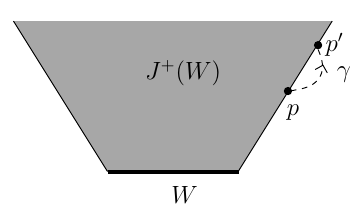}\caption{\small   
   The purpose of this picture is to show that, for any subset $W$ of spacetime, $\partial J^+(W)$ is always achronal.   For if there is a  future-going timelike path  $\gamma$ from $p$ to $p'$
   with $p,p'\in\partial J^+(W)$, then a neighborhood of $p'$ can be reached from $p$, and therefore from $W$, by a timelike path. So $p'$ is in the interior of $J^+(W)$, not in its boundary.  Thus in the diagram, the curve $\gamma$ cannot be everywhere timelike.    An extension of the reasoning shows that if $\gamma$ is a causal curve from $p$ to $p'$, it must be a null geodesic entirely contained in $\partial J^+(W)$.  \label{Fig26M}}\end{center}
\end{figure}

  A similar argument shows that  $\partial J^+(W)$ is always achronal.   For if $p\in \partial J^+(W)$, 
  and $\gamma$ is a future-going timelike path from $p$ to $p'$, then the future of $p$ contains a neighborhood of $p'$   (fig. \ref{Fig26M});
hence $p'$ is in the interior of $J^+(W)$, not in  its boundary.    As an example of this, the future light cone of a point $q$ in Minkowski space is achronal.

\subsection{More On The Boundary Of The Future}\label{bfuture}

The boundary of the future of any compact set $W\subset M$ has an additional property that is important in understanding black holes:  $\partial J^+(W)$ is always a codimension one submanifold of $M$,
though generally not smooth; moreover, if $M$ is globally hyperbolic, then $\partial J^+(W)$ is closed in $M$ (as we know from section \ref{pfp}) so it is actually
a closed submanifold of $M$.   Consider again the example that $q$ is a point in Minkowski
space (fig. \ref{Fig21M}).  $\partial J^+(q)$ is the future light cone, including the vertex of the cone.   It is a closed  submanifold
of Minkowski space but (because of the conical singularity at $q$) it is not smoothly embedded in Minkowski space.  This is the typical situation.

\begin{figure}
 \begin{center}
   \includegraphics[width=3in]{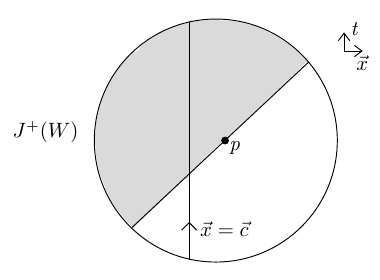}\caption{\small  Here we show a small neighborhood of a point $p\in \partial J^+(W)$, the goal being to demonstrate that $\partial J^+(W)$
   is a manifold near $p$. The shaded region represents $J^+(W)$. 
    In a small enough neighborhood of $p$, we pick a time coordinate $t$ and spatial coordinates $\vec x$ such that $p$ is at $t=\vec x=0$.  Consider now the timelike
   path $\gamma_{\vec c}$ defined by $\vec x=\vec c$, for a constant $\vec c$.  As explained in the text, for sufficiently small $\vec c$, $\gamma_{\vec c}$ intersects $\partial J^+(W)$
   in a unique point.    Hence we can parametrize $\partial J^+(W)$ near $p$ by coordinates $\vec c$, showing that $\partial J^+(W)$ is a manifold near $p$.
   This is so for every $p\in\partial J^+(W)$, so $\partial J^+(W)$ is a manifold.   \label{Fig42M}}\end{center}
\end{figure}

To show that in general $\partial J^+(W)$ is a manifold,  pick an arbitrary point $p\in\partial J^+(W)$.    We want to show that $\partial J^+(p)$ is a manifold near $p$.   We pick
a small ball near $p$ in which $M$ can be approximated by Minkowski space and pick coordinates $t, \vec x$ centered at $p$ (fig. \ref{Fig42M}).
Consider the timelike path $\gamma_{\vec c}$ that is parametrized by $t$ with a fixed value $\vec x=\vec c$ of the spatial coordinates.   
Since $p\in \partial J^+(W)$, a point to the past of $p$ is not in $J^+(W)$ and a point to the future of $p$ is in $J^+(W)$.
 For sufficiently small $\vec c$, the point  $t=t_0,$ $\vec x=\vec c$ is to the past of $p$ and so not in $J^+(W)$ if $t_0$ is sufficiently negative, and it is to the future of $p$ and  is in $J^+(W)$
 if $t_0$ is sufficiently positive.   So as $t$ varies from negative to positive values, there is a first value  $t=t_0(\vec c)$ at which the point $(t,\vec c)$ is in $J^+(W)$, and this point is in $\partial J^+(W)$.
 If $t>t_0$, then the point $(t,\vec c)$ is in the interior of $J^+(W)$.    So near $p$, $\partial J^+(W)$ consists of the points $t_0(\vec c), \vec c$.   
   So $\partial J^+(W)$ can be parametrized near $p$ by $\vec c$,
 and in particular is a manifold, of codimension 1 in $M$.

It is a good exercise to return to the example that $W$ consists of a single point $q$, so that $\partial J^+(W)$ is the future light cone of $q$, and to verify that the reasoning
that we have just given is valid even if we pick $p=q\in\partial J^+(W)$. 
For another example,   let $W$ be a circle or a two-sphere, embedded in Minkowski four-space in the obvious way.    Describe $\partial J^+(W)$, showing
that it is a manifold, though not smoothly embedded in spacetime.  What is the topology of $\partial J^+(W)$?  Answer the same question for an ellipse or an ellipsoid
instead of a circle or sphere.  You should find that  $\partial J^+(W)$ is equivalent
topologically in each case to the initial value surface $t=0$, where $t$ is the time.   In particular, $\partial J^+(W)$ is never compact.  Indeed, Minkowski space
has no compact achronal hypersurface, according to  a general argument
that was sketched in fig. \ref{Fig4}.

\subsection{The Null Raychaudhuri Equation}\label{nullraych}

 \begin{figure}
 \begin{center}
   \includegraphics[width=3in]{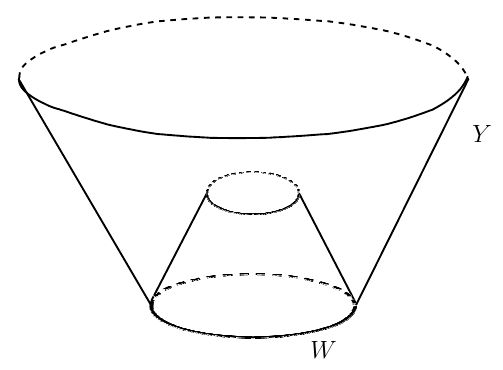}\caption{\small   If $W$ is a spacelike submanifold of spacetime of codimension 2, then two families of future-going null
   geodesics emanate from $W$.   In the diagram, the ``outgoing'' family sweeps out a null hypersurface $Y$.   \label{Fig27M}}\end{center}
\end{figure}
Just as in section \ref{geofocal}, to make progress, we need a reasonable way to predict that null geodesics will develop focal points.   This is provided by the null
 Raychaudhuri equation (this null analog of Raychaudhuri's original equation was first described by Sachs \cite{sachs}).

Let $W$ be a codimension two spacelike submanifold of a spacetime $M$.   For example, if $M$ is Minkowski space, $W$ might be a sphere, embedded in Minkowski space
with line element  $\d s^2=-\d t^2+\d \vec x^2$ in the standard fashion, as the submanifold $\vec x^2=R^2$, $t=0$.  

Emanating from $W$ are two families of  future-going null geodesics orthogonal to $W$ (fig. \ref{Fig27M}).  In the right conditions (for instance, if $W$ is a sphere in Minkowski space), one can naturally call
these families
``outgoing'' and ``incoming.''   In what follows, we focus on just one of these two families, which for convenience we will call outgoing.

The outgoing orthogonal null geodesics emanating from $W$ sweep out a codimension 1 submanifold $Y\subset M$.    To be more precise, $Y$ is a manifold near $W$.
When we extend the orthogonal null geodesics far enough so that they intersect each other or form focal points, $Y$ may fail to be a manifold.   This is actually the phenomenon
that we are interested in.

It is possible to choose a fairly natural set of coordinates
on $Y$.   First we pick any set of local coordinates $x^A$, $A=1,\cdots, \D-2$, on $W$.   In a somewhat similar situation in section \ref{rayeq}, we chose as one additional
coordinate the proper time measured along a timelike geodesic.    For a null geodesic, there is no notion of proper time.   But the affine parameter of a null geodesic
serves as a partial substitute.  

For a geodesic parametrized by an affine parameter $\lambda$, the geodesic equation
reads
\be\label{ziggo}0=\frac{D^2 x^\mu}{ D\lambda^2}=\frac{\d^2 x^\mu}{\d\lambda^2}+\Gamma^\mu_{\alpha\beta}\frac{\d x^\alpha}{\d\lambda}\frac{\d x^\beta}{\d\lambda}. \ee
This affine version of the geodesic equation is invariant under affine transformations
\be\label{higgo}\lambda\to a\lambda+b,~~~a,b\in\R,\ee
and two solutions that differ only by such a transformation are considered equivalent.   

On each of the null geodesics $\ell$ that make up $Y$, we pick an affine coordinate $u$ that increases towards the future
and  vanishes at the point $\ell\cap W$.   This uniquely determines
$u$ up to the possibility of multiplying it by a positive constant.  

Of course, if we let $x^A$ vary, this ``constant'' can be  a function of the $x^A$.
So following the procedure just described
 for every $\ell$ (and making sure to normalize $u$ in a way that varies smoothly with the $x^A$), we get a smooth function $u$ on $Y$ that is uniquely determined up to the possibility of multiplying it by a positive function of the $x^A$, that is,
a positive function on $W$.  We make an arbitrary specific choice.
In addition, we can extend the $x^A$ from functions on $W$ to functions on $Y$ by declaring that they are constant along each  $\ell$.
Thus at this point we have a full system of local coordinates $x^A$ and $u$ on the $\D-1$-manifold $Y$.  To complete this to a full coordinate system on $M$, or at least
on a neighborhood of $Y$ in $M$,
we need one more function $v$, and of course we need to extend $x^A$ and $u$ off of $Y$.

 \begin{figure}
 \begin{center}
   \includegraphics[width=2in]{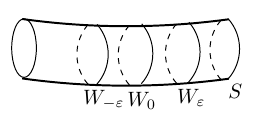}\caption{\small   A spacelike submanifold $W$ of codimension 2 can always be embedded locally in a spacelike hypersurface $S$.
  Moreover, near $W$, $S$ can be swept out by a family of codimension 2 submanifolds $W_\varepsilon$, $\varepsilon\in\R$, 
   such that $W_0$ is the original $W$.   \label{Fig28M}}\end{center}
\end{figure}
One way to do this (fig. \ref{Fig28M}) is to first embed $W$ in a spacelike hypersurface $S$ that is defined at least locally near $W$ (we will only be interested in what happens
near $W$).   Then we locally build $S$ out of a family of codimension 1 submanifolds $W_\varepsilon\subset S$, $\varepsilon\in\R$, where $W_0$ is the original $W$.   For this, we simply
pick a function $v$ on $S$ that vanishes along $W$ but has an everywhere nonzero normal derivative, and define  $W_\varepsilon$ by the condition $v=\varepsilon$.   We extend
the $x^A$ to functions on $S$ in an arbitrary fashion.   Then we consider the outgoing orthogonal null geodesics from $W_\varepsilon$ just as we did for the original $W$.
Letting $\varepsilon$ vary,  we are now allowing any starting point in $S$, so these geodesics together sweep out all of $M$ (or more exactly a neighborhood of $Y$ in $M$).
We parametrize these geodesics by an affine parameter $u$, normalized to vanish on $W_\varepsilon$, as we did at $\varepsilon=0$.  As before, this $u$ is well-defined up to multiplication by
a function on $S$ and we make a specific choice.   
Next, we extend $v$ and $x^A$ to functions on all of $M$ (or more exactly, on the portion of $M$ that is near $Y$) by saying that they
are constant along the outgoing null geodesics.   Finally, $u,v,$ and $x^A$ provide coordinates on a neighborhood of $Y$ in $M$.

In this coordinate system, $g_{uu}$ is identically 0, because we have arranged so that the curves parametrized by $u$, with constant $v$ and $x^A$, are null.   In addition, we have ensured that the geodesic equation is satisfied with $u$ equal to an affine parameter $\lambda$ and with $v$ and $x^A$ constant.
The condition for this is that $\Gamma^\alpha_{uu}=0$ for all $\alpha$.   Given that $g_{uu}=0$, this implies $\partial_u g_{u \beta}=0$ for all $\beta$.
Since the construction was made using null geodesics that are orthogonal to each $W_\varepsilon$, $g_{u A}$ vanishes at $u=0$, and therefore it vanishes identically.
As for $g_{uv}$, we learn that it is independent of $u$. Since $g_{uu}=g_{uA}=0$, $g_{uv}$ must be everywhere nonzero (or we would have $\det g=0$).
So we write $g_{uv}=-2e^{q}$ (the minus sign is a convention which could be reversed by changing the sign of $v$).    The line element therefore takes the general  form
\be\label{fracb}\ds^2 = -2e^q\d v \,\d u + g_{AB}\d x^A\d x^B+ 2c_A \d v \d x^A + g_{vv}\d v^2,\ee
where in general $g_{AB}$ and $c_A$ depend on all coordinates $u,v, x^C$ but $q$ depends only on $x^C$ and $v$.  


One immediate conclusion comes by looking at the metric on $Y$.   Setting $v=0$, we see that this is the degenerate metric $g_{AB}(x^C,u)\d x^A\d x^B$,
with no $\d x^A \d u$ or $\d u ^2$ term.   Thus the signature of the hypersurface $Y$ is $++\cdots +0$, with  one 0 and the rest $+$'s.   Such a $Y$ is called
a null hypersurface.  We have learned that a hypersurface swept out by orthogonal null geodesics from a codimension 2 spacelike manifold $W$ is a null hypersurface.   (This statement
also has a converse, which is explained in Appendix \ref{null}.)

The null Raychaudhuri equation is just the Einstein equation $R_{uu}=8 \pi G T_{uu}$ on the hypersurface $Y$ defined by $v=0$.    Before computing $R_{uu}$, it helps to notice that by adding to $u$ a function that vanishes at $v=0$
-- and so without affecting the metric on $Y$ -- we can adjust $g_{vv}$ as we wish.   In particular, we can put the line element in the form\footnote{Once the line element is in this form modulo terms of order $v^n$,
by shifting $u$ by $v^{n+1}f_n(u,x^A)$ (for some function $f_n$), we put the line element in this form modulo terms of order $v^{n+1}$. We start the process at $n=0$.}
\be\label{racb}\ds^2=-2e^q \d v\,\d u +g_{AB}(\d x^A+c^A\d v)(\d x^B+c^B \d v),\ee
in general with different functions $q$ and $c^A$; however, $q$ is still independent of $u$ at $v=0$.
(For a more elegant route to this form of the metric, see \cite{Sachstwo}.)
In this coordinate system, 
\be\label{acb} g^{uu}=g^{vv}=g^{vA}=0=g_{uu}=g_{ua}=\Gamma^\alpha_{uu}.\ee

 A straightforward calculation of $R_{uu}$
similar to the computation of $R_{tt}$ in section \ref{rayeq} gives a similar result.  Discarding terms that vanish because of (\ref{acb}), we have
\be\label{wacf}R_{uu}=-\partial_u \Gamma^A_{u A}-\Gamma^A_{uB}\Gamma^B_{uA}=-\frac{1}{2}\partial_u\left(g^{AB}\partial_u g_{AB}\right)
-\frac{1}{4} \left(g^{AC}\partial_u g_{BC}\right)\left(g^{BD}\partial_u g_{DA}\right)  .\ee
Here $g_{AB}$ is the ($u$-dependent) metric of $W$ that appears in eqn. (\ref{racb}), and $g^{AB}$ is its inverse.   This formula is in perfect parallel with eqn. (\ref{wobbo});
the only difference is that the $(\D-1)\times (\D-1)$ metric of the hypersurface $S$ has been replaced by the ($\D-2)\times (\D-2)$ metric of the codimension 2 submanifold $W$.
The subsequent discussion therefore  can proceed in close parallel to the timelike case.  

We define the transverse area of a little bundle of orthogonal null geodesics as
\be\label{bongo}A=\sqrt{\det g_{AB}}. \ee 
As in the timelike case, we will look for points at which $A\to 0$. The interpretation of $A\to 0$ is quite analogous to the interpretation of $V\to 0$ in the timelike case.
$A\to 0$ can represent a spacetime singularity, or simply a focal point at which some of the orthogonal null geodesics from $W$ converge.\footnote{For the same reason as in the timelike
case, the condition $\det\,g_{AB}=0$ is necessary as well as sufficient for a focal point.    Indeed, as long as $M$ is smooth, the point in $M$ that is reached by an orthogonal
null geodesic that originates at a point $q\in S$ and propagates up to a value $u$ of its affine parameter is a smooth function of $q$ and $u$.  So $g_{AB}$ remains regular as long as $M$ is 
smooth.  So whenever $g_{AB}$ acquires a zero eigenvalue (corresponding to a focal point), $\det \, g_{AB}$ will vanish.}

The null expansion is defined as\footnote{It hopefully will not cause confusion to use the same symbol $\theta$ for the expansion of a $(\D-2)$-dimensional  bundle
of null  geodesics or of a $(\D-1)$-dimensional bundle of timelike ones.  Normally the context makes clear what is meant.}
\be\label{tongo}\theta=\frac{\dot A}{A}=\frac{1}{2}\Tr \,g^{-1}\dot g. \ee
The dot now represents a derivative with respect to $u$. 
One also defines the shear, the traceless part of $g^{-1}\dot g$:
\be\label{ongo}\sigma^A_B=\frac{1}{2}\left(g^{AC}\dot g_{CB}-\frac{1}{\D-2} \delta^A_B \Tr\,g^{-1}\dot g\right). \ee
So
\be\label{songo}R_{uu}=-\dot\theta-\frac{\theta^2}{\D-2}-\Tr\,\sigma^2.\ee
Hence the Einstein-Raychaudhuri equation $R_{uu}=8\pi G T_{uu}$ reads 
\be\label{kongo}\dot\theta+\frac{\theta^2}{\D-2} =-\Tr\,\sigma^2-8\pi G T_{uu}. \ee

The {\it null energy condition} is the statement that at each point and for any null vector $v^\mu$, the stress tensor satisfies
\be\label{wongox} v^\mu v^\nu T_{\mu\nu}\geq 0.\ee
This condition is satisfied by all reasonable classical matter.\footnote{There is a subtlety here in a theory with an elementary scalar field $\phi$.   In such a theory,
the metric tensor $g$ is in general not uniquely defined, as it could be redefined by a Weyl transformation $g\to e^{f(\phi)}g$ for an arbitrary function $f$.  
 There is then a unique
``Einstein frame'' in which the Ricci scalar $R$ appears in the action with a coefficient $1/16\pi G$, independent of $\phi$.   The statement made in the text
is valid in Einstein frame.   It may not hold in a more general frame.}
  It  is satisfied by all of the usual relativistic classical fields, and is not affected by a cosmological constant.
Thus the null energy condition is a very solid foundation for discussion of classical properties of General Relativity. This stands in contrast to the strong energy condition,
which was used in deducing Hawking's theorem about the Big Bang; it is not universal in classical field theory.
In the rest of this article, we will generally assume the null energy condition,
except that in section \ref{anec} we discuss a weaker condition with some of the same consequences.  

The null energy condition ensures that $T_{uu}$ is nonnegative, and therefore that the right hand side of eqn. (\ref{kongo}) is nonpositive.
Given this, we can reason as we did in the timelike case and predict the occurrence of focal points within a bounded affine distance from $W$.

Indeed, similarly to eqn. (\ref{zeg}), eqn. (\ref{kongo}) implies that
\be\label{rongox}\partial_u \frac{1}{\theta}\geq \frac{1}{\D-2}.\ee
Now suppose that $\theta<0$ at some point on $W$, say $\theta=-w$.
Then $1/\theta\geq -1/w+u/(\D-2)$ or 
\be\label{qongox} \frac{\dot A}{A} \leq -\left(\frac{1}{w}-\frac{u}{\D-2}\right)^{-1}.\ee
Integrating this formula for $\dot A/A$, we learn as in eqn. (\ref{mleg}) that $A\to 0$ at a value of the affine parameter no greater than
$u=(\D-2)/w$.  

Just as in the timelike case, there might be a spacetime singularity when $A\to 0$ (or even before), but in general $A\to 0$ represents only a focal point at which
the coordinate system assumed in the calculation breaks down.   To predict a singularity requires additional input beyond Raychaudhuri's equation.

\subsection{Trapped Surfaces}\label{trapped}

A codimension two spacelike submanifold $W$ of spacetime has two families of future-going orthogonal null geodesics, so it is possible to define two different null expansions.
Each of them is governed by the null Raychaudhuri equation that we have just described. 

If, for example, $W$ is a sphere embedded in Minkowski space in the standard fashion,
then one family of orthogonal null geodesics is ``outgoing,'' in an obvious sense, and one is ``incoming.''   The outgoing ones have positive null expansion, and the incoming
ones have negative null expansion.

\begin{figure}
 \begin{center}
   \includegraphics[width=2.5in]{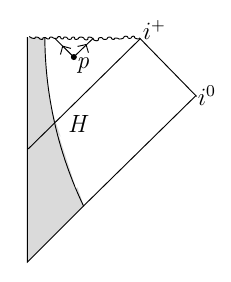}\caption{\small  This is a Penrose diagram describing spherically symmetric collapse to a black hole.   The
   area shaded gray represents matter and radiation that is collapsing to form a black hole; the exterior, unshaded region represents part of the Schwarzschild
   solution.     (See Fig. \ref{Fig40M} in section \ref{cc} for the Penrose diagram of the
   maximally extended Schwarzschild solution.)  A point in the diagram represents a two-sphere (or in $\D$ dimensions, a $(\D-2)$-sphere) whose radius $r$ is a function
   of the two coordinates in the diagram.   Behind the horizon, which is labeled $H$, $r$ decreases along every future-going causal curve.   For example, the point labeled $p$ is behind the horizon
   and represents a sphere $W$ of area $4\pi r^2$ (or the analog of this in $\D-2$ dimensions).    The future-going orthogonal null geodesics from $p$ are represented in the diagram by ingoing and outgoing null geodesics  (the rays
   emerging from $p$ at $\pi/4$ angles to the vertical).   These are future-going causal curves beyond the horizon, so $r$ decreases along each of them and hence the area of $W$ also decreases.
   Thus, $W$ is a trapped surface.   The points in the diagram labeled $i^0$ and $i^+$ represent spatial infinity and the future infinity of an outside observer, respectively.
    \label{Fig43M}}\end{center}
\end{figure} 
Penrose defined a ``trapped surface'' to be a codimension two spacelike
submanifold $W$ such that the expansion of each family of orthogonal future-going null geodesics is everywhere negative.   The motivating example 
 is a spherically symmetric surface behind the horizon of a spherically symmetric black hole.    Such a black hole can be conveniently represented by
a two-dimensional picture, known as a Penrose diagram, in which only the radial and time directions are shown (fig. \ref{Fig43M}). In the diagram, 
ingoing and outgoing radial null geodesics are represented by
straight lines at a $\pm\pi/4$ angle to the vertical.   A point in the diagram  represents a two-sphere (or in $\D$ dimensions,
a $\D-2$-sphere) with a round metric.   The radius of the sphere is a function $r$ on the Penrose diagram; in four dimensions, 
the two-sphere represented by a given point has an area
$4\pi r^2$.    What characterizes a Schwarzschild black hole is that behind the horizon $H$ 
of the hole, $r$ is decreasing along every future-directed 
causal curve.  Now pick a point $p$ in the diagram behind the horizon
of the black hole and let $W$ be the corresponding two-sphere.  The future-going orthogonal geodesics from $W$ are simply future-directed 
outgoing or incoming radial  geodesics, so they are represented in
the diagram by future-going rays that leave $p$ at angles $\pm \pi/4$ from the vertical, as shown.   In particular, these rays represent future-directed 
causal paths, so $r$ decreases along each of them.
Therefore both null expansions of $W$ are negative; $W$ is a trapped surface.

As an exercise, the reader can try to visualize a sphere embedded in Minkowski space in such a way that each of its null expansions is 
negative in one region and positive in another.  There is no difficulty in finding such an example.
But there is no trapped surface in Minkowski space: there is no way to embed a sphere so that each of the two null expansions is everywhere negative.  
This statement is actually a corollary of  Penrose's theorem, to which we come next.

\subsection{Penrose's Theorem}\label{penroseproof}

Penrose's theorem \cite{Penrosesing} was the first  modern singularity theorem, though in the presentation here, we have not followed the historical order.

Penrose's goal was to prove that the formation of a singularity is a generic phenomenon in gravitational collapse.   If spherical symmetry is assumed, one can solve Einstein's equations
for a collapsing star rather explicitly  and show  that formation of a singularity is unavoidable.
  But what happens if the geometry is not quite spherically symmetric?    Does infalling matter still collapse to a singularity, or does it ``miss'' and, perhaps, re-emerge in an explosion?
Penrose wanted a robust criterion for formation of a singularity that would not depend on precise spherical symmetry. This goal was the motivation for the concept of a trapped surface.

 Like the other results that came later (see the discussion at the end of section \ref{hbb}), 
Penrose's ``singularity theorem'' is proved without directly studying singularities.   It is really a statement about completeness.  Penrose proved that a spacetime $M$ satisfying certain conditions
is null geodesically incomplete, meaning that at least one null geodesic in $M$ cannot be continued in the future to an arbitrarily large value of its affine parameter.

Penrose's theorem concerns a  globally hyperbolic spacetime $M$ with a noncompact Cauchy hypersurface $\S$. 
For example, in any spacetime that is asymptotic at spatial infinity to Minkowski space, a Cauchy hypersurface -- if there is one -- is not compact. So Penrose's theorem applies to any spacetime that is asymptotically flat and also globally hyperbolic.
 In the statement and proof of Penrose's theorem,
it is not necessary to assume that $M$ is maximal in the sense of section \ref{maximal}, but the most interesting case is that $M$ is maximal.

The final input to Penrose's theorem, apart from the classical Einstein equations, is the null energy condition.
  Penrose's theorem states that, assuming global hyperbolicity, a noncompact Cauchy hypersurface,  the null energy condition, and the classical Einstein equations,
a spacetime $M$ that contains a compact trapped surface is not null geodesically complete.

The condition for a spacetime to contain a compact trapped surface $W$ is stable against small perturbations of the geometry, since if the null expansions of $W$
are negative, they remain negative after a sufficiently small perturbation.  So Penrose's theorem
applies to any spacetime that is sufficiently close to Schwarzschild, that is, sufficiently close to spherically symmetric collapse to a black hole in a spacetime that is asymptotic to Minkowski
space at spatial infinity.  

Penrose's theorem is commonly
called a ``singularity theorem'' because of a presumption that  generically, the reason that a null geodesic cannot be continued is the same as in the Schwarzschild case: it ends at a singularity.
  However, this goes beyond what is proved, and the exercises below show that on such matters, caution is called for.

Penrose's theorem is proved as follows. 
Let $W$ be a compact trapped surface in $M$. Pick affine parameters for the future-going null geodesics that are orthogonal to $W$.  
As usual, it is natural to normalize 
these affine parameters to vanish on $W$.   For any given point $q\in W$, one is still free to multiply the affine parameter (for either the ingoing or outgoing null geodesic from $q$) by a constant.
There is no natural way to fix the normalization; for what follows, it suffices to use any normalization that varies smoothly with $q$.
 
The  null expansions $\dot A/A$ of the compact surface $W$, being everywhere negative, satisfy a bound
   $$\frac{\dot A}{A}<-w$$
for some constant $w>0$.  Then Penrose proves, to be precise, that at least one of the future-going null geodesics orthogonal to $W$ cannot be extended to a value of
its affine parameter greater than $(\D-2)/w$.   

Suppose on the contrary that every one of the future-going null geodesics  $\ell$ orthogonal to $W$
can be extended to a value of its affine parameter greater than $(\D-2)/w$.  This means,
according to Raychaudhuri's equation, that every such $\ell$ can be extended beyond its first focal point.   

Suppose that $\ell$ leaves $W$ at a point $q\in W$ and continues past its first focal point $p$.  Let $\ell_{qp}$ be the segment of $\ell$ connecting $q$ to $p$; it is compact.
We have $\ell\cap \partial J^+(W)\subset \ell_{qp}$, since $\ell$ is not prompt when continued beyond $p$.     
Moreover,  $\ell\cap\partial J^+(W)$ is closed in $\ell$, since $\partial J^+(W)$ is closed in $M$.   So $\ell\cap \partial J^+(W)$ is a closed subset of
$\ell_{qp}$.     A closed subset of a compact space is compact, so $\ell \cap \partial J^+(W)$ is compact.\footnote{If $\ell$ cannot be continued until reaching a focal point -- for instance because it
 ends at a singularity or leaves the globally hyperbolic spacetime -- then  it is possible for the closed subset $\ell\cap\partial J^+(W)$ of $\ell$ to
 satisfy  $\ell\cap\partial J^+(W)=\ell$, and then $\ell\cap\partial J^+(W)$ can be noncompact.   Penrose's proof
 shows that if $W$ is a trapped surface, this situation will arise for some $\ell$.  (Promptness of $\ell$ can fail without $\ell$ reaching a focal point, 
 but the alternative failure mode, which is described in Appendix \ref{failure}, does not give a new way for the prompt portion of $\ell$ to be noncompact.)   
 If $\ell$ can be continued past its first focal point, then we can be more specific about $\ell\cap\partial J^+(W)$,
 though the details are not needed in the proof of Penrose's theorem.  $\ell\cap \partial J^+(W)$ is either empty (this is only possible if $W$ is not achronal), consists only of the point
 $q\in \ell$ (this is only possible if there exists a causal path to $q$ from some other point $q'\in W$, but any such path is a null geodesic), 
 or consists of a nontrivial initial segment of $\ell$.   The reader may wish
 to try to justify these statements. }

Every point $p\in\partial J^+(W)$ is connected to $W$ by one of these future-going orthogonal null geodesics.  So $p$ is determined  by\footnote{Such a triple determines $p$, but more than one triple may lead to the same
 $p$.} a triple consisting of  the choice of a point $q\in W$
 (at which $\ell$ originates), the choice of whether $\ell$ is ``incoming'' or ``outgoing'' at $q$, and the affine parameter measured along $\ell$. 
 Moreover, from what we have just seen, the affine parameter on each $\ell$ can be restricted to a compact interval.
 Since $W$ itself is compact and the affine parameter measured along each $\ell$
 ranges over a compact interval, this implies that $\partial J^+(W)$ is compact. 

But $\partial J^+(W)$ is an achronal codimension 1 submanifold of $M$, as explained earlier  in sections \ref{pfp} and \ref{bfuture}.
In a globally hyperbolic spacetime $M$ with a noncompact Cauchy hypersurface $\S$, there is no compact achronal submanifold
of codimension 1, as we learned in section \ref{somep}.   Putting this together, our hypothesis was wrong and at least one future-going null geodesic
$\ell$ that is orthogonal to $W$ cannot be extended within $M$ beyond an affine distance $(\D-2)/w$ to the future of $W$.   This completes the proof of Penrose's theorem.

Here are some exercises that aim to help one understand the fine print in Penrose's theorem. (The two forms of the de Sitter space metric are described in Appendix \ref{dSitter}.)

(1) De Sitter space in dimension $\D$ can be described with the line element
\be\label{dS}\ds^2=-\d t^2+R^2\cosh^2(t/R) \d \Omega^2\ee
where $\d\Omega^2$ is the line element of a $\D-1$-sphere.  Convince yourself that this is a globally hyperbolic spacetime with a {\it compact} initial value
surface $S$ (because of the compactness of $S$, Penrose's theorem does not apply).  Also convince yourself that de Sitter space
 is geodesically complete; all geodesics can be continued to infinite affine
parameter in both directions.  Find a compact trapped surface $W$.  Can you describe the boundary of the future of $W$?   You should find that $\partial J^+(W)$
is topologically equivalent to the initial value surface, as the above arguments imply.

(2)  Consider the following line element, which describes just a portion of de Sitter space:
\be\label{dSp}\d \t s^2=-\d t^2+R^2\exp(-2t/R) \d \vec x^2,~~~~\vec x\in \R^{\D-1}.\ee
(A time-reversed version of this line element provides a model of an accelerating universe.  The $t$ coordinate in eqn. (\ref{dSp})
is not simply related to the coordinate of the same name in eqn. (\ref{dS}). See Appendix \ref{dSitter} for a derivation of this form of the line element and an explanation
that it describes just part of de Sitter space.)
Show that this line element describes a globally hyperbolic spacetime  with noncompact initial value surface $t=0$.  Thus Penrose's theorem applies.
Find a compact trapped surface $W$.  Do you see a singularity?
What does Penrose's theorem mean for this
spacetime?  What is the boundary of the future of the trapped surface $W$?   As a hint, look at null geodesics in this spacetime. 
You should find that every null geodesic intersects the initial value surface at (say) $t=0$, as implied by global hyperbolicity.  But
are null geodesics complete in the sense of extending to infinite affine parameter in both directions?

 \begin{figure}
 \begin{center}
   \includegraphics[width=3in]{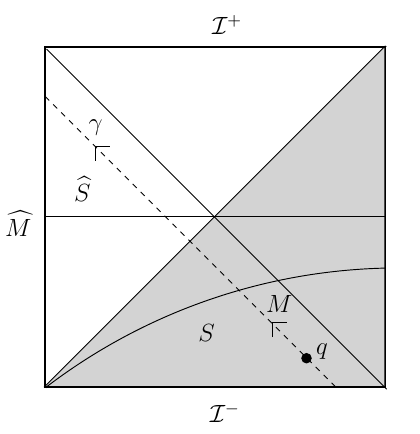}\caption{\small  The square is a Penrose diagram of de Sitter space $\h M$.    $\I^+$ and $\I^-$ represent future and past infinity of $\h M$
   and are not part of $\h M$.    The shaded triangle is a contracting portion $M$
   of $\h M$ ($M$ is actually the causal past of a point in $\I^+$).   $\h M$ is globally hyperbolic, with compact Cauchy hypersurface $\h S$,
   and $M$ is globally hyperbolic with noncompact Cauchy hypersurface $S$ (to compactify $S$, we would have to include in $S$ the lower left corner of the diagram; this point lies
   in $\I^-$ and is not part of $M$ or $\h M$).  $M$ contains compact trapped surfaces, represented by points near the bottom of the diagram.  In particular, the point labeled $q$
   represents a trapped surface $W$.
   $M$ therefore satisfies all the conditions of Penrose's theorems.  The incompleteness implied by Penrose's theorem just means that some null geodesics orthogonal to $W$ --
   such as the one labeled $\gamma$ --  exit $M$ at a finite value of their affine parameters.   Thus $M$ is not null geodesically complete; its completion is $\h M$.  \label{Fig30MNew}}\end{center}
\end{figure}

To understand these examples better,
let $\h M$ be the de Sitter space of eqn. (\ref{dS}), and let $M$ be the spacetime of eqn. (\ref{dSp}).  Then $M$ can be embedded in $\h M$.
Both $M$ and $\h M$ are globally hyperbolic, but a Cauchy hypersurface $\S$ for $M$, when embedded in $\h M$, is not a Cauchy hypersurface
for $\h M$.   A Cauchy hypersurface $S$ for $M$ is a plane $\R^{\D-1}$, while a Cauchy hypersurface $\hat S$ for $\hat M$ is a sphere $S^{\D-1}$, obtained topologically
by adding a point at infinity to the plane.   
When $M$ is embedded in $\h M$, the portion of $\h M$ that is not in $M$ lies beyond a Cauchy horizon.  A Penrose diagram
illustrating these statements is in fig. \ref{Fig30MNew}.    Null geodesics in $\h M$ can be
continued to infinite values of their affine parameter, but they can exit $M$ at a finite value of their affine parameter.  A trapped surface $W$ can be represented by any point
below the main diagonals.  For instance, we can pick the point $q$ on the null geodesic $\gamma$, near the lower right corner of the diagram.
  Applied to this example, the phenomenon described by Penrose's theorem
is that some null geodesics orthogonal to $W$, when viewed as geodesics in $\h M$, leave $M$ at a finite value of their affine parameters and enter the part of $\h M$ that is beyond a Cauchy horizon.
$\gamma$ itself has this behavior.

The second example  illustrates why some careful authors refer to Penrose's theorem as an incompleteness theorem rather than a singularity theorem.
The geodesics that, according to Penrose's theorem, cannot be continued to arbitrary values of their affine parameter might terminate on a singularity,
as in the case of a Schwarzschild black hole.   But the theorem allows the possibility that these geodesics simply leave the globally hyperbolic spacetime $M$
under discussion, without reaching any singularity, as in our second example.   The phenomenon  is a failure of predictivity:   what happens to these
geodesics after they leave $M$ cannot be predicted based on initial data on the Cauchy hypersurface $\S$ of $M$.

\section{Black Holes}\label{blackholes}

What Penrose's theorem actually says about the region inside a black hole is quite limited, as we learned from the examples in section \ref{penroseproof}.  
It therefore comes as a nice surprise that the
ideas on which this theorem is based lead rather naturally to an understanding of important properties of black holes.\footnote{Chapter 12 of the book \cite{Wald} is particularly useful for background to the present section, as well as to further properties of black holes that go beyond the scope of this article.  In this article, we only consider issues directly related to causality.}

To be more exact, these ideas {\it plus one more assumption} lead to such an understanding.

\subsection{Cosmic Censorship}\label{cosmic}

It is not possible to get a good theory of black holes without knowing, or assuming, that something ``worse'' than collapse
to a black hole does not occur.  

Roughly speaking, ``something worse'' might be the formation of what Penrose called a ``naked singularity'' 
\cite{PenroseA,PenroseB,Penrose\S}. This is a singularity
not enclosed by a horizon, and visible to an outside observer.

Formation of a naked singularity might bring the predictive power of classical General Relativity to an end, because the classical theory
would not uniquely determine what comes out of the naked singularity.   As an extreme case, formation of a naked singularity might bring the spacetime itself
to an end (this possibility was part of the motivation for a refinement of Penrose's conjectures \cite{GerochHorowitz}). 
For example, imagine that gravitational collapse creates an outgoing shock wave singularity
that expands to infinity at the speed of light.  If the singularity is severe enough that the classical
Einstein equations break down, and the classical spacetime cannot be continued to the future of the shock wave based only
on information provided by Einstein's classical theory, then one could describe this by saying that spacetime -- or at least
the ability to describe spacetime based only on the classical theory -- has come to an end.

Since this may seem fanciful, perhaps it is worth mentioning that something somewhat similar, though without a singularity, can actually
happen in Kaluza-Klein theory.   Consider a $\D$-dimensional spacetime that is asymptotic
at spatial infinity to $M^{\D-1}\times S^1$ (where $M^{\D-1}$ is Minkowski space of dimension $\D-1$ and $S^1$ is a circle).   At distances long compared to the asymptotic radius of the circle, such a spacetime is effectively $(\D-1)$-dimensional.
The classical equations admit initial conditions in which, from a $(\D-1)$-dimensional point of view, there is a ``hole'' in space of zero or even negative energy  \cite{Witten}.   (From a $\D$-dimensional
point of view; there is no hole in space; the spacetime is topologically different from $M^{\D-1}\times S^1$, but looks like $M^{\D-1}\times S^1$ near spatial infinity.)   The hole expands
to infinity at the speed of light.   There is nothing to the future of this cataclysm; spacetime comes to an end.   So in such a situation, it is not
possible to make arguments -- as we will do in analyzing black holes -- based on what will be seen in the far future by an observer
who remains at a safe distance.

Penrose \cite{PenroseA,PenroseB,Penrose\S} introduced the hypothesis of ``cosmic censorship,'' which (in its simplest form, known as ``weak'' cosmic censorship,\footnote{``Strong'' cosmic censorship says that in  an arbitrary 
spacetime, not necessarily asymptotic to Minkowski space, no observer can see a naked singularity. There is no general claim of an ability to continue the spacetime to the future.  For our
purposes here, ``weak'' cosmic censorship is the relevant statement.  The terminology, though standard, is potentially misleading since ``weak'' cosmic censorship is not a special case of
``strong'' cosmic censorship.} and as elaborated later \cite{GerochHorowitz}) says that in a globally hyperbolic spacetime
asymptotic at spatial infinity to Minkowski space, no such catastrophe occurs: 
in gravitational collapse, and more generally in any localized process in an asymptotically Minkowskian spacetime, the region
in the far distance and the far future continues to exist, just as in Minkowski space.   Moreover,  the evolution seen by an outside observer
is supposed to be predictable based on the classical Einstein equations.   Any singularity is hidden by a horizon, and does not affect the outside
evolution.

If true, this is a quite remarkable and genuinely surprising fact, and possibly a little disappointing.  It is genuinely surprising because
the classical Einstein equations have no obvious stability properties that would tend to guarantee cosmic censorship. It is possibly disappointing because if cosmic censorship
is true, we lose our chance to get observational evidence concerning a hypothetical better theory.   After all, if cosmic censorship is false, and the classical Einstein
equations break down in a way that is visible to a distant observer, we might in principle hope to observe what happens when this breakdown 
occurs.\footnote{In the conclusion of \cite{PenroseB}, Penrose
expresses some ambivalence about the cosmic censorship hypothesis, observing that there appears to have been a Big Bang, which is somewhat
analogous to a naked singularity, so naked singularities might be an inescapable part of physics.}  

 For many years, the evidence
for cosmic censorship appeared meager, at least to the present author. However, by now reasonable evidence for cosmic censorship, at least for $\D=4$, has come from the fact that simulations of black hole
collisions  \cite{Pretorius} have not generated naked singularities.\footnote{For $\D>4$,
it appears that  the Gregory-Laflamme instability \cite{GLa} may violate the usual statement of cosmic censorship \cite{Pret}.   However, see footnote
\ref{welg} below.}
   If it were the case that black hole collisions produce naked singularities rather than bigger black holes,\footnote{On this
question, Penrose wrote in \cite{Penrose}, ``We might, for example, envisage two comparable black holes spiraling into one
another. Have we any reason, other than wishful thinking, to believe that a black
hole will be formed, rather than a naked singularity? Very little, I feel; it is really a
completely open question.''} 
then the recent LIGO/VIRGO observations of colliding black holes,
rather than providing a particularly interesting test of classical General Relativity, 
might have given us information about what happens when classical General Relativity breaks down.

Whether cosmic censorship is true -- and how exactly to formulate it, as there are many important subtleties\footnote{\label{welg}For example, causal propagation governed by the Einstein equations may continue
to make sense in the presence of certain limited types of singularities.    Weak null singularities that propagate outward at the speed of light are a 
candidate.
Above $\D=4$, a more concrete candidate might be the singularity arising from the Gregory-Laflamme instability. 
 So in the statement of
cosmic censorship, one might allow the formation of certain types of singularity that are  visible by an outside observer.}  -- is widely regarded as the outstanding unanswered question about
classical General Relativity.    For our purposes, we will assume that some version of cosmic censorship holds, such that it makes sense to study gravitational collapse
from the point of view of an observer at infinity in the far future.

\subsection{The Black Hole Region}\label{helpme}

If cosmic censorship is assumed, one can make a nice theory of black holes in a globally hyperbolic and asymptotically flat spacetime $M$.

     First of all, the {\it black hole region} in $M$
is the region $B$ that is not visible to an outside observer.    To be somewhat more exact, let $\I$ be the worldline of a timelike observer
who remains more or less at rest at a great distance, in the asymptotically flat region observing whatever happens.   We denote as $J^-(\I)$
the causal past of this observer, i.e. the set of  points from which the observer can receive a signal. 
$J^-(\I)$  is always open, because if a causal path $\gamma$ from a point $	q\in M$  reaches the distant observer at a point $p\in \I$,
then from any point $q'$ that is sufficiently near $q$, there is a causal path $\gamma'$ that reaches $\I$ at a point $p'$ that is near $p$.  One just defines
$\gamma'$ to follow $\gamma$  wherever it goes (somewhat as $\gamma'$ tags along after $\gamma$ in fig. \ref{46M}(b), though in the
present context $\gamma'$ is not necessarily to the past of $\gamma$).  Intuitively, an observer who can see the point $q$ at some time $t$ can
see any point that is sufficiently close to $q$ at a time close to $t$.

The black hole
region $B$ is the complement of $J^-(\I)$ in $M$:
\be\label{wondo}B=M\backslash J^-(\I).\ee
$B$ is closed, since $J^-(\I)$ is open.
 
The {\it black hole horizon} $H$ is defined to be the boundary of $B$:
\be\label{bondo} H =\partial B.\ee  (As usual, a point in $B$ is in the interior of $B$ if it has a neighborhood in $B$, and otherwise it is in the boundary of $B$.)

Let us first prove that these definitions are  sensible by showing that the existence of a black hole region is a generic property of gravitational collapse.
 We will show that {\it any compact trapped surface $W$ is in the black hole region $B$.}    In other words, we will show that a signal from a compact trapped surface
cannot reach the outside observer.

\begin{figure}
 \begin{center}
   \includegraphics[width=2in]{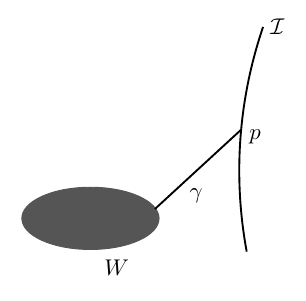} 
 \end{center}
\caption{\small 
If an observer with worldline $\I$ can receive a signal from a compact set $W$ in spacetime, then the earliest possible such signal arrives
on a prompt null geodesic $\gamma$ that connects $W$ to a point $p\in\I$.  If $W$ is a trapped region and $\I$ is sufficiently far away, this is impossible.
 \label{Fig32M}}
\end{figure}

If a causal signal from the compact trapped surface $W$ can reach the worldline $\I$ of the distant observer, there is a first point $p\in\I$ at which this can occur.
The causal path from $W$ to $p$ would be prompt, so it would be a future-going null geodesic  $\ell$ from  $W$ to $p$, orthogonal to $W$,
and without focal points (fig. \ref{Fig32M}).
Since $W$ is a compact  trapped surface, there is a focal point on $\ell$ within a known, bounded affine distance from $W$.    But $\I$, the worldline of the outside
observer, can be arbitrarily far away.  
   This is a contradiction and hence there can be no causal signal from  $W$ to $\I$. 
   
    So assuming cosmic censorship, a black hole forms in any asymptotically flat spacetime that contains a trapped surface -- and thus in any spacetime
that is close enough to the explicit Schwarzschild  solution (or any of the explicit black hole solutions, such as the Kerr solution for a rotating black hole).    
Without cosmic censorship, the conclusion does not apply because the worldline $\I$
that was assumed in the discussion might not exist.

\begin{figure}
 \begin{center}
   \includegraphics[width=2in]{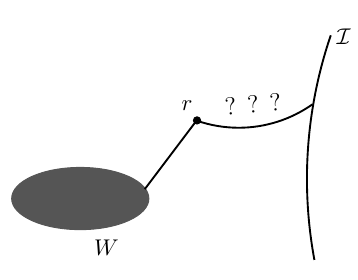} 
 \end{center}
\caption{\small An observer who can receive a signal from a point $r$ in the causal future of a set $W$ can also receive a signal from $W$.   So if $W$ is contained in the black hole
set and $r$ is in its causal future, the curve labeled $?\,?\,?$ connecting $r$ to the worldline $\I$ of an observer at infinity
 must not be future-going causal; $r$ must be contained in the black hole set.  \label{optimal2}}
\end{figure}
An obvious fact is that if a set $W$ is contained in the black hole region $B$, then its future $J^+(W)$ is also in $B$.   For  if $r$ is in the future of
$W$ and an event at $r$ can be seen by the distant observer, then that observer can also receive a signal from $W$ (fig. \ref{optimal2}).

Suppose that a trapped surface $W$ is the boundary of a codimension 1 spacelike submanifold $Z$ (which might be the ``interior'' of $W$ on some Cauchy hypersurface).
As explained in section  \ref{pfp}, a prompt causal path from $Z$ to $\I$ would actually be an orthogonal null geodesic from $W$ to $\I$.   But we already know that no such geodesic exists.
So $Z$ is in the black hole region.

\begin{figure}
 \begin{center}
   \includegraphics[width=3.5in]{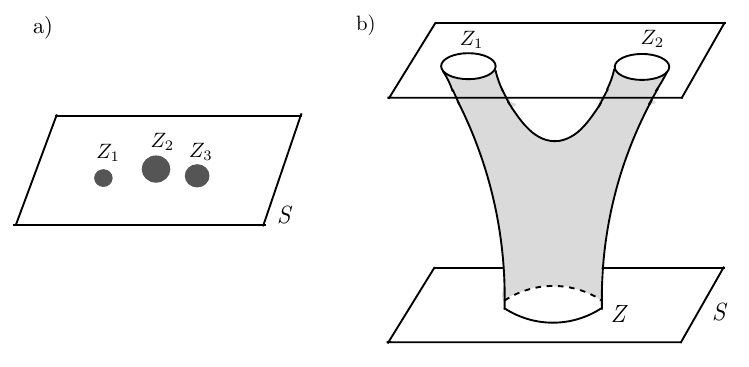} 
 \end{center}
\caption{\small (a) In a spacetime that contains multiple black holes, the black hole region on a given Cauchy hypersurface $\S$ might have several connected components $Z_i$.
(b) An impossible situation, in which a black hole splits into two, or more precisely in which a
component $Z$ of the black hole region on a Cauchy hypersurface $\S$ evolves to two disconnected components $Z_1$ and $Z_2$
on a future Cauchy hypersurface $\S'$.   Why this cannot happen is explained in the text.
 \label{Fig34M}}
\end{figure}
There might be several black holes in spacetime, so on a given Cauchy hypersurface $\S$, the black hole region might have several disconnected components
(fig. \ref{Fig34M}(a)).   Black holes can merge, but a black hole cannot split, in the sense that if $Z$ is a connected component of the black hole region on
a given Cauchy hypersurface $\S$, then the future of $Z$ intersects any Cauchy hypersurface $\S'$ to the future of $\S$ in a connected set.
To prove this, suppose that the black hole region intersects $\S'$ in disconnected components $Z'_i$, $i=1,\cdots, r$ (fig. \ref{Fig34M}(b)). 
Consider first future-going causal geodesics from $Z$.   Any such causal geodesic (like any causal path from $Z$) remains in the black hole region,
so it intersects $\S'$ in one of the $Z'_i$.  But the space of future-going causal geodesics starting at $Z$ is connected, and cannot be continuously divided into
two or more disjoint subsets that would intersect $\S'$ in different components of the black hole region.  So all causal geodesics from $Z$ arrive at $\S'$ in the
same component $Z'_i$ of the black hole region.    Now consider any future-going causal path from  $Z$.   If there is a causal path $\gamma$ from $Z$ to a given component
$Z'_i$ of the black hole region, then by maximizing
the elapsed proper time of such a path, we learn that there is a causal geodesic (lightlike or timelike) from $Z$ to $Z_i'$.  So in fact, precisely one component $Z_i'$ of the
black hole region on $\S'$ is in the future of $Z$. 

\subsection{The Horizon And Its Generators}\label{horizon}

\begin{figure}
 \begin{center}
   \includegraphics[width=2in]{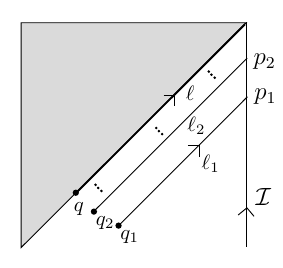} 
 \end{center}
\caption{\small In this Penrose diagram, the black hole region is shaded, and $q$ is a point in its boundary.   So $q$ is the limit of a sequence of points $q_1,q_2,\cdots$ that are outside the black hole region.
Being outside the black hole region, $q_1,q_2,\cdots$ are connected 
 by future-going causal curves to the worldline $\mathcal I$ of an observer
at a great distance. (The part of the spacetime to the right of $\mathcal I$ is not drawn.)  Hence there exist prompt null geodesics $\ell_1,\ell_2,\cdots $ from $q_1,q_2,\cdots$ to $\I$.   In a globally hyperbolic spacetime, the $\ell_i$ converge to a null geodesic
$\ell$ through $q$.   Because $q$ is contained in the black hole region, $\ell$ does not reach the worldline $\I$ of the outside observer.  What happens is rather that as $i$ increases,
the geodesic $\ell_i$ arrives at $\I$ later and later; the proper time at which the distant observer can see the point $q_i$ diverges for $i\to\infty$.   The upper corner of this
diagram, at which $\ell$ and $\I$ appear to meet, is actually at future infinity.   It is not a point in the spacetime.
 \label{Fig35M}}
\end{figure}
 Now we want to discuss the ``horizon generators.''    Let $q$ be a point on the black hole horizon, and let $\I$ be the timelike worldline of an 
 observer who is more or less stationary at infinity.     We recall that the horizon $H$ is the boundary of the black hole region $B$.
A point $q\in H$ is the limit of a sequence of points $q_1,q_2,q_3,\cdots $ that are outside of $B$.     Each of the $q_i$ is connected to the worldline $\mathcal I$ by
a future-going prompt null geodesic $\ell_i$. As $q_i\to q$, the $\ell_i$ approach a future-going null geodesic $\ell$ from $q$.   But since $q\in H\subset B$, it is not true that $\ell$ connects $q$ to a point in $\I$.   Rather, what happens is that as $q_i\to q$, the point at which $\ell_i$
reaches $\I$ goes to $\infty$ along $\I$ (fig. \ref{Fig35M}).  The limiting geodesic $\ell$ does not reach $\mathcal I$ at all. 

$\ell$ remains everywhere in the horizon.  One can see this as follows.
 $\ell$ can never go outside $B$, since in general a causal curve starting at $q\in H=\partial B$  can never reach outside $B$.
On the other hand,  $\ell$ can nowhere be in the interior of $B$, because it is the limit of the prompt null geodesics from $q_i$ that are strictly outside $B$.  So $\ell$ is everywhere in $H=\partial B$.   $\ell$ is called a horizon generator.

 In fact, we can be more precise.   
Any point $r$ that is in the past of $\ell$ (meaning that some  future-going timelike curve from $r$ reaches some point on $\ell$) is also in the past of $\ell_i$, for sufficiently
large $i$, since $\ell$ is the limit of the $\ell_i$.
  Since the $\ell_i$ meet $\I$, it follows that a causal (and in fact strictly timelike)
  path from $r$ can meet $\I$.   Thus any point $r$ that is to the past of $\ell$ is outside of the black hole
region $B$. 

$\ell$ is a prompt null geodesic, in particular with no focal point, no matter how far we continue it into the future.   Indeed, if $\ell$ fails to be prompt, there is a causal
path from $q$ to some point $r$ that is strictly to the past of some $q'\in\ell$.   As we have just discussed, such an $r$ is outside the black hole region,
and the existence of a causal path from $q$ to $r$ would show that $q$ is also outside the black hole region.

Now pick an initial value hypersurface $\S$ that contains $q$ and define $W=\S\cap H$. $\ell$ must be orthogonal\footnote{We assume here that $W$ is differentiable at least at a generic point.} to $W$, 
since a prompt causal path from any submanifold (to any destination) is always orthogonal to that submanifold.      Together, the horizon generators that pass through $W$ sweep out a three-manifold $H'$.   From what we have seen, $H'$ is contained in $H$,
and near $W$, the two are the same (assuming that $W$ is differentiable).    But if we continue into the future, $H'$ might not coincide with $H$,  since, for example, new black holes may form, as a result of which the horizon (even its connected component that contains $W$)
 may not be swept out entirely by the horizon generators that come from $W$.

Since it is swept out locally by orthogonal null geodesics, $H$ must  be a null hypersurface,
that is a hypersurface with a degenerate metric of signature $++\cdots +0$.    This is true for a general reason explained in section \ref{nullraych}:  any hypersurface
swept out by  null geodesics orthogonal to a codimension two manifold, such as $W$, is a null hypersurface.   But in the present case we can be more direct.   
The tangent vector $v$ to $\ell$ at $q$ is orthogonal to itself (since $\ell$ is null) and is orthogonal to the tangent space to $W$ (since $\ell$ is orthogonal to $W$),
so it is orthogonal to the whole tangent space to $H$, showing that the metric of $H$ is degenerate at $q$.   But here $q$ could have been any point on $H$,
so the metric of $H$ is degenerate everywhere.

 The final result that we will discuss about classical black hole horizons is possibly the most important: the Hawking area theorem.  It says that the area of the black
 hole horizon can only increase, in the sense that the area measured on an initial value hypersurface $\S'$ that is to the future of $\S$ is equal to or greater  than the
 area measured on $\S$.   The theorem applies separately to each component of the black hole region; if  two or more black holes merge, 
 the theorem says that the merger produces a black hole with a horizon area at least equal to the sum of the
 horizon areas of the original black holes.

 It suffices to show that the null expansion $\theta=\dot A/A$ of the horizon generators is everywhere nonnegative.   This being so along every horizon generator
 means that the horizon area is everywhere nondecreasing.  (To fully determine the growth of the horizon area, one has to take into account that new horizon generators can come into existence in the future of $S$, for example because of the formation of new black holes.  But that
can only give a further increase in the horizon area.)

If we know that the horizon generators are complete in the sense that they can be continued to arbitrarily positive values of their affine parameters,
we would prove the nonnegativity of $\theta$ as follows.   From Raychaudhuri's equation, if $\theta$ is negative for one of the horizon generators $\ell$,
then there will be a focal point along $\ell$ at some bounded value of its affine parameter.   But we have already shown that the horizon generators have
no focal points.

\begin{figure}
 \begin{center}
   \includegraphics[width=2in]{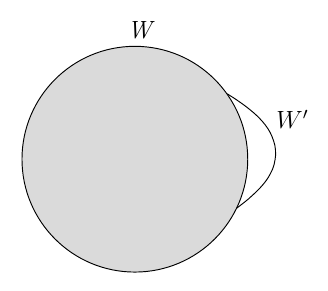} 
 \end{center}
\caption{\small $W$ is  the boundary of a component of the black hole set (shaded) on some Cauchy hypersurface $S$.  Such a $W$ is a codimension
2 spacelike submanifold of spacetime.     If there is a portion of $W$ on
which the  null expansion $\theta$ of the horizon generators is negative, then by pushing $W$ outward slightly in that region, while remaining in $S$,
we get a new submanifold $W'$, still spacelike and of codimension 2,
that is partly outside of the black hole set, and such that the part of $W'$ that is outside the black hole set still has $\theta<0$.  In the figure, the region between $W$ and $W'$ is unshaded, as it is not
part of the black hole  set.
 \label{Fig36M}}
\end{figure}
We might view the claim that the horizon generators are complete (which roughly says that the horizon is nonsingular) as a slight extension of the cosmic censorship hypothesis (which roughly says that the region outside the horizon is nonsingular).   However, it is possible
to prove the area theorem without assuming this.   Imagine that in some portion of $W=H\cap \S$, one has $\theta<0$.
 Then  we define a new surface $W'$  by pushing $W$ out a little in the region with $\theta<0$, leaving $W$ unchanged wherever $\theta\geq 0$.  
$\theta$ varies continuously as $W$ is moved in $M$, so if we do not push too far, $W'$ has $\theta<0$ in the portion outside of $B$ (fig. \ref{Fig36M}).  Since it is not
 entirely contained in $B$, $W'$ is connected to the worldline $\I$ of an observer at infinity by a causal path, which we can choose to be a prompt null geodesic $\ell$.
 $\ell$ must connect $\I$ to a point in $W'$ that is outside $B$, 
 but we have chosen $W'$ so that at such points, $\theta<0$.   Hence $\ell$ must have a focal
point within a bounded affine distance of $W'$.  This contradicts the fact that $\I$ can be arbitrarily far away. So in fact there was nowhere on $W$
with $\theta<0$.

There is certainly much more to say about black holes, but at this point we will leave the reader to continue the journey elsewhere.   We just conclude with a comment on the time-reversal
of a black hole horizon.
Black hole horizons  are {\it future horizons}; an observer at infinity can transmit a signal to the region behind a future horizon, but cannot receive a signal from that region.
Time-reversing these statements, one gets the notion of a {\it past horizon}; an observer at infinity can receive a signal from behind a past horizon, but cannot transmit a signal to that region.  
The region behind
a past horizon in an asymptotically flat spacetime is called a white hole. The time-reversed version of Penrose's theorem says that a white hole region contains a past singularity or at least a failure of past null geodesic completeness.    Idealized solutions of General
Relativity have white holes as well as black holes (the Schwarzschild solution is an example; see fig. \ref{Fig40M} in section \ref{cc}).
 In the real world, black holes form from gravitational collapse
but there is no equivalent mechanism to create white holes; and cosmic censorship suggests that white holes cannot form from good initial data.

There are also horizons in cosmology, but this notion is strongly observer-dependent. The boundary of the region that an observer can see is called the future horizon of that observer,
and the boundary of the region that an observer can influence is called the observer's past horizon.   In cosmology, past and future horizons are both natural.    For example,
the diagonal boundary of the shaded region in fig. \ref{Fig30MNew} is the future horizon of any observer whose timelike worldline ends at the upper right corner of the diagram, and the past
horizon of  any observer whose timelike worldline begins at the lower  left corner.  The shaded region is the region that the former observer can see and that the latter observer cannot influence.

. 

\section{Some Additional Topics}\label{addtop}

We turn now to several additional topics, including topological censorship, the averaged null energy condition,
and the Gao-Wald theorem.

\subsection{Topological Censorship}\label{cc}

\begin{figure}
 \begin{center}
   \includegraphics[width=4in]{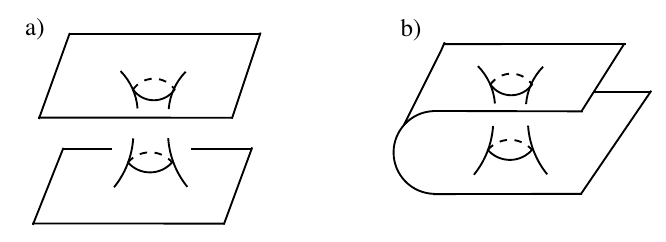} 
 \end{center}
\caption{\small  Two types of wormhole: (a) a connection between two different worlds; (b) a shortcut between distant regions of one world.   In each case, what is depicted
is a Cauchy hypersurface.    (Compare p. 837 of \cite{MTW}.)
 \label{Fig37M}}
\end{figure}

A ``wormhole'' is, for example,\footnote{A similar discussion applies in a spacetime that is asymptotic to Anti de Sitter
space \cite{topocensor2}.} 
 a geometrical connection 
between two different asymptotically flat worlds (fig. \ref{Fig37M}(a)), or 
a shortcut between two distant regions of a single asymptotically flat world (fig. \ref{Fig37M}(b)).   The two cases are closely related. The existence of a wormhole of the second type
means that spacetime is not simply-connected; by taking a cover of spacetime, one can pass to a situation of the first type. 

We recall that, topologically, a globally hyperbolic spacetime $M$  is simply $M=\S\times \R$ where $\S$ is an initial value hypersurface and $\R$ parametrizes the ``time.''
So for $M$ to have a wormhole means simply that  $\S$ has a wormhole.

\begin{figure}
 \begin{center}
   \includegraphics[width=4in]{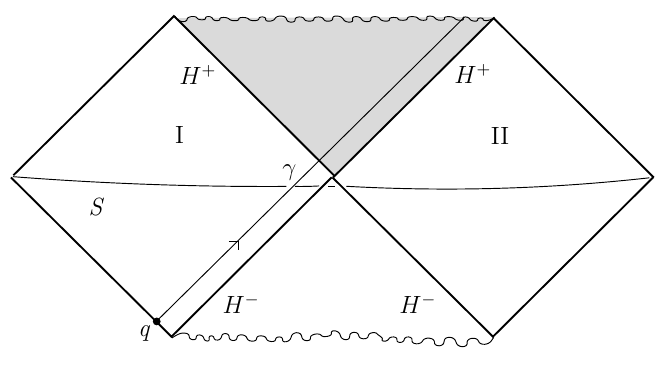}\caption{\small   This is the Penrose diagram of the maximal analytic extension of a Schwarzschild black hole.  It contains
   two asymptotically flat regions, here labeled I and II.   They are spacelike separated and causal communication between them is not possible.   The spacetime is globally
   hyperbolic, with initial value surface $S$.  It connects the two asymptotically flat regions, and has a ``wormhole'' topology, similar to that of fig. \ref{Fig37M}(a).  
   Labeled $H^+$ are the future horizons of observers who
   remain at infinity in regions I or II; beyond $H^+$ and to its future is the black hole region (shaded).  From the point of view of the outside observer,
   there also are past horizons $H^-$; beyond $H^-$ and to its past is the white hole
   region (unshaded).
    Shown in the
   figure is a radial null geodesic $\gamma$ that originates at the point $q$ at past null infinity in the asymptotically flat region on the left.   It crosses $S$ in the ``wormhole'' region, but it does not
   ``traverse the wormhole'' and enter the asymptotically flat region on the right; rather it enters the black hole region and terminates on the future singularity.      \label{Fig40M}}\end{center}
\end{figure} 
The motivating example of topological censorship is the analytically continued Schwarzschild solution (fig. \ref{Fig40M}).  This is a globally hyperbolic spacetime with
 two asymptotically flat ``ends,'' labeled I and II
in the figure.  The initial value surface $\S$ 
is actually quite similar to what was depicted in fig. \ref{Fig37M}(a).  The left and right ends of $\S$ are respectively in the asymptotically  flat regions I and II; the interior part of $\S$
 passes through the wormhole.   But as one can see from the Penrose diagram, a causal signal from one asymptotically flat region cannot travel through the wormhole to the other
region. If one enters the wormhole from the left, hoping to traverse it and to come out on the right, one will instead end up at the black hole singularity.

  Topological censorship \cite{topocensor} says that, in a spacetime that satisfies the null energy condition (and the classical Einstein equations), this is the general situation: there may be a wormhole in space, but
  an observer cannot probe it, in the sense that it is not
  possible for a causal signal to go through the wormhole and come out the other side.   In proving this, it suffices to consider a wormhole that connects two
  distinct asymptotically flat worlds, as in fig. \ref{Fig37M}(a).   The case of fig. \ref{Fig37M}(b) can be reduced to this, without affecting the classical
  null energy condition, by taking a cover of spacetime.
  
  Let $M_1$ and $M_2$ be two asymptotically flat ``ends'' of spacetime.   If it is possible for a causal signal to travel through the wormhole from $M_1$ to $M_2$, 
then such a signal could originate in the far past and at a great distance on $M_1$, travel through the wormhole, and propagate on $M_2$ to an observer
at a great distance.  To exploit this observation, we proceed as follows.

  The line element of $M_1$ is asymptotic at infinity to the Minkowski line element
\be\label{woggo}\ds^2=-\d t^2+\d r^2+r^2\d \Omega^2,\ee
where $\d\Omega^2$ is the line element of a $(\D-2)$-sphere.    Let $W$  be a sphere embedded in $M_1$ by, say, $r=r_0$, $t=-r_0+k$,
where $k$ is a constant and $r_0$ is taken to be very large.   Thus $W$ is embedded in $M_1$ at a very great distance from the wormhole and in the far past, in such
a way that
 the ``advanced time'' $t+r=k$ remains fixed.   The purpose of this is to ensure that a signal can be sent in from $W$  to arrive at the wormhole at a time of order 1, independent of $r_0$.
By varying $k$, we can adjust when the signal will arrive at the wormhole.  We will see that regardless of the choice of $k$, a signal from $W$ cannot
travel through the wormhole.  

In region $M_2$, we take $\I$ to be the worldline of an observer more or less at rest a great distance from the wormhole.\footnote{One could  introduce
Penrose's ``conformal null infinity,'' and  take $W$ to be a two-sphere in past null infinity on $M_1$, while $\I$ could be replaced by a 
 null worldline in future null infinity on $M_2$.   That  gives a simple framework to make the following argument rigorous.
We will not enter into that degree of detail.}

If it is possible for a signal to propagate from $M_1$ through the wormhole and to emerge in $M_2$, then it is possible for such a signal to be emitted from $W$ (perhaps
with $k$ chosen to be sufficiently negative), travel into the interior of $M_1$, pass through the wormhole, and eventually emerge to be detected in $M_2$ by the observer
traveling along $\I$.   If there is a causal path that propagates in this way from $W$ to $\I$, then there is a prompt causal path of 
this type, which arrives on $\I$ as soon as possible.
This prompt causal path will be a future-going null geodesic $\ell$, orthogonal to $W$ and without focal points.

Given how $W$ was defined, the future-going null geodesics orthogonal to $W$ propagate either ``outward,'' to larger $r$, or ``inward,'' to smaller $r$.
The outward-going null geodesics simply propagate outward to $r=\infty$ in the asymptotically flat region.   It is the inward-going null geodesics that might propagate
into the wormhole.    However, the inward-going null geodesics have an initially negative value of the null expansion $\theta=\dot A/A$.   Assuming the null energy condition,
Raychaudhuri's equation implies that these inward-going null geodesics reach focal points within a bounded value of their affine parameters.   As $\I$ can be arbitrarily
far away in $M_2$, it follows that all of these inward-going null geodesics reach focal points before reaching $\I$.    But this shows that there can be no prompt
causal path from $W$ to $\I$, and hence that there can be no such causal path at all.

\subsection{The Averaged Null Energy Condition}\label{anec}

  In our discussion of topological censorship,
we have assumed the null energy condition.   This is a pointwise condition on the stress tensor and was discussed in section \ref{nullraych}.
The null energy  condition is satisfied by reasonable classical matter, but  in quantum field theory,
 it is not satisfied by the expectation value of the quantum stress tensor.   So the question arises of whether topological censorship is valid in a quantum 
 universe.
 
 It turns out that topological censorship is valid under a weaker hypothesis known as the  ``averaged null energy condition'' (ANEC).\footnote{The use of integrated conditions such as the ANEC
 to prove results in General Relativity was introduced in \cite{Tipler}.} 
For a complete null geodesic $\ell$ , with affine parameter $U$ that runs to infinity at both ends, the ANEC asserts that
\be\label{zoffo}\int_\ell \d U \,T_{UU}\geq 0.\ee 
Here
\be\label{moffo} T_{UU}=\frac{\d x^\alpha}{\d U}\frac{\d x^\beta}{\d U} T_{\alpha\beta}. \ee
In the same notation, the null energy condition gives
\be\label{loffo}T_{UU}\geq 0. \ee
Clearly, the ANEC is a strictly weaker condition.

For the applications that we will give, one can think of the ANEC as a statement about the stress tensor of a classical spacetime: a certain integral is
nonnegative.  However, the ANEC is most often considered as a statement about quantum field theory in a background spacetime.
In that context, the ANEC is taken to mean that the operator $\int_\ell \d U \,T_{UU}$ is nonnegative, in the sense that its expectation value in any quantum state $\Psi$ is
nonnegative.   With that understanding of what the  statement means, the ANEC is not true in general, even if the null geodesic $\ell$ is complete (meaning that its affine
parameter extends to infinity in both directions).   For a simple counterexample,
consider the cylindrical spacetime of eqn. (\ref{zorfo}).  We earlier used this spacetime to illustrate the fact that a null geodesic is not necessarily achronal.  
To get a counterexample to the ANEC, consider a conformally invariant quantum field theory in the cylindrical spacetime.
The ground
state of a conformal field theory in this spacetime  has a negative Casimir energy.   Moreover, by translation invariance, the energy density in this state is a constant
in spacetime.    The integrand in the ANEC integral (for any null geodesic $\ell$)  is therefore negative-definite in this example, and the ANEC is not satisfied.

 It is believed that the ANEC may hold under two additional conditions:   the null geodesic in question should be achronal, and the spacetime should be 
self-consistent, meaning roughly that the Einstein equations are obeyed with a source given by the expectation of the stress tensor \cite{WY,PSW,GO}.
This has not been proved.   One basic result is that the ANEC holds for a geodesic in Minkowski space \cite{FLPW,HKT}.    There are partial results for more
general spacetimes \cite{KO}.

For our purposes, what is interesting about  the ANEC is that it suffices for some of the applications of the classical null energy condition, including topological
censorship.    In section \ref{cdp}, we will show that if $\ell$ is a null geodesic that extends in both directions to infinite values of its affine parameter, and if the ANEC is satisfied for
$\ell$ as a strict inequality,
\be\label{zoddo}\int_\ell \d U \,T_{UU}>0, \ee
then a segment of $\ell$ that is sufficiently extended in both directions is not prompt.   In the proof of topological censorship in section \ref{cc}, the key was to consider a hypothetical
prompt null geodesic $\ell$ between a sphere $W$ that is extremely far away in the asymptotically flat region $M_1$ and a timelike path $\I$ that is extremely far away in the asymptotically
flat region $M_2$.  Such an $\ell$ can be continued indefinitely  in the past and future in the two asymptotically flat 
regions, and its segment connecting $W$ to $\I$ can extended arbitrarily in both directions by moving $W$ and
$\I$ off to infinity in their respective regions.     So if we know that the ANEC would hold with a strict inequality, such an $\ell$ cannot exist and topological censorship holds for this spacetime.

What about the possibility that the ANEC would be saturated for $\ell$, in the sense that $\int_\ell \d U \,T_{UU}=0$?    
Then we can reason as follows.   If $M$ is a spacetime in which causal communication is possible from one asymptotically flat region to another, violating
topological censorship, then after a sufficiently small perturbation of $M$, causal communication between the two regions is still possible so topological censorship is
still violated.   If the ANEC is saturated by a complete null geodesic $\ell\subset M$, then after a suitable infinitesimal perturbation of $M$ -- for example, by adding a small amount of 
positive energy nonrelativistic matter
near one end of $\ell$ --
we get a spacetime $M'$ on which the ANEC holds as a strict inequality.   Thus $M'$ satisfies topological censorship, and hence so does $M$.

\subsection{The Gao-Wald Theorem}\label{gaowald}

For the next topic, the reader will need a basic familiarity with the AdS/CFT correspondence,
that is the correspondence beween quantum gravity in an asymptotically Anti de Sitter spacetime and conformal field theory (CFT) on the conformal boundary of the spacetime
(for an overview, see \cite{Overview}).

\begin{figure}
 \begin{center}
   \includegraphics[width=3.8in]{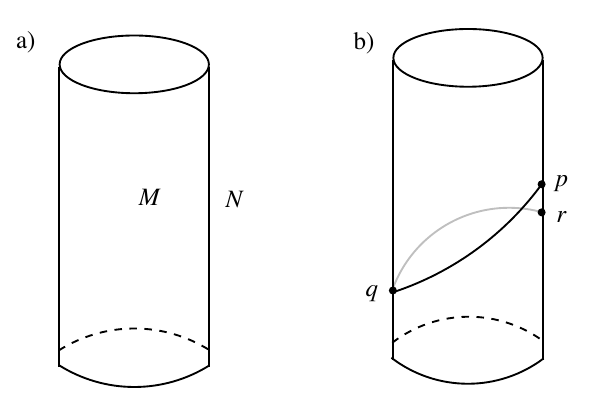} 
 \end{center}
\caption{\small  (a) According to the AdS/CFT correspondence, a quantum gravity theory in the Anti de Sitter spacetime $M$ is equivalent to an ordinary quantum
field theory on a spacetime $N$ of one dimension less which is the conformal boundary of $M$.   (b)   In the boundary theory, an event at a point $q\in N$
can influence an event at $p\in N$ if a future-going causal curve in $N$ can propagate from $q$ to $p$.  
Causality will be violated if it is possible to take a shortcut through the bulk, in the sense
that a future-going causal curve in $M$ can propagate from $q$ to a point $r\in N$ that cannot be reached from $q$
by a causal path in $N$. In the figure, the curve $qp$ is a causal curve in the boundary
while $qr$ is a hypothetical causality-violating shortcut in the bulk. The Gao-Wald theorem asserts that such causality
violation does not occur.
 \label{Fig38M}}
\end{figure}

Let $M$ be an asymptotically AdS spacetime, that is a spacetime that is asymptotic at spatial infinity to Anti de Sitter space. By adding some points at spatial infinity,
one can construct a partial conformal compactification of $M$.   The points at infinity make up a Lorentz signature manifold $N$, whose dimension is one less than the
dimension of $M$, and AdS/CFT duality says that a gravitational theory on $M$ is equivalent to some
conformal field  theory on $N$.   The picture is schematically indicated in fig. \ref{Fig38M}(a). 

The Gao-Wald theorem \cite{Gao-Wald} says that, assuming the null energy condition, or more generally assuming the ANEC,
 the AdS/CFT correspondence is compatible with causality.   This means the following.   In the boundary CFT, an event  $q$ can
influence an event  $p$ only if there is a future-going causal path from $q$ to $p$ in $N$.    But in the bulk gravitational theory,  $q$ can influence
 $p$ if there is a future-going causal path between them in $M$.   AdS/CFT duality violates causality if it is possible to ``take a shortcut through the bulk,'' that
is, if a causal path from $q$ through $M$ can arrive at a boundary point that could not be reached by a causal path  in the boundary.
In more detail, if $p$ can be reached from $q$ by a causal path in the boundary that is prompt among causal paths in the boundary, then the duality violates
causality if there is a bulk causal path
from $q$ to a boundary point $p'$ that is strictly to the past of $p$  (fig. \ref{Fig38M}(b)).

In empty AdS spacetime, every null geodesic is prompt, and a bulk null geodesic from $q$ can arrive on the boundary precisely at $p$, but no earlier.   In other words, the causality condition
is precisely satisfied in empty AdS spacetime.

  What happens if one perturbs away from empty AdS, by adding some matter or gravitational 
perturbations?  The Gao-Wald theorem says that, assuming the null energy condition -- or more generally the ANEC -- 
there is never a causality-violating shortcut through the bulk.

If there is a causality-violating shortcut, then there is one that is prompt in the sense that on some chosen
timelike worldline $\I$ through $p$, it arrives as soon as possible at some point $p'\in\I$ that is strictly to the past of $p$.  
This prompt shortcut will be a null geodesic $\ell$ without focal points.

To show that this situation cannot occur, we need to know that the affine parameter of a null geodesic diverges as it reaches the conformal boundary of an asymptotically
AdS spacetime.   This happens even though the time, as measured on the boundary, does not diverge.    In suitable coordinates, the line element of an asymptotically AdS
spacetime looks near the conformal boundary like
\be\label{wonfo} \ds^2=\frac{R^2}{z^2}\left(-\d t^2+\d z^2+\d \vec x^2\right), \ee
where $R$ is the radius of curvature (see Appendix \ref{ads}).
The spacetime is the region $z>0$; the conformal boundary is at $z=0$.   A typical null geodesic is 
\be\label{bonfo} z=-t, ~~\vec x=0,~~~ t\leq 0,\ee
and clearly reaches the boundary at $z=0$ at a finite value of $t$, namely $t=0$.  However, the affine parameter of this geodesic diverges as $t\to 0^-$.  To see this, one observes
that the equation $D^2 x^\mu/ D\lambda^2=0$ for a geodesic with affine parameter $\lambda$ gives in this case 
\be\label{lonfo}\frac{\d}{\d\lambda}\left(\frac{1}{z^2}\frac{\d t}{\d\lambda}\right)=0,\ee
so 
\be\label{ponfo}\frac{1}{z^2} \frac{\d t}{\d\lambda}=w, \ee
with a constant $w$.  This constant cannot be zero, since $t$ is not constant in the geodesic (\ref{bonfo}).    Setting $z=-t$, we get
$\d t/t^2=w\d \lambda$, so that up to an additive constant, $\lambda = -1/wt$.   Thus $\lambda$ diverges as $t\to 0^-$.  

Now let us return to our hypothetical causality-violating prompt geodesic $\ell$, a shortcut through the bulk from $q$ to a point $p'$ that is strictly to the past of $p$.
If the ANEC is satisfied along $\ell$ with a strict inequality,
\be\label{zonfo} \int_\ell \d U \,T_{UU}>0,\ee
then, since the affine parameter of $\ell$ diverges at both ends,
  the analysis in section \ref{newlook} will show that $\ell$ cannot be prompt and therefore the hypothetical causality-violating shortcut $\ell$ cannot exist.

As in the discussion of topological censorship, we can deal as follows with the possibility that the ANEC might be saturated,
\be\label{zonfox} \int_\ell \d U \,T_{UU}=0.\ee
 We perturb the spacetime slightly, adding an infinitesimal amount of matter or slightly changing the quantum state
so that the ANEC is satisfied with a strict inequality.   For a suitably small perturbation, $\ell$ will arrive on the conformal boundary so close to $p'$ as to be still strictly
to the past of $p$.   Then we get the same contradiction as before.

Of course, as a special case of this, we could have deduced the Gao-Wald theorem from the classical null energy condition $T_{UU}\geq 0$ rather than the ANEC.

\section{Another Look At Promptness}\label{newlook}

\subsection{Overview}\label{overview}

  In sections \ref{geofocal} and  \ref{ng}, we  studied a geodesic $\ell$  that originates at a specified point $q$ or a  suitable specified  submanifold $W$,
and investigated the behavior when $\ell$ is continued into the future.   One of the main ideas was that $\ell$, when continued past a focal point, is not proper time maximizing (in the
timelike case) or prompt (in the null case).    The goals of the present section are to justify these claims in a more precise way for the case of null geodesics,\footnote{We could
study timelike geodesics in a similar (and somewhat simpler)
way, but the ideas we will describe give more value added in the case of null geodesics.}  to explain a claim about the averaged null energy condition (ANEC) that was made
in section \ref{anec}, and to discuss complete achronal null geodesics.

 We explore in sections \ref{cdp} and \ref{igd} the causal paths that originate at a specified point $q$. In section \ref{ong} we will explore in a similar way the causal paths that originate on
 a codimension 2 spacelike submanifold $W$.    In section \ref{sra}, we tie things together by rederiving the Raychaudhuri equation, which was the main tool in section \ref{ng}, from a
 ``Schr\"{o}dinger'' equation -- essentially the equation of geodesic deviation --  which will be the main tool in the present analysis.   In section \ref{cang}, we discuss complete achronal
 null geodesics.

\subsection{Causal Deformations And A Necessary Condition For Promptness}\label{cdp}

Let $\ell$ be a null geodesic.   We essentially already described
in fig. \ref{Fig23M} of section \ref{prompt} a necessary condition for a segment $qp$ of $\ell$ to be prompt:   $qp$ will fail to be prompt if a proper subsegment $rr'$ of $qp$ can be
deformed, at least to first order, to a nearby null geodesic $\gamma$ from $r$ to $r'$ (in which case we say that $r$ and $r'$ are conjugate points along $\ell$). 
For then, by replacing the $rr'$ segment of $\ell$ with $\gamma$, we get a causal path from  $q$ to $p$ which is not a null geodesic, so as usual it can be modified to  make a causal path from $q$ that arrives in the past of $p$.    The reasoning still applies if $q=r$ or $p=r'$ (but not both). 

In this section, we will justify that reasoning in a more precise way, and also deduce some further properties.
It is convenient  to use the fact that along a null geodesic $\ell$, it is possible to pick  coordinates $U,V,X^A$, $A=1,\cdots,\D-2$,  such that $\ell$ lies
at $V=X^A=0$ and along $\ell$ the metric coincides with the Minkowski space metric up to quadratic order in $V$ and $X^A$:
\be\label{kofflox}\d s^2=-2\d U \d V +\sum_{A=1}^{\D-2} (\d X^A)^2+\O(V^2,VX^A,X^A X^B). \ee   Such coordinates are called Fermi normal coordinates.
The reason that such coordinates exist is as follows.  In Riemannian geometry, there is no invariant information in the metric tensor and its first derivative at a given point; all invariant local information
is contained in the Riemann tensor, which depends on the second derivative of the metric, and  covariant derivatives of the Riemann tensor, which depend on higher derivatives.
Similarly,  there is no invariant information in the metric tensor and its first derivative along a given geodesic $\ell$ beyond whether
$\ell$ is timelike, spacelike, or null.  Hence the metric can be put in standard form along $\ell$, up to second order.  The procedure to do so was described for spacelike or timelike
geodesics in \cite{MM}, and extended to null geodesics in \cite{Blau}.
   In Fermi normal coordinates, the geodesic
equation $D^2 X^\mu/D\lambda^2=0$  is satisfied with $V=X^A=0$, $U=\lambda$.  In other words, for an affine parameter for the geodesic $\ell$ at $V=X^A=0$, we can take
simply $\lambda=U$.

Since the first derivative of the metric vanishes along $\ell$ in Fermi normal coordinates, the Riemann tensor along $\ell$ can be expressed in a relatively simple way in terms of
second derivatives of the metric tensor.    The special case of this that we will need is that along $\ell$
\be\label{woflox} R_{AUBU}=-\frac{1}{2}\frac{\partial^2 g_{UU}}{\partial X^A \partial X^B} . \ee
In verifying this, one uses the fact that the first derivatives of the metric vanish along $\ell$, and also that $\partial_U \partial_A g_{BU}=\partial^2_U g_{AB}=0$ along $\ell$ (since $\partial_A g_{BU}
=\partial_U g_{AB}=0$ along
$\ell$ for all $U$).  

We aim to find a useful condition that will ensure that $\ell$ is not prompt as a path from the point $q$ with $\lambda=\lambda_0$ to the point $p$ with $\lambda=\lambda_1$ (for some choices of 
$\lambda_0$ and $\lambda_1$).   For this aim, we can
 consider deformations of $\ell$ that  preserve the fact that it is a causal curve that originates at the point $q$, 
 but do not necessarily satisfy the geodesic equation.  (In  section \ref{igd}, we will see that in a sense, the optimal deformation actually does satisfy the geodesic equation.) 
 The condition for a curve to be causal is simply
that
\be\label{toflox}-g_{\mu\nu}\frac{\d X^\mu}{\d \lambda}\frac{\d X^\nu}{\d\lambda}\geq 0 \ee
for all $\lambda$.   In deforming $\ell$, there is no point in modifying the relation $\lambda=U$, since such a modification would just amount to reparametrizing  $\ell$.    But we do want to modify
the relations $V=X^A=0$.   

The obvious idea might be to expand $V=\veps v(\lambda)+\O(\veps^2)$,  $X^A=\veps x^A(\lambda)+\O(\veps^2)$, with a small parameter $\veps$, which we may as well assume
to be positive.   However, if we proceed in this way,
then  to first order in $\veps,$ the causality condition (\ref{toflox}) depends only on $v$ and just says that
\be\label{noggy}\frac{\d v}{\d\lambda}\geq 0, \ee
so that $V$ is an increasing function of $\lambda$.    But in this case, the deformed curve $\ell'$, if it starts out on $ \ell$ at $\lambda=\lambda_0$,
 enters the future of $\ell$ as soon as $\d v/\d\lambda>0$.   The most prompt deformation
in order $\veps$ is simply the one for which $\d v/\d\lambda=0$, which together with the initial condition $v(\lambda_0)=0$ implies that $v(\lambda)$ is identically 0.

We get a more interesting test of whether $\ell$ is prompt if we set $V=\veps^2 v(\lambda)+\O(\veps^3)$,  $X^A=\veps x^A(\lambda)+\O(\veps^2)$, so that $X^A$ is perturbed in
linear order, but  $V$ is perturbed only in quadratic order.     With this choice, the causality condition (\ref{toflox}) is trivial up to order $\veps^2$, and in that order it gives
\be\label{norfox} 2 \frac{\d v}{\d\lambda}- \sum_{A=1}^{\D-2} \left(\frac{\d x^A}{\d\lambda}\right)^2 +\sum_{A,B=1}^{\D-2} x^A x^B R_{AUBU}\geq 0. \ee
(In expanding the causality condition to second order, one has to take the second derivative of $g_{UU}$; evaluating this via eqn. (\ref{woflox}) leads to the curvature term in eqn. (\ref{norfox}).   Terms
involving the first derivative of $g_{UU}$ do not contribute, because this vanishes along $\ell$ in Fermi normal coordinates.)
For example, consider a perturbed curve $\ell'$ defined by  a perturbation such that $x^A$ vanishes outside the  interval $I$ defined by $\lambda_0\leq \lambda\leq \lambda_1$.  Moreover, suppose that
$v(\lambda_0)=0$, so $\ell'$ and $\ell$ coincide at $\lambda=\lambda_0$.   Then integrating (\ref{norfox}), we get
\be\label{porfox} v(\lambda_1)\geq \frac{1}{2} J(x^A), \ee
where
\be\label{lorfox} J(x^A)=\int_I \d \lambda\left( \sum_{A=1}^{\D-2} \left(\frac{\d x^A}{\d\lambda}\right)^2 -\sum_{A,B=1}^{\D-2} x^A x^B R_{AUBU}\right). \ee

Now to get a useful criterion under which $\ell$ is not prompt, let $p$ be the point on $\ell$ with $U=\lambda_1$, $V=X^A=0$.
Suppose that $x^A(\lambda_1)=0$ and $v(\lambda_1) <0$; then at $\lambda=\lambda_1$, $\ell'$ reaches the point $p'$ with coordinates $U=\lambda_1$, $X^A=0$, $V=-\veps^2 |v(\lambda_1)|<0$.
The point $p'$ is on the past light cone of $p$.    If such a causal path $\ell'$ from $q$ to $p'$ exists, then for $\lambda$ slightly less than $\lambda_1$, $\ell'$ is
inside the past light cone of  $p$. 
Thus, in this case, $\ell$ is not prompt as a path from $q$ to $p$.

In short, $\ell$ is not prompt if it can be deformed   to a causal curve $\ell'$ that coincides with $\ell$ at $\lambda=\lambda_0$ (and therefore has $x^A=0$ there)
and  at $\lambda=\lambda_1$ satisfies $x^A=0$, $v<0$.   In other words, $\ell$ is definitely not prompt if we can have $x^A=0$ at both ends of the interval and $v(\lambda_0)=0$, $v(\lambda_1)<0$.
Since the lower bound on $v(\lambda_1)$ associated to causality is $v(\lambda_1)-v(\lambda_0)\geq J(x^A)/2$,  we can arrange for this if and only if it
is possible for the functional $J(x^A)$ to be negative for a function $x^A$ that vanishes at both endpoints of the interval.  

To state this differently, in order for $\ell$ to be prompt on a given interval $I$, the functional $J(x^A)$, evaluated on functions that vanish at the endpoints of $I$, must be nonnegative.   The condition for a functional such as $J(x^A)$ to be nonnegative 
is of a type that may be familiar from nonrelativistic quantum mechanics.   Think of $x^A$ as the components of a column vector that represents a quantum state $\Psi$.
Define the ``Hamiltonian''
 \be\label{acf}H=-\delta_{AB}\frac{\d^2}{\d\lambda^2}+P_{AB} ,\ee
with the matrix-valued potential
\be\label{hacf} P_{AB}=-R_{AUBU}. \ee
Then 
\be\label{nogod} J(x^A)=\langle\Psi|H|\Psi\rangle, \ee
and the condition for $J(x^A)$ to be nonnegative is simply that the operator $H$, acting on wavefunctions that vanish at the endpoints of the interval $I$, is nonnegative.
The condition for $x^A$ to vanish at each end of the interval represents what we will call Dirichlet boundary conditions.

The ground state energy of $H$ on a sufficiently small interval with Dirichlet boundary conditions is strictly positive, because the ``kinetic energy'' $-\delta_{AB}\frac{\d^2}{\d\lambda^2}$, which
is positive, dominates over the ``potential energy'' $P_{AB}$.
The ground state energy on an interval $[\lambda_0,\lambda_1]$  
is monotonically decreasing as the interval is enlarged, because the ground state on one interval can be
used as a variational trial wavefunction on a larger interval.   So there will be a zero-mode on some interval if and only if the ground state energy is negative on a sufficiently large interval.    A
zero-mode is related to a deformation that satisfies the geodesic equation, as we will discuss in section \ref{igd}.  Verifying this will give a more precise demonstration of the
the relationship between focal points and promptness that we presented in section \ref{ng}.

Now let us specialize to the case that $\ell$ is  a complete null geodesic, one whose affine parameter is unbounded in both directions.  The ground state energy is then negative
on an interval that is extended sufficiently in both directions if and only if it is negative on the whole real line.
An obvious way to prove that the ground state energy is negative on the whole real line is to find a trial wavefunction $x^A(\lambda)$ in which the expectation value of $H$
is negative.   Suppose that there is a constant $c^A$ such that for $x^A=c^A$, the  potential energy integrated over all of $\ell$  is negative:
\be\label{woblo} 0> \int_{\ell}\d\lambda \,c^A P_{AB} c^B=-\int_\ell\d\lambda\, c^A c^B R_{AUBU}. \ee
If we naively use $c^A$ as a variational wavefunction on the whole line, then the kinetic energy vanishes, and eqn. (\ref{woblo}) says that the potential energy is negative.
This suggests that the ground state energy is negative on the whole
line and therefore also on a sufficiently large interval.

We have to be a little more careful because the constant $c^A$ is not square-integrable and so cannot be used as a variational wavefunction.
However, we can get around this by using a variational wavefunction $x^A(\lambda)=c^A\exp(-\alpha\lambda^2)$, for sufficiently small positive $\alpha$.   The kinetic energy vanishes (as $\alpha^{1/2}$)
for $\alpha\to 0$, while the limit of the potential energy for $\alpha\to 0$ is simply\footnote{We assume $R_{AUBU}$ vanishes sufficiently rapidly for $U\to\pm\infty$ to make the integral
converge and to justify this statement. However, see \cite{Ciccone, Borde} for more careful analysis.}
 the original integral $-\int_\ell\d\lambda\, c^A c^B R_{AUBU}$.
Given this, if the  inequality (\ref{woblo}) is satisfied for some constant $c^A$, it follows that the ground state energy is negative on a sufficiently large interval. 

 How can we find a suitable
$c^A$?   If we average the inequality (\ref{woblo}) over all possible choices of $c^A$, we get simply 
\be\label{unvi} 0>-\int_\ell \d U  R_{UU},\ee
where since $\lambda=U$, we write the integration variable as $U$. 
If this condition is satisfied, then certainly (\ref{woblo}) must be satisfied for some $c^A$, and again the ground state energy is negative on a sufficiently large interval.

But via Einstein's equation $R_{UU}=8\pi G T_{UU}$, eqn. (\ref{unvi}) is equivalent to 
\be\label{wunvi}\int_\ell \d U \,T_{UU}>0,  \ee
which is the averaged null energy condition or ANEC  with a strict inequality.   Thus if the ANEC holds for $\ell$ with a strict inequality, 
then the ground state energy is negative on a sufficiently large interval, and $\ell$ is not achronal.

\subsection{Interpretation In Terms Of Geodesic Deviation}\label{igd}

In section \ref{cdp}, we found a necessary condition for $\ell$ to be prompt: the operator $H$ cannot have a zero-mode on any interval $I$ that is properly contained in $\ell$.   On the other hand, in section
\ref{overview}, we claimed as a necessary condition for promptness that the segment $I$ of $\ell$ cannot be displaced, to first order, to a nearby null geodesic segment between the same two endpoints.
The relation between the two statements is s   null geodesic displacements of $\ell$ correspond to zero-modes of $H$.   To explain this, we describe  the equation that governs a linearized perturbation of a geodesic; this is often called the equation of geodesic deviation, and its solutions are called Jacobi fields.  We will see that perturbations of a null geodesic on a given interval,
satisfying the geodesic equation to first order in the perturbation, are in natural correspondence with zero-modes of $H$.

   The path $X^\mu(\lambda)$ is a null geodesic with affine parameter $\lambda$ if 
\be \label{toggo}0=\frac{D^2 X^\mu}{D\lambda^2}=\frac{\d^2 X^\mu}{\d\lambda^2}+\Gamma^\mu_{\alpha\beta}\frac{\d X^\alpha}{\d\lambda}\frac{\d X^\beta}{\d\lambda}
\ee
and
\be\label{woggomm}0=g_{\mu\nu}\frac{\d X^\mu}{\d\lambda}\frac{\d X^\nu}{\d\lambda}.\ee  (For the moment, we write these equations in general without specializing to Fermi normal coordinates.)
We describe a perturbation of the null geodesic by 
$X^\mu_\veps(\lambda)=X^\mu(\lambda)+\veps x^\mu(\lambda)+{\mathcal O}(\veps^2)$, where $\veps$ is a small parameter.    In order for the perturbation to describe a null geodesic
up to first order in $\veps$, it should satisfy the linearized equations
\be \label{toggof}0=\frac{D^2 x^\mu}{D\lambda^2}+R^\mu{}_{\nu\alpha\beta} \frac{\d X^\nu}{\d\lambda}\frac{\d X^\beta}{\d\lambda} x^\alpha\ee
and
\be\label{toffox} 0= g_{\mu\nu}\frac{D x^\mu}{D\lambda}\frac{\d X^\nu}{\d\lambda}.\ee   
These equations always have uninteresting solutions that reflect the possibility of redefining the affine parameter of the original geodesic by $\lambda \to a\lambda+b$, with real constants $a,b$.
Concretely, these solutions take the form
\be\label{noffox}x^\mu=\alpha \frac{\d X^\mu}{\d\lambda}+\beta \lambda \frac{\d X^\mu}{\d\lambda},~~~~~\alpha,\beta\in\R.\ee 
Such solutions just describe a reparametrization of the original geodesic, not a displacement of it.   We are, of course, interested in nontrivial solutions that actually describe displacements.  

To get a null geodesic displacement of the segment $I$, we want  a solution of the linearized equations (\ref{toggof}), (\ref{toffox}) on that segment  that vanishes at the ends of the segment.   
If the endpoints  of the segment are at $\lambda=\lambda_0$ and $\lambda=\lambda_1$, we want to solve the linearized equations on the interval $[\lambda_0,\lambda_1]$ with the boundary condition
\be\label{wigo}x^\mu(\lambda_0)=x^\mu(\lambda_1)=0. \ee
The trivial solutions (\ref{noffox}) do not satisfy the boundary condition (\ref{wigo}), so this boundary condition implicitly includes a fixing of the reparametrization invariance that leads
to the existence of the trivial solutions.  Concretely, the reparametrization invariance has been fixed by specifying that after the perturbation, the values of the affine parameter at the endpoints
of the interval are unchanged.

From eqns. (\ref{toffox})  and (\ref{toggo}), we immediately deduce that
\be\label{offlox} 0 =\frac{\d}{\d\lambda}\left(g_{\mu\nu} x^\mu\frac{\d X^\nu}{\d\lambda}\right).\ee
Therefore $g_{\mu\nu}x^\mu \frac{\d X^\nu}{\d\lambda}$ is a conserved quantity along $\ell$.   This conserved quantity must vanish, since the boundary condition (\ref{wigo}) says that it vanishes
at the endpoints.   Thus we are interested in solutions of the linearized equations with
\be\label{nofflox}0=g_{\mu\nu}x^\mu\frac{\d X^\nu}{\d\lambda}. \ee

Now let us specialize to Fermi normal coordinates, $U,V,X^A$, with $\ell$ being the usual geodesic $U=\lambda$, $V=X^A=0$.   We perturb this to $U=\lambda+\veps u(\lambda)$, $V=\veps v(\lambda)$,
$X^A=\veps x^A(\lambda)$, and work to first order in $\veps$.   (Unlike section \ref{cdp}, we cannot set $u=0$.   The reason is that now we are asking for the perturbation to satisfy the geodesic
equation with affine parameter $\lambda$, while in section \ref{cdp}, there was no such requirement.)
The condition (\ref{nofflox})  just tells us that $v=0$.
Moreover, in Fermi normal  coordinates, since $\Gamma^A_{\mu\nu}=0$ at $v=x^A=0$, the equation (\ref{toggof}) reduces to
\be\label{wacb} \frac{\d^2 x^A}{\d\lambda^2}+R^A{}_{UBU}x^B=0 \ee
and 
\be\label{umab}\frac{\d^2 u}{\d\lambda^2}+R^U{}_{UBU}x^B=0. \ee
Eqn. (\ref{wacb}) is the familiar equation $H\Psi=0$, with the same Hamiltonian $H$ as in section \ref{cdp}.   On the other hand, no matter what $x^A$ may be,  there
is a unique solution of eqn. (\ref{umab}) for $u$ that obeys the boundary condition $u(\lambda_0)=u(\lambda_1)=0$.   To find this solution, pick any solution of eqn. (\ref{umab}) and then
add linear combinations of the trivial solutions (\ref{noffox}) of the homogeneous equation $\d^2 u/\d\lambda^2=0$  in order to satisfy the boundary conditions.

Thus linearized null geodesic deformations on a segment $I$  are in a natural 1-1 correspondence with zero-modes of $H$ on that segment.   Hence the criterion for promptness in terms of linearized null
geodesic deformations is equivalent to the criterion in terms of the spectrum of $H$.

Since solutions of the linear equation (\ref{wacb}) are also known as Jacobi fields,  a solution on an interval $I$ is the same
as a Jacobi field that vanishes at the endpoints of the interval.

\subsection{Orthogonal Null Geodesics}\label{ong}

So far in this section, we have considered a single null geodesic $\ell$.   However, in section \ref{ng} and in subsequent applications, it was important to also consider a family
of null geodesics that are orthogonal to a codimension two spacelike surface $W$.  Here we will adapt the present discussion to this case.   

Without any essential loss of generality, we can work in Fermi normal coordinates centered on $\ell$, and chosen so that $\ell$ intersects $W$ at $U=0$, while $W$ is defined
near $\ell$ by $U=V=0$, modulo terms of quadratic order in the normal coordinates $X^A$.  Thus
\be\label{woggi} V=f(X^A), ~~~ U=h(X^A), \ee
where $f(X^A)=f_2(X^A)+\O(X^3)$, $h(X^A)=h_2(X^A)+\O(X^3)$, and the functions $f_2(X^A)$ and $h_2(X^A)$ are homogeneous and quadratic.   
 Actually, we will see that $h_2$ plays no role and only $f_2$ is important.

  For some $\lambda_1>0$, we let $p$ be the point $V=X^A=0$, $U=\lambda_1$.
Thus we consider $\ell$, restricted to the interval $I=[0,\lambda_1]$, as a causal path from $W$ to $p$.   We want to know whether this path is prompt.
To find a necessary condition for promptness, we consider a small deformation of $\ell$ to a causal path $\ell'$ from $W$.   

We make the same expansion as before, defining $\ell'$ by $U(\lambda)=\lambda$, $V(\lambda)=\veps^2 v(\lambda)$, $X^A(\lambda)=\veps x^A(\lambda)$ modulo higher order terms.   We 
define
\be\label{tongi} \lambda_0=\veps^2 h_2(x^A).\ee 
and 
require 
\be\label{fongo} v(\lambda_0)=f_2(x^A(\lambda_0)).  \ee We consider $\ell'$ to be defined on the interval $[\lambda_0,\lambda_1]$.
The purpose of these 
conditions is to ensure that $\ell'$ starts on $W$ at the endpoint $\lambda=\lambda_0$ of the interval.
 (We can use $f_2$ and $h_2$ instead of $f$ and $h$, because we will only work to order $\veps^2$.)
 We do not put any constraint on $x^A(\lambda_0)$, since we want to allow $\ell'$ to originate at a different
point on $W$ than $\ell$. Suppose that it is possible to arrange so that $\ell'$ is a causal curve, $x^A(\lambda_1)=0$, and $v(\lambda_1)<0$.   Then the $\lambda=\lambda_1$ endpoint of $\ell'$
is the point $p'$ given by $U=\lambda_1$, $V=-\veps^2 |v(\lambda_1)|$, $X^A=0$.   This point is on the past light cone of $p$.   The existence of  a causal curve $\ell'$ from $W$
to a point on this past light cone implies, just as in section \ref{cdp}, that $\ell$, as a causal path from $W$ to $p$, is not prompt.

Though we will not constrain $\ell'$ to be orthogonal to $W$, it will be useful later to know what would be the condition of orthogonality.
A vector field tangent to $W$ has to preserve the conditions (\ref{woggi}), so it takes the
form
\be\label{tonoggo} \frac{\partial}{\partial X^A} +\frac{\partial f(X^A)}{\partial X^A}\frac{\partial}{\partial V} +\frac{\partial h(X^A)}{\partial X^A}\frac{\partial}{\partial U}=
\frac{\partial}{\partial X^A} +\veps\frac{\partial f_2(x^A)}{\partial x^A}\frac{\partial}{\partial V} +\veps\frac{\partial h_2(x^A)}{\partial x^A}\frac{\partial}{\partial U}+\O(\veps^2). \ee
The tangent vector to $\ell'$ at $\lambda=\lambda_0$ is 
\be\label{wonoggo} \frac{\partial}{\partial U}+\veps^2 \frac{\d v(\lambda)}{\d\lambda}\frac{\partial}{\partial V} +\veps \frac{\d x^A}{\d\lambda}\frac{\partial}{\partial X^A}. \ee
The condition of orthogonality in order $\veps$ is
\be\label{onoogo} \left.\left(\frac{\d x^A}{\d \lambda}-\frac{\partial f_2(x^A)}{\partial x^A}\right)\right|_{\lambda=\lambda_0} =0. \ee

Much of the derivation in section \ref{cdp} goes through without change.   In particular, the causality condition is as before
\be\label{porfix} 2 \frac{\d v}{\d\lambda}- \sum_{A=1}^{\D-2} \left(\frac{\d x^A}{\d\lambda}\right)^2 +\sum_{A,B=1}^{\D-2} x^A x^B R_{AUBU}\geq 0. \ee  
Integrating it with $v(\lambda_0)=f_2$, we  get 
\be\label{pikk} v(\lambda_1)\geq \frac{1}{2}\t J(x^A), \ee
where now 
\be\label{colorfix}\t  J(x^A)=2f_2(x^A(\lambda_0))+\int_{\lambda_0}^{\lambda_1} \d \lambda\left( \sum_{A=1}^{\D-2} \left(\frac{\d x^A}{\d\lambda}\right)^2 -\sum_{A,B=1}^{\D-2} x^A x^B R_{AUBU}\right). \ee

Just as before, $\ell$ will fail to be prompt as a path from $W$ to $p$  if it is possible to have $\t J(x^A)<0$ along with $x^A(\lambda_1)=0$.  
The strategy to determine if $\t J(x^A)$ can be negative is the same as before.     A key is that if the $x^A$ are viewed as the components of a ``quantum state'' $\Psi$, then as before (after imposing
the right boundary condition, as we explain in a moment)
\be\label{nogodd} \t J(x^A)=\langle\Psi|H|\Psi\rangle, \ee
where $H$ is the ``Hamiltonian'' 
 \be\label{acof}H=-\delta_{AB}\frac{\d^2}{\d\lambda^2}+P_{AB}, ~~~~~~ P_{AB}=-R_{AUBU}.\ee
 The only novelty in the derivation of this statement
 is that we have to be careful with surface terms at $\lambda=\lambda_0$.    There is an explicit surface term $f_2(x^A(\lambda_0))$ in eqn. (\ref{colorfox}), and in addition when one
 integrates by parts to relate $\t J(x^A)$ to the expectation value of $H$, one picks up an additional surface term.
 In our previous discussion, $x^A$ satisfied Dirichlet boundary
conditions on the interval $I$, that is it vanished
at both ends.   This being so, there was no problem in the integration by parts.  In our present discussion, $x^A$ still satisfies
Dirichlet boundary conditions at one endpoint $\lambda=\lambda_1$,  but we do not want to constrain it in that way at the other end.
To make eqn. (\ref{nogodd}) true, we need to impose on $x^A$ a boundary condition at $\lambda=\lambda_0$ that will make the surface terms vanish.  
The  condition we need is
\be\label{wacof}\left.\left( 2f_2(x^A)-\sum_A x^A \frac{\d x^A}{\d\lambda}\right)\right|_{\lambda=\lambda_0}=0. \ee
Since $f_2(x^A)$ is homogeneous and quadratic in the $x^A$, this condition follows from (\ref{onoogo}).  In other words,
we can take the boundary condition to be that $\ell'$ is orthogonal to $W$ at the point $\lambda=\lambda_0$ where they meet.   We can think of the boundary condition
at $\lambda=\lambda_0$ as a Neumann boundary condition, while at $\lambda=\lambda_1$, $\ell'$ satisfies a Dirichlet boundary condition.    

At this point,  since $\lambda_0$ is of order $\veps^2$ and this is the only place that $\veps$ appears in the definition of $\t J(x^A)$, we can set $\lambda_0=0$
and work on the original interval $I=[0,\lambda_1]$:
\be\label{colorfox}\t  J(x^A)=2f_2(x^A(0))+\int_I \d \lambda\left( \sum_{A=1}^{\D-2} \left(\frac{\d x^A}{\d\lambda}\right)^2 -\sum_{A,B=1}^{\D-2} x^A x^B R_{AUBU}\right). \ee
Likewise in eqn. (\ref{nogodd}), we consider $H$ to be defined on that interval.
The point here is that if $\t J(x^A)$ can be negative with mixed Neumann and Dirichlet boundary conditions on the full interval $[0,\lambda_1]$, then the same is true on the interval $[\lambda_0,\lambda_1]$
if $\veps$ is sufficiently small.  

    With mixed  Neumann and Dirichlet boundary
conditions at the left and right endpoints of the interval $I=[0,\lambda_1]$, the operator $H$ is self-adjoint, and the identity (\ref{nogodd}) is satisfied for every eigenfunction of $H$.   
If the ground state energy of $H$ with mixed Neumann-Dirichlet boundary
conditions is negative,
then the ground state wavefunction gives an example of a perturbation $x^A(\lambda)$ satisfying the appropriate boundary conditions and with
$\t J(x^A)<0$, showing that $\ell$ is not prompt as a path from $W$ to $p$.   Conversely, though we will not use this, if the
spectrum of $H$ is nonnegative, then it can be shown that $\t J(x^A)$ is nonnegative, for perturbations satisfying Dirichlet conditions at $\lambda=\lambda_1$ (and without assuming
any particular boundary condition on $x^A(0)$).

Just as in section \ref{cdp}, on a sufficiently small interval, the operator $H$ with mixed Neumann-Dirichlet boundary conditions has positive ground state energy.
Moreover, the ground state energy of $H$ on an interval $[0,\lambda_1]$ is monotonically decreasing as a function of $\lambda_1$, since the ground state on one interval can
be used as a variational wavefunction on a larger interval.    Therefore, the ground state energy is negative on a sufficiently large interval if and only if there is some $\lambda_1$ such
that the ground state energy on the interval $[0,\lambda_1]$ vanishes.  If there is such a $\lambda_1$, then for every $\lambda_2>\lambda_1$, the ground state energy is
negative on the interval $[0,\lambda_2]$, and $\ell$ is not prompt on that interval.

Finally, the argument of section \ref{igd} carries over without change for the case of mixed Neumann-Dirichlet boundary conditions, and shows that a zero-mode of $H$ on an interval
$[0,\lambda_1]$ corresponds to a first-order deformation of $\ell$ as a null geodesic from $W$ to $p$ that is orthogonal to $W$.    In other words, the existence of such a zero-mode is
equivalent, in the language of section \ref{pfp},  to $p$ being a focal point of the orthogonal null geodesics from $W$.    In section \ref{pfp}, we claimed that when continued past such
a focal point, an orthogonal null geodesic is no longer prompt.   From our present standpoint, this is so because the ground state energy of $H$ becomes negative.

\subsection{The Raychaudhuri Equation and Geodesic Deviation}\label{sra}

In section 
\ref{ng}, we used Raychaudhuri's equation as a way to predict the occurrence of focal points of null geodesics that are orthogonal to a given codimension two spacelike surface $W$.  In the present section, we have done the same using the equation of geodesic
deviation, which reduced to a Schr\"{o}dinger-like equation.   This second treatment is more elaborate but gives some more detailed information.   Here we will explain
how the Raychaudhuri equation can be recovered from the more complete treatment.  

 In doing this, we will make a small generalization of our previous discussion:
we consider null geodesics that originate at a given point $q$ as well as null geodesics that are orthogonal to a given surface $W$.  (The two cases can be unified in the language
of ``twist-free null congruences,'' as explained for example in \cite{Wald}; alternatively, the first case can be put in the language of the second
by defining $W$ as a  section of the future light cone of $q$.)

\def\tr{{\mathrm {tr}}}
As before, we use Fermi normal coordinates $U,V,X^A$, $A=1,\cdots ,\D-2$,
 along a geodesic $\ell$ that is defined by $V=X^A=0$.  We consider null geodesics that are constrained either to originate from the point $q$ with  $U=V=X^A=0$
 or to be orthogonal to a specified  codimension two submanifold $W$ defined by $V=f(X^A)$, $U=h(X^A)$ (where $f,h$ and their first derivatives vanish at $X^A=0$).  We use the affine parameter $\lambda=U$, which vanishes at $q$.   
 
It will suffice to look at perturbations of $\ell$ to first order.   This means that we can set $V=0$, $X^A=\veps x^A$, and set to 0 what was called $\lambda_0$ in section \ref{ong}.
  
For $i=1,\cdots, \D-2$, there is a unique solution of the equation
\be\label{zolgo} \frac{\d^2}{\d \lambda^2}x^A + R^A{}_{UBU}x^B=0 \ee
on the half-line $\lambda\geq 0$  provided that we impose suitable initial conditions.   To study null geodesics that are constrained to originate at $q$, we choose 
the initial conditions 
\be\label{molgo} x^A(0)=0,~~~~ \left.\frac{\d x^A}{\d\lambda}\right|_{\lambda=0}=\delta^A_i, \ee
for some $i\in \{1,2,\cdots,\D-2\}$.  
For null geodesics that are constrained to originate on and be orthogonal to $W$, we choose the initial conditions
\be\label{plook} x^A(0)=\delta^A_i,   ~~~\left.\left(\frac{\d x^A}{\d\lambda}-\frac{\partial f_2(x^A)}{\partial x^A}\right)\right|_{\lambda=0}=0.  \ee
The derivation will proceed in much the same way in the two cases.

Let us denote the solution that satisfies the given initial condition as $E^A{}_i(\lambda)$.    Thus, $E^A{}_i(\lambda)$ is a $(\D-2)\times (\D-2)$ matrix and
\be\label{moood} E^A{}_i(0)=0, ~~ \left.\frac{\d E^A{}_i}{\d\lambda}\right|_{\lambda=0}=\delta^A_i \ee
in the first case
or
\be\label{woood} E^A{}_i(0)=\delta^A_i,~~~   \left.\left(\frac{\d E^A{}_i}{\d\lambda}-\left.\frac{\partial f_2(x^A)}{\partial x^A}\right|_{x^A=E^A{}_i}\right)\right|_{\lambda=0}=0 \ee
in the second case.
The equation (\ref{zolgo}) implies that
\be\label{olgo}\frac{\d^2}{\d\lambda^2}E^A{}_i +R^A
{}_{UBU}E^B{}_i =0. \ee
Multiplying on the right by the inverse matrix $(E^{-1})^i{}_C$, which satisfies $E^B{}_i (E^{-1})^i{}_C=\delta^B{}_C$, we get
\be\label{holgo}\left(\frac{\d^2}{\d\lambda^2}E^A{}_i\right) (E^{-1})^i{}_C +R^A{}_{UCU}=0. \ee
Taking the trace, 
\be\label{orgo}\Tr\, \frac{\d^2 E}{\d\lambda^2}E^{-1}+R_{UU}=0.   \ee
Equivalently,
\be\label{worgo} \frac{\d}{\d\lambda}\left(\Tr \frac{\d E}{\d\lambda} E^{-1}\right) + \Tr \left[\left(\frac{\d E}{\d\lambda} E^{-1}\right)^2\right]+R_{UU}=0. \ee

This is actually equivalent to  Raychaudhuri's equation.  To explain this, first note that the area element of a little bundle of null geodesics emanating from $q$  or of
orthogonal null geodesics emanating from $W$ is
\be\label{wacko} A= \det E. \ee    So the null expansion is
\be\label{zacko}\theta =\frac{1}{A}\frac{\d A}{\d \lambda} = \Tr\,\frac{\d E}{\d \lambda} E^{-1}. \ee

The ``Schr\"{o}dinger'' equation (\ref{olgo}) implies that
\be\label{zackor}0=\frac{\d}{\d\lambda}\left( E^A{}_i \frac{\d E^A{}_j}{\d\lambda}-   E^A{}_j \frac{\d E^A{}_i}{\d\lambda} \right) ,\ee
where $E^A{}_i \frac{\d E^A{}_j}{\d\lambda}-   E^A{}_j \frac{\d E^A{}_i}{\d\lambda}$ is called the Wronskian.  The initial conditions (\ref{moood}) or (\ref{woood}) imply that the Wronskian vanishes at $\lambda=0$ 
and therefore this is true for all $\lambda$.  
This vanishing can be written\be\label{acko}  E^\tr E'-(E^\tr)' E =0, \ee
where $E^\tr$ is the transpose of the matrix $E$, and  the prime represents $\d/\d\lambda$.  Multiplying by $(E^\tr)^{-1}$ on the left and by $E^{-1}$ on the right, we get
\be\label{xacko} E' E^{-1}- (E^\tr)^{-1} (E^\tr)'=0. \ee 
This says that the matrix $E' E^{-1}$ is symmetric.    

Because $E' E^{-1}$ is  symmetric and its trace is $\theta$, we have
\be\label{nacko} E' E^{-1}=\frac{1}{\D-2}\theta + \sigma, \ee
where $\sigma $ is symmetric and traceless.    Eqn. (\ref{worgo}) now becomes 
\be\label{placko}\frac{\d\theta}{\d\lambda} + \frac{\theta^2}{\D-2} +\Tr\,\sigma^2 +R_{UU}=0, \ee
and this, after imposing the Einstein equation $R_{UU}=8 \pi G T_{UU}$, is the null Raychaudhuri equation (\ref{kongo}).

 From the traceless part of eqn. (\ref{holgo}), one can extract an equation that governs the evolution of $\sigma$:
\be\label{olacko} \frac{\d\sigma_{AB}}{\d\lambda}+\frac{2}{\D-2}\theta \sigma_{AB}+\sigma^2_{AB} +{\sf W}_{AUBU}=0 . \ee  
Here ${\sf W}$ is the Weyl tensor, the traceless part of the Riemann tensor: ${\sf W}_{AUBU}=R_{AUBU}-{\delta_{AB}R_{UU}}/(\D-2)$.
(In deriving eqn. (\ref{olacko}), it helps to observe that
$E'' E^{-1}=(E' E^{-1})'+(E'E^{-1})^2$.) We will see soon why this result,  which can also be obtained by
directly computing $R_{AUBU}$ in the metric (\ref{racb}), is useful.

The matrix-valued equation (\ref{olgo}) also has solutions (with different initial conditions) for which $E' E^{-1}$ is not symmetric.   In general, the antisymmetric part of $E' E^{-1}$ is
called the twist, and a solution for which $E' E^{-1}$ is symmetric is said to be twist-free.   The derivation of Raychaudhuri's equation (\ref{placko}) and of the evolution equation
(\ref{olacko}) for $\sigma$ is valid for any twist-free solution, that is, any solution whose Wronskian vanishes.

\subsection{Complete Achronal Null Geodesics}\label{cang}

Finally,  let us discuss what happens if we assume the classical null energy condition, a pointwise condition $T_{UU}\geq 0$, and not just its integrated version.

Unless $T_{UU}$ vanishes identically along $\ell$, 
the inequality (\ref{wunvi}) is satisfied and implies that $\ell$ is not achronal.   In the presence of classical fields that satisfy the null energy condition, it is rather special for $T_{UU}$ 
to vanish identically along a null geodesic, so it is very exceptional for a complete null geodesic to be achronal.   

But even in pure gravity, complete achronal null geodesics are highly nongeneric.    In the language of Hawking and Penrose, the generic condition for a null geodesic $\ell$ asserts that $R_{AUBU}$
is not identically zero along $\ell$.   As they originally argued in \cite{HP}, a complete null geodesic that satisfies the generic condition is never achronal.  
For the proof of this, first observe that in view of the evolution equation (\ref{olacko}), the generic condition implies that $\sigma$ cannot vanish identically along $\ell$.   The claimed
result will be obtained by using this fact in conjunction with Raychaudhuri's equation.

Consider the matrix-valued solution $E(\lambda)$ of the equation (\ref{olgo}) of geodesic deviation, with the initial conditions
\be\label{moowod} E^A{}_i(\lambda_0)=0, ~~ \left.\frac{\d E^A{}_i}{\d\lambda}\right|_{\lambda=\lambda_0}=\delta^A_i .\ee
Here $\lambda_0$ is a constant that we will eventually take to be very negative.

If $R_{AUBU}=0$, the exact solution for $E^A_i(\lambda)$ is
\be\label{ooood} E^A_i(\lambda)=\delta^A_i(\lambda-\lambda_0). \ee
This leads to
\be\label{noood}\theta(\lambda)= \Tr\, E' E^{-1} =\frac{\D-2}{\lambda-\lambda_0}, \ee
as well as $\sigma=0$.
Thus, $\theta(\lambda)$ is positive for all $\lambda>\lambda_0$, but it becomes very small for large $\lambda$.   More to the point, $\theta(\lambda)$ is very small for
any given $\lambda$ if $\lambda_0$ is taken to be very negative.

In general, the initial conditions (\ref{ooood}) for $E^A_i$ correspond to an initial behavior of $\theta$
\be\label{pombo}\theta\sim \frac{\D-2}{\lambda-\lambda_0}, ~~~~\lambda-\lambda_0\to 0^+.\ee
In particular
\be\label{kunombo}\left.\frac{1}{\theta}\right|_{\lambda=\lambda_0}=0.\ee
We can write Raychaudhuri's equation in the form
\be\label{knombo}\frac{\d}{\d\lambda}\frac{1}{\theta}=\frac{1}{\D-2}+ \frac{\Tr\,\sigma^2+R_{UU}}{\theta^2}. \ee
Integrating this and assuming that $R_{UU}\geq 0$, we find  in general $1/\theta\geq (\lambda-\lambda_0)/(\D-2)$, and therefore (as long as $\theta>0$)
\be\label{blombo}\theta\leq \frac{\D-2}{\lambda-\lambda_0}.\ee

We want to prove that $\theta(\lambda) $ becomes negative for some $\lambda>\lambda_0$.  In the evolution equation for $\sigma$, we have the initial condition  $\sigma=0$ at
$\lambda=\lambda_0$; moreover, the bound (\ref{blombo}) means that the $\theta\sigma$ term in the equation is not important if $\lambda_0$ is sufficiently negative.
 In any region with ${\sf W}_{AUBU}\not=0$, $\sigma$ will  be nonzero.
Suppose that, for sufficiently negative $\lambda_0$,  $\sigma\not=0$ in some interval $\lambda'\leq \lambda\leq \lambda''$. 
More specifically, assume that 
 $\Tr\,\sigma^2\geq w$ in that interval, for some $w>0$.   We assume that $\theta$ is nonnegative for $\lambda\leq \lambda'$, or there is nothing to prove.
 This being so, we can use
 the inequality (\ref{blombo}).   It  shows that by taking $\lambda_0$ to be sufficiently negative, we can ensure that $\theta(\lambda')<w(\lambda''-\lambda')$.
Since the Raychaudhuri equation implies that $\d\theta/\d\lambda\leq -\Tr\,\sigma^2$,  $\d\theta/\d\lambda$ will then be sufficiently  negative for $\lambda'\leq \lambda\leq \lambda''$  as to ensure that
$\theta(\lambda'')<0$.   Given this, as usual the Raychaudhuri equation implies that $\theta(\lambda)\to -\infty$ at some value $\lambda_1\leq \lambda''+(\D-2)/|\theta(\lambda'')|$.

The fact that $\theta\to -\infty$ for $\lambda\to \lambda_1$ means that a linear combination of the columns of $E$ is a Jacobi field that vanishes at $\lambda=\lambda_1$.
This Jacobi field also vanishes at $\lambda=\lambda_0$, because of the initial condition (\ref{moowod}).   So we have found a Jacobi field with two zeroes, showing that $\ell$ is not
achronal.

Clearly, then, complete achronal null geodesics are very scarce.  
The main known examples of spacetimes with complete achronal null geodesics are homogeneous spaces such as Minkowski space or
Anti de Sitter space  (in these spacetimes, every null geodesic is  complete and achronal) and the Schwarzschild, Kerr, and Reissner-Nordstr\"{o}m black hole
solutions (the horizon generators are complete and achronal).
Complete achronal null geodesics actually have remarkable properties.  For example, a splitting theorem of Galloway
 \cite{Galloway2} says that any complete achronal null geodesic in a spacetime that satisfies the classical null energy condition is contained in a null hypersurface $H$ that has vanishing null expansion and
  can be defined as the boundary of the closure\footnote{The causal future or past of a null geodesic is not necessarily closed, even in a globally hyperbolic spacetime.
  What was proved in section \ref{pfp} was that  the causal future (or past) of a {\it compact} set $W$ is closed in such a spacetime.}
   of either 
 the causal future or causal past of $\ell$. $H$ is ``ruled'' by achronal null geodesics,'' meaning that every point $p\in H$ is contained in some achronal null geodesic $\ell'\subset H$.  If $\ell$ is a horizon generator of one of the usual black hole solutions, then $H$ is the black hole horizon and the $\ell'$ are the other generators.

\def\AdS{{\mathrm{AdS}}}

\begin{appendix}

\section{Anti de Sitter Spacetime}\label{ads}

The purpose of this appendix is to briefly describe some facts about Anti de Sitter (AdS) spacetime  that may help the reader in understanding examples
described in the text.   For simplicity, except at the end, we consider  AdS spacetime of dimension 2, denoted AdS$_2$.  This is the case that was used for
illustration in sections \ref{classic} and \ref{loran}.

An ordinary two-sphere of radius $R$ is described by the familiar equation
\be\label{fameq} u^2+v^2+w^2=R^2.\ee
The line element is
\be\label{lameq} \ds^2=\d u^2+\d v^2+\d w^2. \ee
An SO(3) symmetry is manifest.    The ``antipodal point'' of the point $q$
with coordinates $\vec X=(u,v,w)$ is the point $p$ with coordinates $\vec X'=-\vec X=(-u,-v,-w)$.
We also then say that $q$ and $p$ are a pair of antipodal points.   SO(3) maps a pair of antipodal
points to another pair of antipodal points.

A typical geodesic is the circle $w=0$, or in parametric form
\be\label{wamteq} u=R\cos\theta, ~~v=R\sin\theta,~~ w=0.\ee
The arc length along this geodesic is $R\d\theta$.     Let $q$ be the point $(u,v,w)=(R,0,0)$
and $p$ the antipodal point $(u,v,w)=(-R,0,0)$.    The geodesic in eqn. (\ref{wamteq}) passes through $q$
at $\theta=0$ and, after traversing a distance $\pi R$, it  arrives at $p$ at $\theta=\pi$.   
More generally, any geodesic through $q$ is equivalent to this one by an SO(3) rotation, and similarly arrives
at  $p$ after a distance $\pi R$.   By SO(3) symmetry, this statement holds for any point $q$ in the two-sphere:
the geodesics through $q$ have as a focal point the antipodal point $p$ to $q$, which they reach after a distance $\pi R$.

To describe Anti de Sitter space of dimension 2, we start by changing a few signs in the preceding formulas.   The equation (\ref{fameq}) is replaced by
\be\label{ameq}u^2+v^2-w^2=R^2, \ee
and the line element by
\be\label{wameq} \ds^2=-\d u^2-\d v^2+\d w^2. \ee   Because of the constraint (\ref{ameq}), this formula describes a two-dimensional spacetime
with Lorentz signature $-+$;  we will call this spacetime AdS$_2^{(0)}$.
The symmetry is now SO(2,1) rather than SO(3).   
Points with coordinates $\vec X=(u,v,w)$ and $\vec X'=-\vec X=(-u,-v,-w)$ are still defined to be antipodal.
SO(2,1) maps a pair of antipodal points to another pair of antipodal points.

 A typical timelike geodesic is the circle $w=0$, or in parametric form
\be\label{pameq} u=R\cos \theta,~~~v=R\sin\theta,~~~w=0. \ee
The elapsed proper time $\tau$ along this geodesic satisfies $\d\tau=R\d\theta$.     Let $q$ be the point $(u,v,w)=(R,0,0)$
and $p$ the antipodal point $(u,v,w)=(-R,0,0)$.    The geodesic in eqn. (\ref{wameq}) passes through $q$
at $\theta=0$ and, after a proper time $\pi R$, it  arrives at $p$ at $\theta=\pi$.   Every geodesic through $q$
is equivalent to this one by an SO(2,1) rotation and therefore all timelike geodesics through $q$ focus at the antipodal point $p$ after a proper time $\pi R$, just as in the case of the two-sphere.

The spacetime $\AdS_2^{(0)}$ that we have just described has closed timelike curves, and is not what is usually called $\AdS_2$.    Indeed, the geodesic (\ref{pameq})  in $\AdS_2^{(0)}$
is a closed timelike curve; it returns to its starting point after a proper time $2\pi R$.    A spacetime AdS$_2$ that is locally equivalent to AdS$_2^{(0)}$ but
has no closed timelike curves can be obtained by passing to the universal cover of AdS$_2^{(0)}$.     One may do this explicitly by solving the equation (\ref{ameq}) with
\be\label{wobboc} u= (R^2+w^2)^{1/2} \cos t,~~~ v=(R^2+w^2)^{1/2}\sin t.\ee
If we view $t$ as an angular variable with $t\cong t+2\pi$, these formulas give a possible description of AdS$_2^{(0)}$ with coordinates $t,w$.   Passing to the universal cover and eliminating the closed
timelike curves can be accomplished by simply regarding $t$ as a real variable.   
In terms of $t,w$, the line element takes the form 
\be\label{yotto}\ds^2=-(R^2+w^2)\d t^2 + \frac{R^2\,d w^2}{R^2+w^2}. \ee
This is one explicit description of AdS$_2$ spacetime.  

There is a map $\varphi:\AdS_2\to \AdS_2^{(0)}$ that  tells us to forget that $t$ is a real variable
and just regard it as an angle. In other words, the map $\varphi$ imposes the equivalence $t\cong t+2\pi$.  For $q\in\AdS_2$, we write $\bar q$ for $\varphi(q)$.  (Up to this
point, we have more loosely written $q$ or $p$ for a point in either $\AdS_2^{(0)}$ or its universal cover.)
A point $\bar q\in \AdS_2^{(0)}$ has infinitely many ``lifts'' to $\AdS_2$, differing by $t\to t+2\pi n$, $n\in \Z$.   

Consider the future-going  timelike geodesics through a point $q\in \AdS_2$.  When projected to $ \AdS_2^{(0)}$, they map to future-going timelike geodesics through $\bar q\in\AdS_2^{(0)}$.  After a proper
time $\pi R$, the projected geodesics focus at the antipodal point $\bar p$ of $\bar q$.   Upon lifting the picture back to $\AdS_2$, this means that future-going timelike geodesics
through $q\in\AdS_2$ focus after a proper time $\pi R$ at a  point $ p$ that is a lift of $\bar p$ to $\AdS_2$.  This focusing was depicted in fig. \ref{Fig1} of section \ref{classic}.
  What happens when we further continue the geodesics up to a proper time $2\pi R$?
In $\AdS_2^{(0)}$, they return to the starting point $\bar q$, as is evident in eqn. (\ref{pameq}).    However, in $\AdS_2$, there are no closed timelike curves, and the picture is instead that the future-going timelike geodesics
through $q$ arrive, after a proper time $2\pi R$, at another focal point $q'$ with the property that $\varphi(q')=\varphi(q)=\bar q$.  In other words, $q'$ and $q$ are two different lifts to $\AdS_2$
of the same point $\bar q\in \AdS_2^{(0)}$, differing by $t\to t+2\pi$.  If we continue the timelike geodesics through $q$ into either the future or the past, the story keeps repeating:
after a proper time $\pi R n$, for any integer $n$, one arrives at a focal point that is a lift to $\AdS_2$ of $\bar q\in \AdS_2^{(0)}$ if $n$ is even, and of $\bar p\in\AdS_2^{(0)}$ if $n$ is odd.
This repeating picture is depicted in fig. \ref{Fig1}.  

    For $\vec X=(u,v,w)$, $\vec X'=(u',v',w')$, let us define the SO(2,1) invariant $\vec X\cdot \vec X'=uu'+vv'-ww'$.    
This invariant is related  as follows to the proper time elapsed along a timelike geodesic.   Consider for example the timelike geodesic (\ref{pameq}) and let $\bar q$ and $\bar p$ be two
points on this geodesic, such as the two points that correspond respectively to 
$\vec X=(R,0,0)$ and to $\vec X'=(R\cos\theta,R\sin\theta,0)$.    The proper time elapsed from $\bar q$ to $\bar p$, assuming that one traverses this geodesic in the direction of increasing $\theta$, is
$\tau = R\theta$.    On the other hand, $\vec X\cdot \vec X'=R^2\cos \theta$.  So
\be\label{polym}\tau=R\arccos(\vec X\cdot \vec X'/R^2). \ee
By  symmetry,\footnote{The symmetry of AdS$_2$ is actually not SO(2,1) but the universal cover $\widetilde {{\mathrm{SO}}}(2,1)$
of this group.}  this formula holds for the arclength of any timelike geodesic between  points in AdS$_2$.  
For any choices of lifts of 
 $\bar q, \,\bar p\in \AdS_2^{(0)}$ to $q,p\in \AdS_2$,  the proper time elapsed on a timelike geodesic from $q$ to $p$ (if one exists) is given by eqn. (\ref{polym}), but the appropriate
 branch of the arccos function depends on the choices of $q$ and $p$.

Now let us ask the following question:   what points in $\AdS_2$ can be reached from a given point $q$ by a timelike geodesic?     It is actually equivalent to ask what points in $\AdS_2^{(0)}$ can be reached
from $\bar q$ by a timelike geodesic, since $p\in \AdS_2$ can be reached from $q$ by a timelike geodesic in AdS$_2$ if and only if $\bar p\in \AdS_2^{(0)}$ can be reached from $\bar q$
in that fashion. (If there is a timelike geodesic from $\bar q$ to $\bar p$, then its lift to AdS$_2$ gives a timelike geodesic from $q$ that, if continued far enough in the past and
future, eventually reaches $p$.)  For the example considered in the last paragraph, with the geodesic (\ref{pameq}) and with  $\bar q$ and $\bar p$ corresponding to $\vec X=(R,0,0)$ and to $\vec X'=(R\cos\theta,R\sin\theta,0)$,
we see that
\be\label{tolmo} R^2\geq \vec X\cdot \vec X'\geq -R^2,\ee
with equality in the upper or lower bound if and only if $\vec X=\vec X'$ or $\vec X=-\vec X'$.  By SO(2,1) symmetry, this result is general:   $\bar p$ can be reached from $\bar q$
by a timelike geodesic if and only if $R^2\geq \vec X\cdot \vec X'\geq -R^2$.    What happens if we consider a spacelike geodesic instead?   An example of a spacelike geodesic is
given by $u=0$, or in parametric form
\be\label{wombo}u=0,~~~ v=R\cosh \theta,~~~w=R\sinh\theta. \ee
Typical 
 points (up to the action of $\widetilde {{\mathrm{SO}}}(2,1)$)  now correspond to $\vec X=(0,R,0)$, $\vec X'=(0,R\cosh \theta, R\sinh\theta)$, so $\vec X\cdot \vec X'=R^2\cosh\theta\geq R^2$.  From the symmetry, it follows that in general
 two points in $\AdS_2^{(0)}$ with
coordinates $\vec X$, $\vec X'$ are
connected by a spacelike geodesic if and only if $\vec X\cdot \vec X'\geq R^2$.  A similar analysis shows that pairs of points in $\AdS_2^{(0)}$ are connected by a null geodesic if and only if
$\vec X\cdot \vec X'=R^2$.  
A pair of points in $\AdS_2^{(0)}$ with $\vec X\cdot \vec X'<-R^2$ are not connected by any geodesic at all. If such a pair of points is lifted to $\AdS_2$ in a suitable fashion, one
gets a pair of points   $q,p\in \AdS_2$ (such as the pair $q,p$ in fig. \ref{Fig1}) such that $p$ is in the future of $q$ but there is no geodesic from $q$ to $p$.   These remarks
account for some statements in section \ref{classic}.

Finally, let us discuss spatial infinity in AdS$_2$.   In the coordinates $t,w$ of eqn. (\ref{yotto}), spatial infinity corresponds to $w\to \pm \infty$.    Let us focus on one end, say $w\to +\infty$.  We note
that $w=\infty$  is infinitely far away along the spacelike hypersurface $t=0$, and similarly along any spacelike hypersurface that is close enough to this one.   However,
a null geodesic can reach $w=\infty$ at finite $t$.   We have actually already demonstrated this in a different guise in section \ref{gaowald}.   Let us set $z=R/w$, so that $w\to\infty$ becomes
$z\to 0$.   From eqn. (\ref{yotto}), we see that near $z=0$, the AdS$_2$ line element is
\be\label{onno}\ds^2=\frac{R^2}{z^2}(-\d t^2+\d z^2). \ee
We analyzed null geodesics in this metric in section \ref{gaowald}, showing that they reach $z=0$ at finite $t$ but at an infinite value of their affine parameter. 

It is convenient to make the change of variables $\sin\sigma=R/\sqrt{R^2+w^2}$, where $0<\sigma<\pi$ for $-\infty<w<\infty$.  The AdS$_2$ line element becomes
\be\label{onco}\ds^2=\frac{R^2}{\sin^2\sigma}\left(-\d t^2+\d \sigma^2\right). \ee
The Weyl factor $R^2/\sin^2\sigma$ does not affect the causal structure of this spacetime, which therefore  is the same as that of the strip $0<\sigma<\pi$, $-\infty<t<\infty$
in a Minkowski spacetime with line element $\ds^2 =-\d t^2+\d \sigma^2$.  This strip is depicted in the Penrose diagram of fig. \ref{Fig1}.  The boundaries of the strip are not part of the spacetime, because
of the $1/\sin^2\sigma$ factor in the line element. The boundaries are at an infinite distance along a spacelike geodesic, or at an infinite value
of the affine parameter along a null geodesic.    But it is sometimes convenient to make a partial compactification of the spacetime by including the boundaries of the strip.

To describe in a similar fashion an Anti de Sitter spacetime of dimension $\D$, denoted AdS$_\D$, we start with the spacetime AdS$_\D^{(0)}$ defined by
\be\label{domo} u^2+v^2-\vec w^2=R^2,\ee
where now $\vec w=(w_1,w_2,\cdots, w_{\D-1})$ is a $(\D-1)$-vector.   The line element
is
\be\label{womo}\ds^2=R^2(-\d u^2-d v^2+\d \vec w^2). \ee
This spacetime has closed timelike curves as before.   To eliminate them, one again passes to the universal cover by writing
\be\label{minno} u=(R^2+\vec w^2)^{1/2}\cos t,~~~~ v=(R^2+\vec w^2)^{1/2}\sin t, \ee
and viewing $t$ as a real variable. This gives a spacetime known as AdS$_\D$, parametrized by $t$ and $\vec w$. Everything we said for AdS$_2$ has a fairly close analog for AdS$_\D$.    Topologically, while AdS$_2$ is $\R\times I$, where $\R $ is parametrized
by the ``time'' $t$ and $I$ is the open interval $0<\sigma<\pi$, AdS$_\D$ is $\R\times B_{\D-1}$, where $\R$ is as before but $B_{\D-1}$ is now an open ball of dimension $\D-1$.   As in the case $\D=2$,
it is convenient to make a partial compactification by adding boundary points to $B_{\D-1}$, replacing it by a closed ball $\bar B_{\D-1}$.    The partial compactification is thus $\R\times \bar B_{\D-1}$,
as  schematically depicted in fig. \ref{Fig38M} of section \ref{gaowald}.  

We can introduce a new variable $\sigma$ by $\sin\sigma=R/\sqrt{R^2+|\vec w|^2}$.  $\sigma$
varies from $0$ to $\pi/2$ as $|\vec w|$ varies from  $\infty$ to 0.    A Penrose diagram of AdS$_\D$ that depicts the coordinates $t$ and $\sigma$ looks like just half of fig. \ref{Fig1}, for instance the left half with $\sigma\leq \pi/2$.  
A point in the diagram represents a $(\D-1)$-sphere parametrized by $\vec w$ with fixed $|\vec w|$. These $(\D-1)$-spheres sweep out all of AdS$_\D$.
 Alternatively, one can define a Penrose diagram for AdS$_\D$ that consists of a two-dimensional
slice of AdS$_\D$ with the whole range $0<\sigma<\pi$ included.  From this point of view, the Penrose diagram of AdS$_\D$ looks just like fig. \ref{Fig1}.   

\section{Two Coordinate Systems For de Sitter Space}\label{dSitter}

Here we will derive  the two forms of the de Sitter space metric that were used in eqns. (\ref{dS}) and (\ref{dSp}) in illustrating the fine print in Penrose's theorem.

De Sitter space of dimension $\D$ and radius of curvature $R$ can be described by coordinates $X_0,X_1,\cdots ,X_\D$ satisfying
\be\label{bozo} -X_0^2+\sum_{i=1}^\D X_i^2=R^2, \ee
with the line element
\be\label{nozo}\d s^2=-\d X_0^2+\sum_{i=1}^\D\d X_i^2.\ee
An SO(1,$\D$) symmetry is manifest.

To get one convenient form for the metric, let $\vec Y$ be a unit $\D$-vector, satisfying $\vec Y\cdot \vec Y=1$.   Then we can solve the constraint (\ref{bozo})
by
\be\label{wozo} X_0=R\sinh t/R, ~~\vec X=R\vec Y \cosh t/R .  \ee
The line element is
\be\label{poazo}\d s^2=-\d t^2+R^2\cosh^2 t/R \,\d\Omega^2,~~~~\d\Omega^2=\d\vec Y\cdot \d\vec Y. \ee
This accounts for the form (\ref{dS}) of the de Sitter metric.   Clearly, these coordinates cover all of de Sitter space.

For an alternative coordinate system, define
\be\label{plozo} X_0-X_\D=\sqrt 2 \, V,~~~X_0+X_\D=\sqrt 2 U. \ee
Let $\vec X$ be the $(\D-1)$-vector $\vec X=(X_1,X_2,\cdots, X_{\D-1})$,
and set
\be\label{mottob} V=- \exp(-t/R),~~~\vec X = V\vec x. \ee
The line element turns out to be
\be\label{roszo} \d s^2= -\d t^2+R^2 e^{-2t/R}\d \vec x^2. \ee This accounts for the second form of the metric in eqn. (\ref{dSp}).
Clearly, these coordinates describe only part of de Sitter space, namely the part with $V<0$.   This part is shaded gray in fig. \ref{Fig30MNew} (in the figure, $U$ increases to the upper
right and $V$ to the upper left).

The condition $V=0$ defines a hypersurface $Y$ that bounds the region described by our second set of coordinates $t, \vec x$.   This hypersurface
has interesting properties.   Let $W$ be the codimension 2 submanifold $U=V=0$.    The hypersurface $Y$ is swept out by one of the two families of null
geodesics orthogonal to $W$.   (The other family sweeps out the hypersurface $U=0$.)  The line element of $Y$ is of  the degenerate form $\d \vec X^2$.
  The null geodesics orthogonal to $W$ and tangent to $Y$ are simply defined by $V=0$,
$\vec X=\vec a$ (where $\vec a$ is a constant of length $R$).      These geodesics are complete and achronal; their expansion is 0, since a cross section
of $Y$ at any value of $U$ is a round $(\D-2)$-sphere of radius $R$, independent of $U$.   $Y$ is a cosmological horizon for some observers,
as we noted at the end of section \ref{horizon}.  The second family of null geodesics orthogonal to $W$ is defined by $U=0$, $\vec X=\vec a$, also with vanishing
expansion.   Because its null expansions vanish,
 $W$ is called a marginally trapped surface.    A point with $U,V<0$ -- such as the point $q$ in fig. \ref{Fig30MNew} -- defines a trapped surface, with
negative expansions.   

\section{Existence Of Complete Metrics Of Euclidean Signature}\label{moredetail}

 In section \ref{cpct}, we used the fact \cite{NO}  that any manifold $M$ admits a complete  metric of Euclidean signature (a metric in which every inextendible
 curve has infinite arclength).   Here we explain this fact in more detail.   
 
 First of all, the reason that every $M$ admits some Euclidean signature metric is that such metrics can be added.   A Euclidean metric  is a symmetric
 tensor field $h_{ij}(x)$  that is positive-definite for each $x$; it satisfies no other general constraint.   Any $M$ can be covered by small open sets $U_\alpha$ each of which is
 isomorphic to an open ball in $\R^n$.   To define a Euclidean signature
 metric tensor $h$ on $M$, we simply pick  for each $\alpha$ a  tensor field $h_\alpha$ that is positive-definite on $U_\alpha$ and vanishes outside $U_\alpha$,
 and add them
 up to get a tensor field $h=\sum_\alpha h_\alpha$ on $M$.   Such an $h$ is everywhere positive-definite.
  By picking the $h_\alpha$ to vanish appropriately near the boundary of the closure of $U_\alpha$, 
 one can ensure that $h$ is smooth. 
 
In general, this construction will not give a {\it complete} metric.    In constructing a complete metric, we may as well assume that $M$ is connected,
so that there is a path in $M$ between any two points $p,q\in M$.  Otherwise, we make the following argument for each connected component
in $M$.

To find a complete metric, begin with any Euclidean signature metric $h_0$.
If this metric is not complete, then there is a point $q\in M$ and an inextendible path $\gamma_0$ from $q$ that has finite length in the metric $h_0$. (The
path $\gamma_0$ ``goes to infinity in $M$ in finite length.'')
This implies that there is such a finite length inextendible path $\gamma$ from any $p\in M$: one can just define $\gamma$ as the composition
$\gamma_0 *\gamma_1$, where $\gamma_1$ is any path from $p$ to $q$ (paths are composed by joining them end to end).
So we can 
define on $M$ the following function $\upphi$  valued in the positive real numbers $\R_+$: for a point  $p\in  M$, $\upphi(p)$  is the greatest lower bound of the length  of any inextendible
 curve in $M$ that starts at $p$.  ($\upphi(p)$ is strictly positive for each $p$ because each $p$ has a small neighborhood $B$ that can be approximated by
 a ball in Euclidean space; the radius of such a ball is a lower bound for $\upphi(p)$, since any path from $p$ has to begin with a path in $B$.)   
   If $|p-q|$ is the shortest distance between points $p,q\in M$ in the  metric $h_0$, then $|\upphi(p)-\upphi(q)|\leq |p-q|$, showing that the function $\upphi$
 is continuous.
   Define  a  new metric  $h$ by $h=\exp(1/\upphi(x)) h_0$.
 Since  $1/\upphi(x)> 0$ everywhere, the length of any curve in the metric $h$  exceeds its length  in the metric $h_0$.   So  to show that  in the metric  $h$, every inextendible curve has infinite length,
 we just have to prove that if $\gamma$ is a finite length inextendible curve starting at $p$ in the metric $h_0$,
 then it has infinite length in the metric $h$.   Suppose that $\gamma$ has length $a$ in the  metric $h_0$ and parametrize it by an arclength parameter $t$ in that metric, so $t$
 ranges over a semi-open interval $[0,a)$.   For any point $t=t_0$ in this interval,  the segment $[t_0,a)$ of $\gamma$ is an inextendible
 curve from the point $x(t_0)$ that has length $a-t_0$ in the metric $h_0$.   So $\upphi(x(t_0))\leq a-t_0$ and  $1/\upphi(x(t_0)) \geq 1/(a-t_0)$. Hence $\exp(1/\upphi(x(t)))$ blows up rapidly
 for $t\to a$ and $\gamma$  has
 infinite length in the metric $h$.  Thus every inextendible curve in the metric $h$ has divergent arclength, and $M$ with this metric
 is a complete Riemannian manifold.  In particular, in the metric $h$,  inextendible
 geodesics have infinite arclength in both directions  and $M$  is geodesically complete. 
 (Conversely, geodesic completeness implies that
 all inextendible curves have infinite  length.)

\section{Some Details About Compactness}\label{detailedproof}

Let $M$ be a globally hyperbolic spacetime, with Cauchy hypersurface $\S$, and let $q$ be a point to the past of $\S$.   Let $\CC_q^\S$ be the space of causal
paths from $q$ to $\S$.   An important fact is that $\CC_q^\S$ is compact; a corollary is that $D_q^\S$, the set of points in $M$ that can be visited by a causal path
from $q$ to $\S$, is also compact.
In section \ref{cpct}, we described some initial steps of a proof of compactness of $\CC_q^\S$, deferring details to this appendix.   We will explain two versions of the proof,
adapted respectively from \cite{Wald} (see the proof of  Lemma 8.1.5) and \cite{BEE} (see the proof of Proposition 3.31).

Let $\gamma_1,\gamma_2,\cdots$ be a sequence of causal curves from $q$ to $\S$.   It is convenient to extend the $\gamma_i$ into the future as inextendible causal curves;
for example, we can continue $\gamma_i$ to the future of $S$ as a timelike geodesic orthogonal to $S$, continued into the future for as far as it will go.

Let $\t \gamma$ be some future-going causal curve from $q$.  We do not  assume that $\t\gamma$ is inextendible.  $\t\gamma$ has $q$ as a past endpoint.\footnote{We use the term ``endpoint'' in
this familiar sense, and not in a more technical sense
that the term is used in mathematical relativity.   See footnote \ref{technical} of section \ref{def}, and the accompanying discussion.}   It is convenient
in the following to assume that $\t\gamma$ does not have a future endpoint
 and so is topologically a semi-open interval $[0,1)$.  (If $\t\gamma$ has a future endpoint, we omit this point.)
We say that $\t\gamma$ is a partial limit curve of the $\gamma_i$ if, after possibly passing to a subsequence, the $\gamma_i$  have initial segments
that converge to $\t\gamma$.    
  We call such a $\t\gamma$ a {\it partial} limit curve,
because  the $\gamma_i$ may possibly  continue to the future of $\t\gamma$.   Only initial segments of the $\gamma_i$ are assumed to converge to $\t\gamma$.

The argument in section \ref{cpct} shows that partial limit curves exist, and moreover that if $\t\gamma$ is a partial limit curve that can be extended as a curve, then it can be
extended as a partial limit curve.    To clarify this last point, suppose that $\t\gamma$ is a partial limit curve that is extendible as a curve.   This in particular
means that it is possible to add to $\t\gamma$
a future endpoint $p$. 
 As the curves $\gamma_i$ are causal curves that converge to $\t\gamma$ everywhere in the past of $p$,
and as causality prevents the $\gamma_i$ from having wild fluctuations near $p$, the $\gamma_i$ also converge to $\t\gamma$ at $p$.  (Essentially this argument
was the starting point in section \ref{classic}.)  If the point $p$ is in $S$, then we have found a convergent subsequence of the original sequence $\{\gamma_i\}$, showing that $\CC_q^S$ is compact.
Otherwise,   arguing as in  section \ref{cpct}, 
 we learn that a subsequence of the $\gamma_i$ converges to $\t\gamma$ also to the future of $p$, at least for a limited time.  
 
Thus we can keep improving partial limit curves to find better ones that reach closer to $\S$.  But to finish a proof along these lines is still tricky.  

Let $\L$ be the set of all partial limit curves of the $\gamma_i$.   If $\t\gamma,\t\gamma'\in\L$, we say that $\t\gamma<\t\gamma'$ if $\t\gamma'$ is an extension of $\t\gamma$.
This defines a partial order on $\L$.   It is only a partial order, because in general if $\t\gamma,\t\gamma'\in\L$, neither one is an extension of the other.

\def\W{{\mathcal W}}
A {\it chain} is a subset $\W$ of $\L$ that is totally ordered, meaning that if $\t\gamma,\t\gamma'\in\L$ (and $\t\gamma\not=\t\gamma'$), then one of them is an extension of the other, so that either $\t\gamma<\t\gamma'$
or $\t\gamma'<\t\gamma$.   If $\W$ is a chain, then we can define a future-going causal curve $\bar\gamma$ from $q$ by simply taking the union of all $\t\gamma\in \W$.
We will show in a moment that $\bar\gamma$ is a partial limit curve.    In this case, since every $\t\gamma\in \W$ is contained in $\bar\gamma$, any such $\t\gamma$ obeys
$\t\gamma< \bar\gamma$ (or $\t\gamma=\bar\gamma$), so that $\bar\gamma$ is an ``upper bound'' for the chain $\W$.

An abstract statement of set theory known as Zorn's Lemma\footnote{Zorn's Lemma is equivalent to the Axiom of Choice. To some, it is an essential
part of the foundations of mathematics. To others, it is an unnecessary adornment.}
 asserts that if $\L$ is a partially ordered set in which every chain has an upper bound, then $\L$ has a maximal
element $\gamma^*$.   In the present context, a maximal element of $\L$ is a partial limit curve $\gamma^*$ that is inextendible as a partial limit curve.   But this means that $\gamma^*$
is actually inextendible as a curve (since we have seen that if a partial limit curve can be extended as a curve, then it can be extended as a partial limit curve).   Now we invoke
the assumption that $M$ is globally hyperbolic.    As $q$ is to the past of $M$ and $\gamma^*$ is an inextendible future-going causal curve from $q$, it meets $\S$ and so defines
a point in $\CC_q^\S$.  Thus any sequence $\gamma_1,\gamma_2,\cdots\in \CC_q^S$ has a subsequence that converges to $\gamma^*\in \CC_q^\S$;
in other words, $\CC_q^\S$ is compact.

To complete this argument, we need to justify the claim that $\bar\gamma$ is a partial limit curve.   Since it has no future endpoint,  $\bar\gamma$ is isomorphic topologically to the
semi-open interval $[0,1)$; pick a specific isomorphism.   As
 $\bar\gamma$ is the union of the curves in the totally ordered set $\W$, there is a sequence $\ell_1,\ell_2,\cdots \in \W$ whose union is $\bar\gamma$.  If one of the $\ell_i$ equals $\bar\gamma$,
 we are done.   Otherwise,
after possibly passing to a subsequence, we can assume that the image of the closed interval $[0,1-1/n]$ is contained in $\ell_n$ for every positive integer $n$.   Pick an arbitrary Euclidean signature metric
 $h$ on $M$,
and for any positive integer $n$, let $\Phi_n$ be the subset
of the original sequence $\gamma_1,\gamma_2,\cdots$ consisting of curves that are everywhere within a Euclidean distance $1/n$ of $\bar\gamma$ over the whole interval $[0,1-1/n]$.
Each $\Phi_n$ is nonempty, since each $\ell_n$ is a partial limit curve.   Let $\gamma'_1,\gamma'_2,\cdots $ be a subsequence of the original sequence $\gamma_1,\gamma_2,\cdots$
such that $\gamma'_n\in \Phi_n$ for all $n$.   Then $\gamma'_1,\gamma'_2,\cdots $ is a subsequence of the original sequence that converges to $\bar\gamma$ everywhere,
so $\bar\gamma$ is a partial limit curve.

\def\SS{{\sf S}}
\def\BB{{\sf B}}
As Wald notes (see the footnote on p. 194 of \cite{Wald}), one would prefer not to base a concrete statement of physics on subtleties of set theory and assumptions
about the foundations of mathematics unless this becomes unavoidable.   
The following, adapted from \cite{BEE}, is one way to avoid\footnote{The fundamentals of calculus might depend on Zorn's Lemma, depending on one's point of view.  Still it seems that the use of Zorn's Lemma in the argument  just described is the sort of thing that one would wish to avoid.}   Zorn's Lemma.
Pick again a Riemannian metric $h$ on $M$, that is a metric of Euclidean signature.   
After possibly making a conformal rescaling of $h$ by a factor that blows up at infinity in a suitable fashion,
one can assume that the metric $h$ is complete \cite{NO}.   (For the  proof, see Appendix \ref{moredetail}.)   Let now $q$ be any point in $M$.   Because the metric $h$ on $M$ is complete, any point $p\in M$ is connected to $q$
by a geodesic of the shortest possible length.\footnote{Let $b$ be the greatest lower bound of the lengths of paths from $p$ to $q$ and let $\gamma_1,\gamma_2,\cdots $ be
a sequence of paths from $p$ to $q$ whose lengths approach $b$.  In a complete Euclidean signature metric, a subsequence of $\gamma_1,\gamma_2,\cdots$ will
converge to a geodesic from $p$ to $q$ of the minimum possible length $b$.} The length of this geodesic gives a function $f(p)$.    The generalized sphere $\SS_\rho\subset M$ consisting of points with $f(p)=\rho$ is always
compact.\footnote{A geodesic $\ell$ (in the metric $h$) that originates at $q$ is determined by its initial direction, which takes values in an ordinary sphere $\SS$, which is certainly compact.
Mapping $\ell$ to the point on $\ell$ whose arclength from $q$ is $\rho$ gives a continuous and surjective map from $\SS$ to $\SS_\rho$, showing that $\SS_\rho$ is compact.   This argument has been stated assuming that
every geodesic is length-minimizing up to distance $\rho$.   If not, before making this argument, one replaces
$\SS$ by its compact subset consisting of initial directions of geodesics that are length-minimizing up to that
distance.}   Likewise, the generalized ball $\BB_\rho\subset M$ consisting of points with $f(p)\leq \rho$ is compact.   

Now let $\gamma_1,\gamma_2,\cdots $ be any sequence of inextendible curves that originate at $q\in M$.  (For the moment, there is no need to assume that the $\gamma_i$ are causal
curves in the Lorentz signature metric of $M$, or that $M$ is globally hyperbolic.)   Because the Riemannian metric of $M$ is complete and the $\gamma_i$ are inextendible, they
all have infinite length in the Riemannian metric.   We parametrize them by their Riemannian arclength $t$, which runs over the full range $[0,\infty)$.   If we restrict the $\gamma_i$
to $t\leq 1$, they are contained in the compact set $\BB_1$.   Therefore, by the argument explained in section \ref{classic}, the $\gamma_i$ have a subsequence $\gamma_{i,1}$ that
converges for $t\leq 1$.   Similarly, for $t\leq 2$, the $\gamma_{i,1}$ are contained in $\BB_2$, so they have a subsequence $\gamma_{i,2}$ that converges for $t\leq 2$.    Continuing this way, we define for every positive
integer $n$ a subsequence $\gamma_{i,n}$ of a previously defined sequence $\gamma_{i,n-1}$, such that $\gamma_{i,n}$ converges for $t\leq n$.   Then the ``diagonal sequence''
$\gamma_{1,1},\gamma_{2,2},\cdots$ is a subsequence of the original sequence $\gamma_1,\gamma_2,\cdots$ that converges for all $t\geq 0$.    Thus we have shown, with no assumption
about causality or Lorentz signature, that every sequence $\gamma_1,\gamma_2,\cdots$
 of inextendible curves from $q$ has a subsequence that converges to a limit curve $\bar\gamma$.   Moreover, $\bar\gamma$
has infinite Euclidean length, like the $\gamma_i$, so it is inextendible.  

Now let us remember the Lorentz signature metric of $M$.   If the $\gamma_i$ are causal curves, then, as a limit of causal curves, $\bar\gamma$ is also causal.  If the $\gamma_i$
are future-going curves from a point $q\in M$, then $\bar\gamma$ is also a future-going curve from $q$.  If $M$ is globally hyperbolic and $q$ is to the past of a Cauchy hypersurface 
$\S$, then $\bar\gamma$
 intersects $\S$ somewhere.   Thus, $\bar\gamma$ is the desired limit of a subsequence of the $\gamma_i$, and  we learn again that the space of causal curves from $q$ to $\S$ is compact.

We can also use this second argument as the starting point to show that globally hyperbolic spacetimes are strongly causal, as claimed in section \ref{causality}. (We cannot use the first
argument for this purpose, since strong causality was assumed at the outset in section \ref{cpct}.)    Let $M$ be a globally hyperbolic
spacetime with Cauchy hypersurface $\S$.   Let $q$ be a point to the past of $\S$.
 A point to the future of $\S$ can be treated similarly; later, we consider the case $q\in \S$.   If strong causality
is violated at $q$, this means that there are sequences of points $p_1,p_2,\cdots$ and $p'_1,p'_2,\cdots$, with both sequences converging to $q$, and future-directed causal curves $\gamma_i$
from $p_i$ to $p'_i$, which make large excursions away from $q$ even though the $p_i$ and $p'_i$ are near $q$.   The precise meaning of ``large excursions'' is that a sufficiently 
small open set $U$ containing $q$ does not contain the curves $\gamma_i$, for large $i$.  
After possibly discarding the first few pairs of points, we can
assume that the points $p_i$ and $p'_i$ are all to the past of $\S$, and therefore that the curves $\gamma_i$ are to the past of $\S$.   

We consider two cases: (1) the arclengths of the curves $\gamma_i$
in the complete Riemannian metric introduced earlier are bounded; (2) those arclengths are unbounded.
In case (1), the curves $\gamma_i$ are all contained in a generalized ball $\BB_N$, for sufficiently large $N$.    $\BB_N$ is compact and the $\gamma_i$ are causal, so reasoning
as before, we find that a subsequence of the $\gamma_i$ converges to a causal curve $\bar\gamma$.    Since the sequences $\{p_i\}$
and $\{p'_i\}$ both converge to $q$, $\bar\gamma$ is a closed causal curve from $q$ to itself. ($\bar\gamma$ is a nontrivial closed causal curve, not
just consisting of the single point $q$, since there is an open neighborhood
$U$ of $q$ that does not contain the $\gamma_i$, for large $i$.)    In a globally hyperbolic spacetime, this is not possible.

In case (2), after possibly passing to a subsequence, we can assume that, for all $k\geq 1$,  the curve $\gamma_k$ has Euclidean arclength at least $k$.   We parametrize
the $\gamma_k$ by the Euclidean arclength $t$, with $t=0$ corresponding to the past endpoint $p_k$.    Arguing as before, if we restrict the $\gamma_k$ to $0\leq t\leq 1$,
then a subsequence $\{\gamma_{k,1}\}$ of the sequence $\{\gamma_k\}$ converges to a limit curve $\bar \gamma_1$.  $\bar\gamma_1$ is a future-going causal curve that originates at $q$, since the $p_k$ converge to $q$.
 Restricting to $0\leq t\leq 2$, a subsequence \{$\gamma_{k,2}\}$ of the sequence $\{\gamma_{k,1}\}$
converges on the interval $0\leq t\leq 2$ to a limit curve $\bar\gamma_2$ that is an extension of $\bar\gamma_1$. It is again a future-going causal curve from $q$.
   Continuing in this way, at the $n^{th}$ step, we define a subsequence $\{\gamma_{k,n}\}$
of $\{\gamma_{k,n-1}\}$ that converges for $0\leq t\leq n$ to a limit curve $\bar\gamma_n$ that is an extension of $\bar\gamma_{n-1}$.   All of the $\bar\gamma_{n}$ are future-going causal
curves from $q$ and each one is an extension of the previous one.   Each one is in the past of $\S$, since the original curves $\gamma_k$ are all in the past of $\S$.    The diagonal sequence
whose $n^{th}$ element is $\gamma_{n,n}$ converges to a limit curve $\bar\gamma$ that is the union of the $\bar\gamma_n$.    $\bar\gamma$ is to the past of $\S$, since all of the $\bar\gamma_n$ are.
$\bar\gamma$ is inextendible, since the Euclidean arclength of the curve $\bar\gamma_n$ diverges for $n\to\infty$.   Thus we have found, from a point $q$ to the past of $\S$, an inextendible
future-going causal curve that remains forever to the past of $\S$.   In a globally hyperbolic spacetime, this is not possible.

The reader might want to consider the example (\ref{vunxo}) of a spacetime that has no closed causal curves, but violates strong causality, to see how case (2) comes about.   Taking $q$ to
be a point at $v=0$, describe
suitable pairs $p_i$, $p'_i$ and corresponding causal curves $\gamma_i$, and try to identify the inextendible causal curve $\bar\gamma$  that will arise from this construction.

Finally, we have to consider the case that $q\in \S$.   By displacing $\S$ slightly forward in time in a local Minkowski neighborhood of $q$, we can replace it with a different Cauchy hypersurface
$\S'$ that does not contain $q$, and thus reduce to the previous case.

\section{Geometry Of A Null Hypersurface}\label{null}

In section \ref{nullraych}, we showed that a hypersurface $Y$ that is swept out by orthogonal null geodesics from some codimension 2 spacelike submanifold $W$ is null, that is its metric has signature 
$+ + \cdots +0$.  This statement has a rather surprising converse, which will be explained here.
Let $Y$ be any null hypersurface, that is any hypersurface of signature $+ + \cdots + 0$, in a spacetime $M$. Let $K^\mu$ be a vector field that is tangent to $Y$ and null.
In view of the signature of $Y$, if $Z^\mu$  is any vector field tangent to $Y$, then 
\be\label{zoff} g_{\mu\nu}K^\mu Z^\nu = 0.\ee
These conditions do not determine $K$ uniquely, as we may transform \be\label{delbo}K^\mu\to e^f K^\mu\ee
for any real-valued function $f$  on $Y$.

The integral curves of $K$ are the null curves found by solving the equation $\frac{\d X^\mu}{\d s}=K^\mu(X(s))$, 
 with some choice of an initial point on $Y$. Note that if one transforms $K$ as in eqn. (\ref{delbo}), this simply leads to a reparametrization of the integral curves.
  We want to show that the integral curves are null geodesics. Every point in $Y$ lies on a unique integral curve, so this will give a family of null geodesics with each point in $Y$ contained in a unique member of the family. One describes this by saying that $Y$ is  ``ruled by null geodesics.''

To show that the integral curves are geodesics, it suffices to show that
\be\label{zon}K^\mu D_\mu K^\nu = w K^\nu,\ee
for some function $w$.  
 For then, by a transformation $K^\mu\to e^f  K^\mu$, we can reduce to the case that
\be\label{plon} K^\mu D_\mu K^\nu = 0,\ee
which is equivalent to the geodesic equation for the integral curves.

For $p\in Y$, let $TM|_p$ be the tangent space to $p$ in $M$, and let $TY|_p$ be the tangent space to $p$ in $Y$. A vector in $TM|_p$ is a multiple of $K$ if and only if it is orthogonal to $TY|_p$. So to verify eqn. (\ref{zon})  at $p$, it suffices to show that if $Z$ is any vector in $TY|_p$, then
\be\label{lfelgo} g_{\alpha\beta} (K^\mu D_\mu K^\alpha) Z^\beta  = 0\ee at $p$. We extend $Z$ away from $ p$ in an arbitrary fashion, requiring only that it is tangent to $Y$ . Then $K$
and $Z$ are both vector fields tangent to $Y$, so their commutator, which is 
\be\label{pleffo} Q^\alpha = K^\mu D_\mu Z^\alpha - Z^\mu D_\mu K^\alpha,\ee
is also tangent to $Y$ . So eqn. (\ref{zoff}) holds if $Z$ is replaced by $Q$:
\be\label{izoff} g_{\mu\nu}K^\mu Q^\nu = 0.\ee

Differentiating eqn. (\ref{zoff}), we get
\be\label{droff} 0=K^\mu D_\mu(g_{\alpha\beta}K^\alpha Z^\beta)= g_{\alpha\beta}(K^\mu D_\mu K^\alpha) Z^\beta +g_{\alpha\beta}K^\alpha K^\mu D_\mu Z^\beta.\ee
But the last term on the right hand side vanishes, since
\be\label{miroof} g_{\alpha\beta}K^\alpha K^\mu D_\mu Z^\beta=g_{\alpha\beta}K^\alpha Q^\beta+\frac{1}{2}Z^\mu D_\mu(g_{\alpha\beta}K^\alpha K^\beta)=0, \ee
where we use the fact that $Q$ satisfies eqn. (\ref{izoff})  and $K$ is everywhere null. Thus eqn. (\ref{droff})  is equivalent to the desired result (\ref{lfelgo}).

If $W$ is any spacelike surface that is of codimension 1 in $Y$, then the integral curves are orthogonal to $W$ and so $Y$ is ruled by 
a family of null geodesics that are orthogonal to $W$, the situation considered in section \ref{nullraych}.

\section{How Promptness Fails}\label{failure}

Let $\gamma$ be a future-going null geodesic that originates at a point $q$ in a globally hyperbolic spacetime $M$.    As explained in section \ref{prompt}, in a strongly causal spacetime, an initial segment
of $\gamma$ is prompt.   It may be that $\gamma$ remains prompt no matter how far it is continued into the future.   Otherwise, the prompt portion of $\gamma$ is
an initial segment.   Let $p$ be the endpoint of this segment.    We would like to know what is happening at $p$.  An analogous question concerns the case that $\gamma$ is a null geodesic
orthogonal to a compact codimension 2 spacelike surface $W$.   These questions have been analyzed respectively in  \cite{Ehr} and \cite{ab}.    We will sketch the arguments, starting with the case of 
geodesics that originate at a specified point $q$.

Let $p_1,p_2,\cdots$ be a sequence of points that are to the future of $p$ along $\gamma$, and approach $p$.   The segment $qp_i$ of $\gamma$ is non-prompt for each $i$, so there
exists a strictly timelike curve $\gamma_i$ from $q$ to $p_i$.   In a globally hyperbolic spacetime, we can choose the $\gamma_i$ to maximize the elapsed proper time, and thus
they are timelike geodesics.    Moreover, after possibly passing to a subsequence, we can assume that the $\gamma_i$ converge to a causal curve $\bar\gamma$
from $q$ to $p$.  There are two cases: (1) $\bar\gamma\not=\gamma$; (2) $\bar\gamma=\gamma$.

In case (1), $\bar\gamma$ is a null geodesic, since otherwise it could be deformed to a causal path from $q$ that arrives to the past of $p$, implying that $\gamma$ does not remain prompt
up to $p$.   So case (1) means that at the point $p$, $\gamma$ intersects a second null geodesic that also originates at $q$.  This is perfectly possible.   For example, when gravitational
lensing produces multiple images of the same supernova explosion, there will be some observers who see the first two  images  appear simultaneously.  Note that in case (1), the geodesic
$\gamma$ is always nonprompt when continued to any point $p'$ that is beyond $p$, because by going from $q$ to $p$ along $\bar\gamma$ and then from $p$ to $p'$ along $\gamma$, one gets
a causal path from $q$ to $p'$ that is not a null geodesic.   We leave it to the reader to verify that in case (1), $\bar\gamma$ is also prompt up to $p$, but no farther.

We want to show that case (2) is the main failure mode that has been studied in this paper: it means that to first order, there is a null geodesic deformation of the segment $qp$,
or in other words there is a Jacobi field along $\gamma$ that vanishes at $q$ and $p$.  

To do this, we will construct a convenient set of $\D$ functions on $M$, and ask if they are a good set of coordinates in a small neighborhood of $p$.   We will make use of the Fermi normal coordinates of eqn. (\ref{kofflox}),
with $\gamma$ being the usual geodesic $V=X^A=0$, and $q$ the point $U=0$ on this geodesic.  We consider the geodesic equation with affine parameter $\lambda$
\be\label{uncup} \frac{D^2 X^\mu}{D\lambda^2}=0 .\ee
Thus $\gamma$ is the solution $U=\lambda$, $V=X^A=0$.  If we expand $U=U+\veps u$, $V=\veps v$, $X^A=\veps x^A$, then to first order in $\veps$, the equation of geodesic deviation gives
just
\be\label{wuncup} \frac{\d^2 v}{\d\lambda^2}=0. \ee  There is no curvature term here, because along $\gamma$, $R^V{}_{UAU}=R_{UUAU}=0$.   
(In section \ref{igd}, we did not consider this equation, because we considered only null deformations that were constrained to have $v=0$.)

We consider the geodesic equation (\ref{uncup}) with the initial conditions
\begin{align}\label{guncup} U(0)=V(0)=X^A(0)& = 0 \cr
 \left.\frac{\d U}{\d\lambda}\right|_{\lambda=0}&=1  \cr
\left. \frac{\d V}{\d\lambda}\right|_{\lambda=0}&= \sv\cr
 \left.\frac{\d X^A}{\d\lambda}\right|_{\lambda=0}&=\sx^A. \end{align}
The geodesic with $\sv=\sx^A=0$ is the original $\gamma$, and reaches the point $p$ at some value $\lambda=\lambda_1$.  We map a triple $\lambda, \sv, \sx^B$ to the point with coordinates
 $V(\lambda;\sv,\sx^B)$, $U(\lambda;\sv,\sx^B)$, $X^A(\lambda;\sv,\sx^B)$, where those functions are determined by solving the geodesic equation with the initial conditions (\ref{guncup}).
   We ask whether $\lambda, \sv, $ and $\sx^B$ are a good set of coordinates for $M$ in a neighborhood of $p$, that is in a neighborhood
of $(\lambda,\sv, \sx^B)=(\lambda_1,0,0)$, which corresponds to $p$.

In case (2), the answer to this question is ``no,'' because of the following facts.  For each $i$, $\gamma$ and $\gamma_i$ are two different geodesics from $q$ to $p_i$. 
As the $\gamma_i$ converge to $\gamma$ for $i\to\infty$, the initial conditions of $\gamma_i$ converge for large $i$  to the initial conditions $\sv=\sx^B=0$ of $\gamma$.  Likewise, the
affine parameter at which $\gamma_i$ reaches $p_i$  converges to the corresponding value for $\gamma$.   Thus there are pairs of triples $(\lambda,\sv,\sx^B)$ that are arbitrarily
close to each other and to $(\lambda_1,0,0)$ and represent the same point in $M$; hence there is no neighborhood of $(\lambda_1,0,0)$ in which $(\lambda,\sv,\sx^B)$ are good coordinates.

On the other hand, in general, suppose that $X^\mu,~\mu=1,\cdots,\D$ are a good set of coordinates on a manifold 
$M$ in a neighborhood of a point $p$, and that we parametrize them by variables $f^\alpha,\,
\alpha=1,\cdots ,\D$,
specifying $X^\mu=X^\mu(f^\alpha)$ as smooth functions of the $f^\alpha$.   
The condition for the $f^\alpha$ to be a good set of coordinates on $M$ in
a neighborhood of  $p$ is that $\Delta=\det \partial X^\mu/\partial f^\alpha$ is nonzero at  $p$.    Let us compute this
determinant for the case that the $X^\mu$ are $(U,V,X^A)$, the $f^\alpha$ are $(\lambda,\sv,\sx^B)$, and $p$ is the point $(\lambda_1,0,0)$.

First of all, at $p$, we have 
\be\label{govo}\frac{\partial V}{\partial \sv}\not=0, ~~~~~~\frac{\partial V}{\partial\lambda}=\frac{\partial V}{\partial \sx^A}=0, \ee
Eqn. (\ref{govo}) is true because equation (\ref{wuncup}) together with the initial conditions give simply $V=\lambda \sv$ to first order in $\sv$ and $\sx^B$.   So at $p$, $\partial V/\partial \sv\not=0$
and the other derivatives of $V$ vanish.
Therefore, the determinant $\Delta$ vanishes at $p$ if and only if a reduced determinant with $V$ and $\sv$ omitted vanishes at $p$.
  In other words, we can set $V=\sv=0$, and take  the $X^\mu$ to be just $U$ and $X^A$ and the $f^\alpha$ to be just $\lambda$ and $\sx^B$. 

Similarly, at $\sx^B=0$, we have 
\be\label{lovo} \frac{\partial U}{\partial\lambda}=1, ~~~\frac{\partial X^A}{\partial\lambda}=0, \ee
since the solution of the geodesic equation with $\sv=\sx^B=0$ is $U=\lambda$, $V=X^A=0$.    This means that the original determinant $\Delta$ vanishes if and only
if a reduced determinant vanishes in which we consider only $X^A$ and $\sx^B$, at $\sv=0$, $\lambda=\lambda_1$.

In other words, the condition for vanishing of $\Delta$ is that
\be\label{wackor}\left.\det\frac{\partial X^A(\lambda_1,0,\vec \sx)}{\partial \sx^B}\right|_{\vec\sx=0} =0 .\ee
But the vanishing of this determinant is equivalent to the existence of a Jacobi field on the interval $[0,\lambda_1]$ that vanishes at the endpoints of the interval.  Indeed, vanishing of the determinant of
 a matrix
means that some linear combination of its columns vanishes, say
\be\label{ackor}\sum_B\left.\frac{\partial X^A(\lambda_1,0,\vec\sx)}{\partial \sx^B}\right|_{\vec \sx=0}c^B =0 ,\ee
with constants $c^B$.   If we set $\sx^A=\veps c^A,\,\sv=0$ in eqn. (\ref{guncup}), then to first order in $\veps$, the solution of the geodesic equation will be a Jacobi field that vanishes
at $\lambda=0$ because of the initial conditions  and at $\lambda=\lambda_1$ because of eqn. (\ref{ackor}).

One can study in a similar way a timelike, rather than null, geodesic $\gamma$ from a point $q$.   A short initial segment of $\gamma$ maximizes the elapsed proper time.  As $\gamma$ is continued
into the future, it may fail to be proper time maximizing beyond some point $p$. An argument similar to what we have just described shows that for
 this to happen, either $\gamma$ meets at $p$ another timelike geodesic from $q$ with the same proper time,
or there is a Jacobi field on $\gamma$ that vanishes at $q$ and at $p$, so that $p$ is a focal point of the timelike geodesics from $q$.

Now let us consider the case of a null geodesic $\gamma$ that is orthogonal to a compact codimension 2 spacelike surface $W$ \cite{ab}.   Again, if $\gamma$ does not remain prompt indefinitely,
we define $p$ as the future boundary of the largest prompt segment of $\gamma$, and consider a sequence of points $p_i\in\gamma$ that approach $p_i$ from the future.   The $p_i$ can
now be reached from $W$ by timelike geodesics $\gamma_i$ that are orthogonal to $W$.   After possibly passing to a subsequence, we can assume that the $\gamma_i$ converge 
to an orthogonal null geodesic
$\bar\gamma$ from $W$ to $p$.  We have the same alternatives as before: (1) $\bar\gamma\not=\gamma$; (2) $\bar\gamma=\gamma$. Case (1) now means that there is some other orthogonal
null geodesic $\bar\gamma$ from $W$ to $p$.    This is perfectly possible, and when it happens, $\gamma$ is not prompt when continued beyond $p$.  (Moreover, $\bar\gamma$ is prompt up to
$p$ and no farther.)

As before, case (2) is the main failure mode that has been studied in this paper.   To show this, one repeats the previous argument, now with different initial conditions on the geodesics.
As before, let $\gamma$ be the geodesic $U=\lambda$, $V=X^A=0$,
 and suppose that as in section \ref{ong}, $W$ is described near $\gamma$ by $V=f(X^A)$,  $U=h(X^A)$ where $f$ and $h$ vanish near $X^A=0$ up to second order.   Thus
$\gamma\cap W$ is the point $q$ with $U=0$.    Now we introduce a $(\D-1)$-parameter
family of geodesics orthogonal to $W$.   For initial conditions on these geodesics, we take
\be\label{unv} X^A(0)=\sx^A,~~V(0)=f(\sx^A),~~ U(0) = h(\sx^A), \ee
so that the geodesics originate on $W$ at $\lambda=0$.    We also set
\be\label{punv}\left. \frac{\partial U}{\partial \lambda}\right|_{\lambda=0}=1,  ~~~\left.\frac{\partial V}{\partial \lambda}\right|_{\lambda=0}=\zeta, \ee
where $\zeta$ is a free parameter, and constrain $\d X^A/\d\lambda|_{\lambda=0}$ so that the geodesics in this family are orthogonal to $W$.
Similarly to (\ref{onoogo}), except that now we have to include $\zeta$, the  condition of orthogonality is
\be\label{lunv}\left.\frac{\d  X^A}{\d\lambda}\right|_{\lambda=0} -\partial_A  f(\sx)-\zeta \,\partial_A h(\sx)=0. \ee
We now ask whether we can use $\sx^A$, $\zeta$, and $\lambda$ as a good set of coordinates near $p$.   For the same reason as before, the answer is ``no'' in case (2).   On the other hand,
an analysis similar to the previous one shows that a ``no'' answer to this question means that there is a Jacobi field along $\gamma$ that obeys the boundary
condition at $q$ and vanishes at $p$, or in other words
that $p$ is a focal point for the null geodesics orthogonal to $W$.

An analog of this problem for timelike geodesics is to consider a future-going timelike geodesic $\gamma$ orthogonal to a codimension 1 spacelike surface $S$.    A short initial segment of $\gamma$
maximizes the elapsed proper time to its endpoint.    If this fails when $\gamma$ is continued into the future, let $p$ be the endpoint of the maximal segment on which $\gamma$ is proper time
maximizing.   Then either there is a second timelike orthogonal geodesic $\bar\gamma$ from $S$ to $p$, with the same proper time as $\gamma$, or there is a Jacobi field along $\gamma$
that preserves the orthogonality at  $\gamma\cap S$ and vanishes at $p$, so that $p$ is a focal point of the orthogonal timelike geodesics from $S$.

\end{appendix}

\vskip1cm
\noindent{\it Acknowledgments}

Lectures on some of this material were  presented at the 2018 summer program Prospects in Theoretical Physics
 at the Institute for Advanced Study in Princeton, NJ, and the January, 2020 summer
school of the New Zealand Mathematical Research Institute (NZMRI), held in Nelson, New Zealand.  Both audiences provided useful feedback.
 The NZMRI summer school lectures were  supported  by the Clay Mathematical Institute.
Research was also supported in part by NSF Grant PHY-1606531.  I thank G. Galloway, A. Kar, R. Wald, and A. Wall   for helpful 
comments and suggested improvements and for pointing out useful references, M. Turansick for assistance with the manuscript, and Tong Chen and  Zi-hao Li for pointing out some typographical errors.
\bibliographystyle{unsrt}

\end{document}